\documentclass[runningheads,a4paper,final]{llncs}
\usepackage{stfloats}
\fnbelowfloat
\usepackage[T1]{fontenc}
\usepackage[utf8]{inputenc}
\usepackage{textcomp}
\usepackage{cite}
\usepackage{color}

\usepackage[final]{graphicx}
\graphicspath{{figures/}}
\usepackage{realboxes}

\usepackage{hyperref}
\hypersetup{%
    final=true,
    colorlinks=true,
    urlcolor=black,
    linkcolor=black,
    citecolor=black,
    pdfborder={0 0 0},
    pdfauthor={Timothy Bourke \& Robert J.\ van Glabbeek \& Peter H\"ofner},
    pdftitle={Showing invariance compositionally for a process algebra for 
    network protocols},
    pdfsubject={Mechannized compositional invariant proofs of process 
    algebra models},
    pdfkeywords={Isabelle/HOL, process algebra, reactive systems, invariance 
    proofs, network protocols},
}
\usepackage{breakurl} % for URL-hyphenation when running in dvi-mode.

\usepackage{amssymb}
\usepackage{latexsym}
\usepackage{stmaryrd}

\usepackage{mathpartir}
\mprset{sep=1.10em}  % hor. sep. between premises

\usepackage{url}
\usepackage{subfig}
\usepackage{extarrows}
\usepackage{relsize}

\usepackage[printonlyused]{acronym}
\newcommand{\acworkaround}[1]{{\acsfont {#1}}}

\usepackage{isabelle, isabellesym}
\isabellestyle{tim}
\renewcommand{\isasymAnd}{\isamath{{\mathsmaller{\bigwedge}}}}

\definecolor{gray}{rgb}{0.6,0.6,0.6}
\newcommand{\gray}[1]{\textcolor{gray}{#1}}

\newcommand{\snip}[4]
  {\expandafter\newcommand\csname snippet--#1\endcsname{#4}}
\newcommand{\snippet}[1]{\csname snippet--#1\endcsname}
\newcommand{\msnippet}[1]{\mbox{\csname snippet--#1\endcsname}}

\snip{toy_init}{0}{0}{\isa{{\isasymlparr}ip\ {\isacharequal}\ i,\ no\ {\isacharequal}\ {\isadigit{0}},\ nhip\ {\isacharequal}\ i,\ msg\ {\isacharequal}\ SOME\ x{\isachardot}\ True,\ num\ {\isacharequal}\ SOME\ x{\isachardot}\ True,\ sip\ {\isacharequal}\ SOME\ x{\isachardot}\ True{\isasymrparr}}}
\snip{ptoy_lhs}{0}{0}{\isa{ptoy\ i}}
\snip{ptoy_rhs}{0}{0}{\isa{{\isasymlparr}init\ {\isacharequal}\ {\isacharbraceleft}{\isacharparenleft}toy{\isacharunderscore}init\ i,\ \gammatoy\ PToy{\isacharparenright}{\isacharbraceright},\ trans\ {\isacharequal}\ \seqpsos\ \gammatoy{\isasymrparr}}}

\snip{optoy_lhs}{0}{0}{\isa{optoy\ i}}
\snip{optoy_rhs}{0}{0}{\isa{{\isasymlparr}init\ {\isacharequal}\ {\isacharbraceleft}{\isacharparenleft}toy{\isacharunderscore}init,\ \gammatoy\ PToy{\isacharparenright}{\isacharbraceright},\ trans\ {\isacharequal}\ \oseqpsos\ \gammatoy\ i{\isasymrparr}}}

\snip{pkt}{0}{0}{\isa{pkt\ {\isacharparenleft}data,\ src{\isacharparenright}}}
\snip{newpkt}{0}{0}{\isa{newpkt\ {\isacharparenleft}data,\ dst{\isacharparenright}}}

\snip{automaton_type}{0}{0}{\isa{{\isacharparenleft}{\isacharprime}s,\ {\isacharprime}a{\isacharparenright}\ automaton}}

\snip{seqp_choice}{0}{0}{\isa{p\isactrlsub {\isadigit{1}}\ {\isasymoplus}\ p\isactrlsub {\isadigit{2}}}}
\snip{seqp_call}{0}{0}{\isa{call{\isacharparenleft}pn{\isacharparenright}}}

\snip{lseqp_guard}{0}{0}{\isa{\gray{{\isacharbraceleft}l{\isacharbraceright}}{\isasymlangle}g{\isasymrangle}\ p}}
\snip{lseqp_assign}{0}{0}{\isa{\gray{{\isacharbraceleft}l{\isacharbraceright}}{\isasymlbrakk}u{\isasymrbrakk}\ p}}
\snip{lseqp_ucast}{0}{0}{\isa{\gray{{\isacharbraceleft}l{\isacharbraceright}}unicast{\isacharparenleft}\selip,\ \selmsg{\isacharparenright}\ {\isachardot}\ p\ {\isasymtriangleright}\ q}}
\snip{lseqp_bcast}{0}{0}{\isa{\gray{{\isacharbraceleft}l{\isacharbraceright}}broadcast{\isacharparenleft}\selmsg{\isacharparenright}\ {\isachardot}\ p}}
\snip{lseqp_gcast}{0}{0}{\isa{\gray{{\isacharbraceleft}l{\isacharbraceright}}groupcast{\isacharparenleft}\selips,\ \selmsg{\isacharparenright}\ {\isachardot}\ p}}
\snip{lseqp_send}{0}{0}{\isa{\gray{{\isacharbraceleft}l{\isacharbraceright}}send{\isacharparenleft}\selmsg{\isacharparenright}\ {\isachardot}\ p}}
\snip{lseqp_deliver}{0}{0}{\isa{\gray{{\isacharbraceleft}l{\isacharbraceright}}deliver{\isacharparenleft}\seldata{\isacharparenright}\ {\isachardot}\ p}}
\snip{lseqp_receive}{0}{0}{\isa{\gray{{\isacharbraceleft}l{\isacharbraceright}}receive{\isacharparenleft}\updmsg{\isacharparenright}\ {\isachardot}\ p}}

\snip{seqp_guard}{0}{0}{\isa{{\isasymlangle}g{\isasymrangle}\ p}}
\snip{seqp_assign}{0}{0}{\isa{{\isasymlbrakk}u{\isasymrbrakk}\ p}}
\snip{seqp_ucast}{0}{0}{\isa{unicast{\isacharparenleft}\selip,\ \selmsg{\isacharparenright}\ {\isachardot}\ p\ {\isasymtriangleright}\ q}}
\snip{seqp_bcast}{0}{0}{\isa{broadcast{\isacharparenleft}\selmsg{\isacharparenright}\ {\isachardot}\ p}}
\snip{seqp_gcast}{0}{0}{\isa{groupcast{\isacharparenleft}\selips,\ \selmsg{\isacharparenright}\ {\isachardot}\ p}}
\snip{seqp_send}{0}{0}{\isa{send{\isacharparenleft}\selmsg{\isacharparenright}\ {\isachardot}\ p}}
\snip{seqp_deliver}{0}{0}{\isa{deliver{\isacharparenleft}\seldata{\isacharparenright}\ {\isachardot}\ p}}
\snip{seqp_receive}{0}{0}{\isa{receive{\isacharparenleft}\updmsg{\isacharparenright}\ {\isachardot}\ p}}

\snip{seqp_guard_map}{0}{0}{\isa{{\isacharprime}s\ {\isasymRightarrow}\ {\isacharprime}s\ set}}
\snip{seqp_assign_map}{0}{0}{\isa{{\isacharprime}s\ {\isasymRightarrow}\ {\isacharprime}s}}
\snip{seqp_cast_ip_map}{0}{0}{\isa{{\isacharprime}s\ {\isasymRightarrow}\ ip}}
\snip{seqp_cast_msg_map}{0}{0}{\isa{{\isacharprime}s\ {\isasymRightarrow}\ {\isacharprime}m}}
\snip{seqp_cast_ip_set_map}{0}{0}{\isa{{\isacharprime}s\ {\isasymRightarrow}\ ip\ set}}

\snip{gamma_fun}{0}{0}{\isa{{\isasymGamma}{\isasymColon}{\isacharprime}p\ {\isasymRightarrow}\ {\isacharparenleft}{\isacharprime}s,\ {\isacharprime}m,\ {\isacharprime}p,\ {\isacharprime}l{\isacharparenright}\ seqp}}
\snip{wellformed_gamma}{0}{0}{\isa{wellformed\ {\isasymGamma}}}

\snip{broadcastT}{0}{0}{\isa{{\isacharparenleft}{\isacharparenleft}{\isasymxi},\ \gray{{\isacharbraceleft}l{\isacharbraceright}}broadcast{\isacharparenleft}\selmsg{\isacharparenright}\ {\isachardot}\ p{\isacharparenright},\ broadcast\ {\isacharparenleft}\selmsg\ {\isasymxi}{\isacharparenright},\ {\isacharparenleft}{\isasymxi},\ p{\isacharparenright}{\isacharparenright}{\isasymin}\seqpsos\ {\isasymGamma}}}
\snip{groupcastT}{0}{0}{\isa{{\isacharparenleft}{\isacharparenleft}{\isasymxi},\ \gray{{\isacharbraceleft}l{\isacharbraceright}}groupcast{\isacharparenleft}\selips,\ \selmsg{\isacharparenright}\ {\isachardot}\ p{\isacharparenright},\ groupcast\ {\isacharparenleft}\selips\ {\isasymxi}{\isacharparenright}\ {\isacharparenleft}\selmsg\ {\isasymxi}{\isacharparenright},\ {\isacharparenleft}{\isasymxi},\ p{\isacharparenright}{\isacharparenright}{\isasymin}\seqpsos\ {\isasymGamma}}}
\snip{unicastT}{0}{0}{\isa{{\isacharparenleft}{\isacharparenleft}{\isasymxi},\ \gray{{\isacharbraceleft}l{\isacharbraceright}}unicast{\isacharparenleft}\selip,\ \selmsg{\isacharparenright}\ {\isachardot}\ p\ {\isasymtriangleright}\ q{\isacharparenright},\ unicast\ {\isacharparenleft}\selip\ {\isasymxi}{\isacharparenright}\ {\isacharparenleft}\selmsg\ {\isasymxi}{\isacharparenright},\ {\isacharparenleft}{\isasymxi},\ p{\isacharparenright}{\isacharparenright}{\isasymin}\seqpsos\ {\isasymGamma}}}
\snip{notunicastT}{0}{0}{\isa{{\isacharparenleft}{\isacharparenleft}{\isasymxi},\ \gray{{\isacharbraceleft}l{\isacharbraceright}}unicast{\isacharparenleft}\selip,\ \selmsg{\isacharparenright}\ {\isachardot}\ p\ {\isasymtriangleright}\ q{\isacharparenright},\ {\isasymnot}unicast\ {\isacharparenleft}\selip\ {\isasymxi}{\isacharparenright},\ {\isacharparenleft}{\isasymxi},\ q{\isacharparenright}{\isacharparenright}{\isasymin}\seqpsos\ {\isasymGamma}}}
\snip{sendT}{0}{0}{\isa{{\isacharparenleft}{\isacharparenleft}{\isasymxi},\ \gray{{\isacharbraceleft}l{\isacharbraceright}}send{\isacharparenleft}\selmsg{\isacharparenright}\ {\isachardot}\ p{\isacharparenright},\ send\ {\isacharparenleft}\selmsg\ {\isasymxi}{\isacharparenright},\ {\isacharparenleft}{\isasymxi},\ p{\isacharparenright}{\isacharparenright}{\isasymin}\seqpsos\ {\isasymGamma}}}
\snip{deliverT}{0}{0}{\isa{{\isacharparenleft}{\isacharparenleft}{\isasymxi},\ \gray{{\isacharbraceleft}l{\isacharbraceright}}deliver{\isacharparenleft}\seldata{\isacharparenright}\ {\isachardot}\ p{\isacharparenright},\ deliver\ {\isacharparenleft}\seldata\ {\isasymxi}{\isacharparenright},\ {\isacharparenleft}{\isasymxi},\ p{\isacharparenright}{\isacharparenright}{\isasymin}\seqpsos\ {\isasymGamma}}}
\snip{receiveT}{0}{0}{\isa{{\isacharparenleft}{\isacharparenleft}{\isasymxi},\ \gray{{\isacharbraceleft}l{\isacharbraceright}}receive{\isacharparenleft}\updmsg{\isacharparenright}\ {\isachardot}\ p{\isacharparenright},\ receive\ msg,\ {\isacharparenleft}\updmsg\ msg\ {\isasymxi},\ p{\isacharparenright}{\isacharparenright}{\isasymin}\seqpsos\ {\isasymGamma}}}
\snip{assignT}{0}{0}{\isa{{\isacharparenleft}{\isacharparenleft}{\isasymxi},\ \gray{{\isacharbraceleft}l{\isacharbraceright}}{\isasymlbrakk}u{\isasymrbrakk}\ p{\isacharparenright},\ {\isasymtau},\ {\isacharparenleft}u\ {\isasymxi},\ p{\isacharparenright}{\isacharparenright}{\isasymin}\seqpsos\ {\isasymGamma}}}
\snip{assignT'}{0}{0}{\isa{\mbox{}\inferrule{\mbox{{\isasymxi}{\isacharprime}\ {\isacharequal}\ u\ {\isasymxi}}}{\mbox{{\isacharparenleft}{\isacharparenleft}{\isasymxi},\ \gray{{\isacharbraceleft}l{\isacharbraceright}}{\isasymlbrakk}u{\isasymrbrakk}\ p{\isacharparenright},\ {\isasymtau},\ {\isacharparenleft}{\isasymxi}{\isacharprime},\ p{\isacharparenright}{\isacharparenright}{\isasymin}\seqpsos\ {\isasymGamma}}}}}
\snip{callT}{0}{0}{\isa{\mbox{}\inferrule{\mbox{{\isacharparenleft}{\isacharparenleft}{\isasymxi},\ {\isasymGamma}\ pn{\isacharparenright},\ a,\ {\isacharparenleft}{\isasymxi}{\isacharprime},\ p{\isacharprime}{\isacharparenright}{\isacharparenright}{\isasymin}\seqpsos\ {\isasymGamma}}}{\mbox{{\isacharparenleft}{\isacharparenleft}{\isasymxi},\ call{\isacharparenleft}pn{\isacharparenright}{\isacharparenright},\ a,\ {\isacharparenleft}{\isasymxi}{\isacharprime},\ p{\isacharprime}{\isacharparenright}{\isacharparenright}{\isasymin}\seqpsos\ {\isasymGamma}}}}}
\snip{choiceT1}{0}{0}{\isa{\mbox{}\inferrule{\mbox{{\isacharparenleft}{\isacharparenleft}{\isasymxi},\ p{\isacharparenright},\ a,\ {\isacharparenleft}{\isasymxi}{\isacharprime},\ p{\isacharprime}{\isacharparenright}{\isacharparenright}{\isasymin}\seqpsos\ {\isasymGamma}}}{\mbox{{\isacharparenleft}{\isacharparenleft}{\isasymxi},\ p\ {\isasymoplus}\ q{\isacharparenright},\ a,\ {\isacharparenleft}{\isasymxi}{\isacharprime},\ p{\isacharprime}{\isacharparenright}{\isacharparenright}{\isasymin}\seqpsos\ {\isasymGamma}}}}}
\snip{choiceT2}{0}{0}{\isa{\mbox{}\inferrule{\mbox{{\isacharparenleft}{\isacharparenleft}{\isasymxi},\ q{\isacharparenright},\ a,\ {\isacharparenleft}{\isasymxi}{\isacharprime},\ q{\isacharprime}{\isacharparenright}{\isacharparenright}{\isasymin}\seqpsos\ {\isasymGamma}}}{\mbox{{\isacharparenleft}{\isacharparenleft}{\isasymxi},\ p\ {\isasymoplus}\ q{\isacharparenright},\ a,\ {\isacharparenleft}{\isasymxi}{\isacharprime},\ q{\isacharprime}{\isacharparenright}{\isacharparenright}{\isasymin}\seqpsos\ {\isasymGamma}}}}}
\snip{guardT}{0}{0}{\isa{\mbox{}\inferrule{\mbox{{\isasymxi}{\isacharprime}{\isasymin}g\ {\isasymxi}}}{\mbox{{\isacharparenleft}{\isacharparenleft}{\isasymxi},\ \gray{{\isacharbraceleft}l{\isacharbraceright}}{\isasymlangle}g{\isasymrangle}\ p{\isacharparenright},\ {\isasymtau},\ {\isacharparenleft}{\isasymxi}{\isacharprime},\ p{\isacharparenright}{\isacharparenright}{\isasymin}\seqpsos\ {\isasymGamma}}}}}

\snip{parleft}{0}{0}{\isa{\mbox{}\inferrule{\mbox{{\isacharparenleft}s,\ a,\ s{\isacharprime}{\isacharparenright}{\isasymin}S}\\\ \mbox{{\isasymAnd}m{\isachardot}\ a\ {\isasymnoteq}\ receive\ m}}{\mbox{{\isacharparenleft}{\isacharparenleft}s,\ t{\isacharparenright},\ a,\ {\isacharparenleft}s{\isacharprime},\ t{\isacharparenright}{\isacharparenright}{\isasymin}\parpsos\ S\ T}}}}
\snip{parright}{0}{0}{\isa{\mbox{}\inferrule{\mbox{{\isacharparenleft}t,\ a,\ t{\isacharprime}{\isacharparenright}{\isasymin}T}\\\ \mbox{{\isasymAnd}m{\isachardot}\ a\ {\isasymnoteq}\ send\ m}}{\mbox{{\isacharparenleft}{\isacharparenleft}s,\ t{\isacharparenright},\ a,\ {\isacharparenleft}s,\ t{\isacharprime}{\isacharparenright}{\isacharparenright}{\isasymin}\parpsos\ S\ T}}}}
\snip{parboth}{0}{0}{\isa{\mbox{}\inferrule{\mbox{{\isacharparenleft}s,\ receive\ m,\ s{\isacharprime}{\isacharparenright}{\isasymin}S}\\\ \mbox{{\isacharparenleft}t,\ send\ m,\ t{\isacharprime}{\isacharparenright}{\isasymin}T}}{\mbox{{\isacharparenleft}{\isacharparenleft}s,\ t{\isacharparenright},\ {\isasymtau},\ {\isacharparenleft}s{\isacharprime},\ t{\isacharprime}{\isacharparenright}{\isacharparenright}{\isasymin}\parpsos\ S\ T}}}}

\snip{par_comp}{0}{0}{\isa{s\ {\isasymlangle}{\isasymlangle}\ t\ {\isacharequal}\ {\isasymlparr}init\ {\isacharequal}\ init\ s\ {\isasymtimes}\ init\ t,\ trans\ {\isacharequal}\ \parpsos\ {\isacharparenleft}trans\ s{\isacharparenright}\ {\isacharparenleft}trans\ t{\isacharparenright}{\isasymrparr}}}
\snip{par_comp_term}{0}{0}{\isa{s\ {\isasymlangle}{\isasymlangle}\ t}}
\snip{par_comp_type}{0}{0}{\isa{{\isacharparenleft}{\isacharprime}s\ {\isasymtimes}\ {\isacharprime}p,\ {\isacharprime}m\ seq{\isacharunderscore}action{\isacharparenright}\ automaton}}

\snip{node_bcast}{0}{0}{\isa{\mbox{}\inferrule{\mbox{{\isacharparenleft}s,\ broadcast\ m,\ s{\isacharprime}{\isacharparenright}{\isasymin}S}}{\mbox{{\isacharparenleft}\NodeS\ i\ s\ R,\ R{\isacharcolon}{\isacharasterisk}cast{\isacharparenleft}m{\isacharparenright},\ \NodeS\ i\ s{\isacharprime}\ R{\isacharparenright}{\isasymin}\nodesos\ S}}}}
\snip{node_gcast}{0}{0}{\isa{\mbox{}\inferrule{\mbox{{\isacharparenleft}s,\ groupcast\ D\ m,\ s{\isacharprime}{\isacharparenright}{\isasymin}S}}{\mbox{{\isacharparenleft}\NodeS\ i\ s\ R,\ {\isacharparenleft}R\ {\isasyminter}\ D{\isacharparenright}{\isacharcolon}{\isacharasterisk}cast{\isacharparenleft}m{\isacharparenright},\ \NodeS\ i\ s{\isacharprime}\ R{\isacharparenright}{\isasymin}\nodesos\ S}}}}
\snip{node_ucast}{0}{0}{\isa{\mbox{}\inferrule{\mbox{{\isacharparenleft}s,\ unicast\ d\ m,\ s{\isacharprime}{\isacharparenright}{\isasymin}S}\\\ \mbox{d{\isasymin}R}}{\mbox{{\isacharparenleft}\NodeS\ i\ s\ R,\ {\isacharbraceleft}d{\isacharbraceright}{\isacharcolon}{\isacharasterisk}cast{\isacharparenleft}m{\isacharparenright},\ \NodeS\ i\ s{\isacharprime}\ R{\isacharparenright}{\isasymin}\nodesos\ S}}}}
\snip{node_notucast}{0}{0}{\isa{\mbox{}\inferrule{\mbox{{\isacharparenleft}s,\ {\isasymnot}unicast\ d,\ s{\isacharprime}{\isacharparenright}{\isasymin}S}\\\ \mbox{d\ {\isasymnotin}\ R}}{\mbox{{\isacharparenleft}\NodeS\ i\ s\ R,\ {\isasymtau},\ \NodeS\ i\ s{\isacharprime}\ R{\isacharparenright}{\isasymin}\nodesos\ S}}}}
\snip{node_deliver}{0}{0}{\isa{\mbox{}\inferrule{\mbox{{\isacharparenleft}s,\ deliver\ d,\ s{\isacharprime}{\isacharparenright}{\isasymin}S}}{\mbox{{\isacharparenleft}\NodeS\ i\ s\ R,\ i{\isacharcolon}deliver{\isacharparenleft}d{\isacharparenright},\ \NodeS\ i\ s{\isacharprime}\ R{\isacharparenright}{\isasymin}\nodesos\ S}}}}
\snip{node_receive}{0}{0}{\isa{\mbox{}\inferrule{\mbox{{\isacharparenleft}s,\ receive\ m,\ s{\isacharprime}{\isacharparenright}{\isasymin}S}}{\mbox{{\isacharparenleft}\NodeS\ i\ s\ R,\ {\isacharbraceleft}i{\isacharbraceright}{\isasymnot}{\isasymemptyset}{\isacharcolon}arrive{\isacharparenleft}m{\isacharparenright},\ \NodeS\ i\ s{\isacharprime}\ R{\isacharparenright}{\isasymin}\nodesos\ S}}}}
\snip{node_tau}{0}{0}{\isa{\mbox{}\inferrule{\mbox{{\isacharparenleft}s,\ {\isasymtau},\ s{\isacharprime}{\isacharparenright}{\isasymin}S}}{\mbox{{\isacharparenleft}\NodeS\ i\ s\ R,\ {\isasymtau},\ \NodeS\ i\ s{\isacharprime}\ R{\isacharparenright}{\isasymin}\nodesos\ S}}}}
\snip{node_arrive}{0}{0}{\isa{{\isacharparenleft}\NodeS\ i\ s\ R,\ {\isasymemptyset}{\isasymnot}{\isacharbraceleft}i{\isacharbraceright}{\isacharcolon}arrive{\isacharparenleft}m{\isacharparenright},\ \NodeS\ i\ s\ R{\isacharparenright}{\isasymin}\nodesos\ S}}
\snip{node_connect1}{0}{0}{\isa{{\isacharparenleft}\NodeS\ i\ s\ R,\ connect{\isacharparenleft}i,\ i{\isacharprime}{\isacharparenright},\ \NodeS\ i\ s\ {R\ {\isasymunion}\ {\isacharbraceleft}i{\isacharprime}{\isacharbraceright}}{\isacharparenright}{\isasymin}\nodesos\ S}}
\snip{node_connect2}{0}{0}{\isa{{\isacharparenleft}\NodeS\ i\ s\ R,\ connect{\isacharparenleft}i{\isacharprime},\ i{\isacharparenright},\ \NodeS\ i\ s\ {R\ {\isasymunion}\ {\isacharbraceleft}i{\isacharprime}{\isacharbraceright}}{\isacharparenright}{\isasymin}\nodesos\ S}}
\snip{node_disconnect1}{0}{0}{\isa{{\isacharparenleft}\NodeS\ i\ s\ R,\ disconnect{\isacharparenleft}i,\ i{\isacharprime}{\isacharparenright},\ \NodeS\ i\ s\ {\isacharparenleft}R\ {\isacharminus}\ {\isacharbraceleft}i{\isacharprime}{\isacharbraceright}{\isacharparenright}{\isacharparenright}{\isasymin}\nodesos\ S}}
\snip{node_disconnect2}{0}{0}{\isa{{\isacharparenleft}\NodeS\ i\ s\ R,\ disconnect{\isacharparenleft}i{\isacharprime},\ i{\isacharparenright},\ \NodeS\ i\ s\ {\isacharparenleft}R\ {\isacharminus}\ {\isacharbraceleft}i{\isacharprime}{\isacharbraceright}{\isacharparenright}{\isacharparenright}{\isasymin}\nodesos\ S}}
\snip{node_connect_other}{0}{0}{\isa{\mbox{}\inferrule{\mbox{i\ {\isasymnoteq}\ i{\isacharprime}}\\\ \mbox{i\ {\isasymnoteq}\ i{\isacharprime}{\isacharprime}}}{\mbox{{\isacharparenleft}\NodeS\ i\ s\ R,\ connect{\isacharparenleft}i{\isacharprime},\ i{\isacharprime}{\isacharprime}{\isacharparenright},\ \NodeS\ i\ s\ R{\isacharparenright}{\isasymin}\nodesos\ S}}}}
\snip{node_disconnect_other}{0}{0}{\isa{\mbox{}\inferrule{\mbox{i\ {\isasymnoteq}\ i{\isacharprime}}\\\ \mbox{i\ {\isasymnoteq}\ i{\isacharprime}{\isacharprime}}}{\mbox{{\isacharparenleft}\NodeS\ i\ s\ R,\ disconnect{\isacharparenleft}i{\isacharprime},\ i{\isacharprime}{\isacharprime}{\isacharparenright},\ \NodeS\ i\ s\ R{\isacharparenright}{\isasymin}\nodesos\ S}}}}

\snip{node_comp}{0}{0}{\isa{{\isasymlangle}i\ {\isacharcolon}\ np\ {\isacharcolon}\ \Ri{\isasymrangle}\ {\isacharequal}\ {\isasymlparr}init\ {\isacharequal}\ {\isacharbraceleft}\NodeS\ i\ s\ \Ri\ {\isacharbar}\ s{\isasymin}init\ np{\isacharbraceright},\ trans\ {\isacharequal}\ \nodesos\ {\isacharparenleft}trans\ np{\isacharparenright}{\isasymrparr}}}
\snip{node_comp_term}{0}{0}{\isa{{\isasymlangle}i\ {\isacharcolon}\ S\ {\isacharcolon}\ \Ri{\isasymrangle}}}
\snip{node_comp_type}{0}{0}{\isa{{\isacharparenleft}{\isacharprime}a\ net{\isacharunderscore}state,\ {\isacharprime}b\ node{\isacharunderscore}action{\isacharparenright}\ automaton}}

\snip{pnet_cast1}{0}{0}{\isa{\mbox{}\inferrule{\mbox{{\isacharparenleft}s,\ R{\isacharcolon}{\isacharasterisk}cast{\isacharparenleft}m{\isacharparenright},\ s{\isacharprime}{\isacharparenright}{\isasymin}S}\\\ \mbox{{\isacharparenleft}t,\ H{\isasymnot}K{\isacharcolon}arrive{\isacharparenleft}m{\isacharparenright},\ t{\isacharprime}{\isacharparenright}{\isasymin}T}\\\ \mbox{H\ {\isasymsubseteq}\ R}\\\ \mbox{K\ {\isasyminter}\ R\ {\isacharequal}\ {\isasymemptyset}}}{\mbox{{\isacharparenleft}\subnets{s}{t},\ R{\isacharcolon}{\isacharasterisk}cast{\isacharparenleft}m{\isacharparenright},\ \subnetsadj{s{\isacharprime}}{t{\isacharprime}}{\isacharparenright}{\isasymin}\pnetsos\ S\ T}}}}
\snip{pnet_cast2}{0}{0}{\isa{\mbox{}\inferrule{\mbox{{\isacharparenleft}s,\ H{\isasymnot}K{\isacharcolon}arrive{\isacharparenleft}m{\isacharparenright},\ s{\isacharprime}{\isacharparenright}{\isasymin}S}\\\ \mbox{{\isacharparenleft}t,\ R{\isacharcolon}{\isacharasterisk}cast{\isacharparenleft}m{\isacharparenright},\ t{\isacharprime}{\isacharparenright}{\isasymin}T}\\\ \mbox{H\ {\isasymsubseteq}\ R}\\\ \mbox{K\ {\isasyminter}\ R\ {\isacharequal}\ {\isasymemptyset}}}{\mbox{{\isacharparenleft}\subnets{s}{t},\ R{\isacharcolon}{\isacharasterisk}cast{\isacharparenleft}m{\isacharparenright},\ \subnets{s{\isacharprime}}{t{\isacharprime}}{\isacharparenright}{\isasymin}\pnetsos\ S\ T}}}}
\snip{pnet_arrive}{0}{0}{\isa{\mbox{}\inferrule{\mbox{{\isacharparenleft}s,\ H{\isasymnot}K{\isacharcolon}arrive{\isacharparenleft}m{\isacharparenright},\ s{\isacharprime}{\isacharparenright}{\isasymin}S}\\\ \mbox{{\isacharparenleft}t,\ H{\isacharprime}{\isasymnot}K{\isacharprime}{\isacharcolon}arrive{\isacharparenleft}m{\isacharparenright},\ t{\isacharprime}{\isacharparenright}{\isasymin}T}}{\mbox{{\isacharparenleft}\subnets{s}{t},\ {\isacharparenleft}H\ {\isasymunion}\ H{\isacharprime}{\isacharparenright}{\isasymnot}{\isacharparenleft}K\ {\isasymunion}\ K{\isacharprime}{\isacharparenright}{\isacharcolon}arrive{\isacharparenleft}m{\isacharparenright},\ \subnetsadj{s{\isacharprime}}{t{\isacharprime}}{\isacharparenright}{\isasymin}\pnetsos\ S\ T}}}}
\snip{pnet_deliver1}{0}{0}{\isa{\mbox{}\inferrule{\mbox{{\isacharparenleft}s,\ i{\isacharcolon}deliver{\isacharparenleft}d{\isacharparenright},\ s{\isacharprime}{\isacharparenright}{\isasymin}S}}{\mbox{{\isacharparenleft}\subnets{s}{t},\ i{\isacharcolon}deliver{\isacharparenleft}d{\isacharparenright},\ \subnets{s{\isacharprime}}{t}{\isacharparenright}{\isasymin}\pnetsos\ S\ T}}}}
\snip{pnet_deliver2}{0}{0}{\isa{\mbox{}\inferrule{\mbox{{\isacharparenleft}t,\ i{\isacharcolon}deliver{\isacharparenleft}d{\isacharparenright},\ t{\isacharprime}{\isacharparenright}{\isasymin}T}}{\mbox{{\isacharparenleft}\subnets{s}{t},\ i{\isacharcolon}deliver{\isacharparenleft}d{\isacharparenright},\ \subnets{s}{t{\isacharprime}}{\isacharparenright}{\isasymin}\pnetsos\ S\ T}}}}
\snip{pnet_tau1}{0}{0}{\isa{\mbox{}\inferrule{\mbox{{\isacharparenleft}s,\ {\isasymtau},\ s{\isacharprime}{\isacharparenright}{\isasymin}S}}{\mbox{{\isacharparenleft}\subnets{s}{t},\ {\isasymtau},\ \subnetsadj{s{\isacharprime}}{t}{\isacharparenright}{\isasymin}\pnetsos\ S\ T}}}}
\snip{pnet_tau2}{0}{0}{\isa{\mbox{}\inferrule{\mbox{{\isacharparenleft}t,\ {\isasymtau},\ t{\isacharprime}{\isacharparenright}{\isasymin}T}}{\mbox{{\isacharparenleft}\subnets{s}{t},\ {\isasymtau},\ \subnets{s}{t{\isacharprime}}{\isacharparenright}{\isasymin}\pnetsos\ S\ T}}}}
\snip{pnet_connect}{0}{0}{\isa{\mbox{}\inferrule{\mbox{{\isacharparenleft}s,\ connect{\isacharparenleft}i,\ i{\isacharprime}{\isacharparenright},\ s{\isacharprime}{\isacharparenright}{\isasymin}S}\\\ \mbox{{\isacharparenleft}t,\ connect{\isacharparenleft}i,\ i{\isacharprime}{\isacharparenright},\ t{\isacharprime}{\isacharparenright}{\isasymin}T}}{\mbox{{\isacharparenleft}\subnets{s}{t},\ connect{\isacharparenleft}i,\ i{\isacharprime}{\isacharparenright},\ \subnets{s{\isacharprime}}{t{\isacharprime}}{\isacharparenright}{\isasymin}\pnetsos\ S\ T}}}}
\snip{pnet_disconnect}{0}{0}{\isa{\mbox{}\inferrule{\mbox{{\isacharparenleft}s,\ disconnect{\isacharparenleft}i,\ i{\isacharprime}{\isacharparenright},\ s{\isacharprime}{\isacharparenright}{\isasymin}S}\\\ \mbox{{\isacharparenleft}t,\ disconnect{\isacharparenleft}i,\ i{\isacharprime}{\isacharparenright},\ t{\isacharprime}{\isacharparenright}{\isasymin}T}}{\mbox{{\isacharparenleft}\subnets{s}{t},\ disconnect{\isacharparenleft}i,\ i{\isacharprime}{\isacharparenright},\ \subnets{s{\isacharprime}}{t{\isacharprime}}{\isacharparenright}{\isasymin}\pnetsos\ S\ T}}}}

\snip{pnet}{0}{0}{\isa{pnet}}
\snip{pnet_node_term}{0}{0}{\isa{{\isasymlangle}i{\isacharsemicolon}\ \Ri{\isasymrangle}}}
\snip{pnet_node_type}{0}{0}{\isa{net{\isacharunderscore}tree}}
\snip{pnet_par_term}{0}{0}{\isa{p\isactrlsub {\isadigit{1}}\parallelcomp{}p\isactrlsub {\isadigit{2}}}}
\snip{pnet_par_type}{0}{0}{\isa{net{\isacharunderscore}tree}}
\snip{net_tree}{0}{0}{\isa{net{\isacharunderscore}tree}}
\snip{net_tree_ips}{0}{0}{\isa{net{\isacharunderscore}tree{\isacharunderscore}ips}}

\snip{act_arrive}{0}{0}{\isa{H{\isasymnot}K{\isacharcolon}arrive{\isacharparenleft}m{\isacharparenright}}}
\snip{act_not_arrive}{0}{0}{\isa{{\isasymemptyset}{\isasymnot}{\isacharbraceleft}i{\isacharbraceright}{\isacharcolon}arrive{\isacharparenleft}m{\isacharparenright}}}
\snip{act_connect}{0}{0}{\isa{connect{\isacharparenleft}i,\ i{\isacharprime}{\isacharparenright}}}
\snip{act_disconnect}{0}{0}{\isa{disconnect{\isacharparenleft}i,\ i{\isacharprime}{\isacharparenright}}}
\snip{act_receive}{0}{0}{\isa{receive\ m}}
\snip{act_cast}{0}{0}{\isa{R{\isacharcolon}{\isacharasterisk}cast{\isacharparenleft}m{\isacharparenright}}}

\snip{net_state}{0}{0}{\isa{{\isacharprime}s\ net{\isacharunderscore}state}}
\snip{net_state_nodes}{0}{0}{\isa{\NodeS\ i\ s\ R}}
\snip{net_state_subnets}{0}{0}{\isa{\subnetsadj{s\isactrlsub {\isadigit{1}}}{s\isactrlsub {\isadigit{2}}}}}

\snip{wf_net_tree}{0}{0}{\isa{wf{\isacharunderscore}net{\isacharunderscore}tree\ {\isacharparenleft}p{\isadigit{1}}{\isachardot}{\isadigit{0}}\parallelcomp{}p{\isadigit{2}}{\isachardot}{\isadigit{0}}{\isacharparenright}\ {\isacharequal}\ {\isacharparenleft}net{\isacharunderscore}tree{\isacharunderscore}ips\ p{\isadigit{1}}{\isachardot}{\isadigit{0}}\ {\isasyminter}\ net{\isacharunderscore}tree{\isacharunderscore}ips\ p{\isadigit{2}}{\isachardot}{\isadigit{0}}\ {\isacharequal}\ {\isasymemptyset}\ {\isasymand}\ wf{\isacharunderscore}net{\isacharunderscore}tree\ p{\isadigit{1}}{\isachardot}{\isadigit{0}}\ {\isasymand}\ wf{\isacharunderscore}net{\isacharunderscore}tree\ p{\isadigit{2}}{\isachardot}{\isadigit{0}}{\isacharparenright}\isasep\isanewline%
wf{\isacharunderscore}net{\isacharunderscore}tree\ {\isasymlangle}i{\isacharsemicolon}\ R{\isasymrangle}\ {\isacharequal}\ True}}
\snip{pnet_rules}{0}{0}{\isa{pnet\ np\ {\isasymlangle}i{\isacharsemicolon}\ \Ri{\isasymrangle}\ {\isacharequal}\ {\isasymlangle}i\ {\isacharcolon}\ np\ i\ {\isacharcolon}\ \Ri{\isasymrangle}\isasep\isanewline%
pnet\ np\ {\isacharparenleft}p\isactrlsub {\isadigit{1}}\parallelcomp{}p\isactrlsub {\isadigit{2}}{\isacharparenright}\ {\isacharequal}\ {\isasymlparr}init\ {\isacharequal}\ {\isacharbraceleft}\subnets{s\isactrlsub {\isadigit{1}}}{s\isactrlsub {\isadigit{2}}}\ {\isacharbar}\ s\isactrlsub {\isadigit{1}}{\isasymin}init\ {\isacharparenleft}pnet\ np\ p\isactrlsub {\isadigit{1}}{\isacharparenright}\ {\isasymand}\ s\isactrlsub {\isadigit{2}}{\isasymin}init\ {\isacharparenleft}pnet\ np\ p\isactrlsub {\isadigit{2}}{\isacharparenright}{\isacharbraceright},\ trans\ {\isacharequal}\ \pnetsos\ {\isacharparenleft}trans\ {\isacharparenleft}pnet\ np\ p\isactrlsub {\isadigit{1}}{\isacharparenright}{\isacharparenright}\ {\isacharparenleft}trans\ {\isacharparenleft}pnet\ np\ p\isactrlsub {\isadigit{2}}{\isacharparenright}{\isacharparenright}{\isasymrparr}}}

\snip{pnet1_lhs}{0}{0}{\isa{pnet\ np\ {\isasymlangle}i{\isacharsemicolon}\ \Ri{\isasymrangle}}}
\snip{pnet1_rhs}{0}{0}{\isa{{\isasymlangle}i\ {\isacharcolon}\ np\ i\ {\isacharcolon}\ \Ri{\isasymrangle}}}
\snip{pnet2_lhs}{0}{0}{\isa{pnet\ np\ {\isacharparenleft}p\isactrlsub {\isadigit{1}}\parallelcomp{}p\isactrlsub {\isadigit{2}}{\isacharparenright}}}
\snip{pnet2_rhs}{0}{0}{\isa{{\isasymlparr}init\ {\isacharequal}\ {\isacharbraceleft}\subnetsadj{s\isactrlsub {\isadigit{1}}}{s\isactrlsub {\isadigit{2}}}\ {\isacharbar}\ s\isactrlsub {\isadigit{1}}{\isasymin}init\ {\isacharparenleft}pnet\ np\ p\isactrlsub {\isadigit{1}}{\isacharparenright}\ {\isasymand}\ s\isactrlsub {\isadigit{2}}{\isasymin}init\ {\isacharparenleft}pnet\ np\ p\isactrlsub {\isadigit{2}}{\isacharparenright}{\isacharbraceright},}\\ && \isa{\phantom{\isasymlparr}trans {\isacharequal}\ \pnetsos\ {\isacharparenleft}trans\ {\isacharparenleft}pnet\ np\ p\isactrlsub {\isadigit{1}}{\isacharparenright}{\isacharparenright}\ {\isacharparenleft}trans\ {\isacharparenleft}pnet\ np\ p\isactrlsub {\isadigit{2}}{\isacharparenright}{\isacharparenright}{\isasymrparr}}}

\snip{opnet1_lhs}{0}{0}{\isa{opnet\ onp\ {\isasymlangle}i{\isacharsemicolon}\ \Ri{\isasymrangle}}}
\snip{opnet1_rhs}{0}{0}{\isa{{\isasymlangle}i\ {\isacharcolon}\ onp\ i\ {\isacharcolon}\ \Ri{\isasymrangle}\isactrlsub o}}
\snip{opnet2_lhs}{0}{0}{\isa{opnet\ onp\ {\isacharparenleft}p\isactrlsub {\isadigit{1}}\parallelcomp{}p\isactrlsub {\isadigit{2}}{\isacharparenright}}}
\snip{opnet2_rhs}{0}{0}{\isa{{\isasymlparr}init\ {\isacharequal}\ {\isacharbraceleft}{\isacharparenleft}{\isasymsigma},\ \subnetsadj{s\isactrlsub {\isadigit{1}}}{s\isactrlsub {\isadigit{2}}}{\isacharparenright}\ {\isacharbar}\ {\isacharparenleft}{\isasymsigma},\ s\isactrlsub {\isadigit{1}}{\isacharparenright}{\isasymin}init\ {\isacharparenleft}opnet\ onp\ p\isactrlsub {\isadigit{1}}{\isacharparenright}}\\ && \isa{\phantom{\isasymlparr init\ = \ {\isacharbraceleft}{\isacharparenleft}{\isasymsigma}{\isacharcomma}\ \subnetsadj{s\isactrlsub {\isadigit{1}}}{s\isactrlsub {\isadigit{2}}}{\isacharparenright}\ {\isacharbar}}{\isasymand}\ {\isacharparenleft}{\isasymsigma},\ s\isactrlsub {\isadigit{2}}{\isacharparenright}{\isasymin}init\ {\isacharparenleft}opnet\ onp\ p\isactrlsub {\isadigit{2}}{\isacharparenright}}\\ && \isa{\phantom{\isasymlparr init\ = \ {\isacharbraceleft}{\isacharparenleft}{\isasymsigma}{\isacharcomma}\ \subnetsadj{s\isactrlsub {\isadigit{1}}}{s\isactrlsub {\isadigit{2}}}{\isacharparenright}\ {\isacharbar}}{\isasymand}\ net{\isacharunderscore}ips\ s\isactrlsub {\isadigit{1}}\ {\isasyminter}\ net{\isacharunderscore}ips\ s\isactrlsub {\isadigit{2}}\ {\isacharequal}\ {\isasymemptyset}{\isacharbraceright},}\\ && \isa{\phantom{\isasymlparr}trans {\isacharequal}\ \opnetsos\ {\isacharparenleft}trans\ \hspace{-2pt}{\isacharparenleft}opnet\ onp\ p\isactrlsub {\isadigit{1}}{\isacharparenright}{\isacharparenright}\ \hspace{-1pt}{\isacharparenleft}trans\ \hspace{-2pt}{\isacharparenleft}opnet\ onp\ p\isactrlsub {\isadigit{2}}{\isacharparenright}{\isacharparenright}{\isasymrparr}}}

\snip{oclosed'}{0}{0}{\isa{oclosed\ A}}
\snip{oclosed_term}{0}{0}{\isa{A{\isasymlparr}trans\ {\isacharcolon}{\isacharequal}\ \ocnetsos\ {\isacharparenleft}trans\ A{\isacharparenright}{\isasymrparr}}}

\snip{netglobal}{0}{0}{\isa{netglobal\ P\ {\isacharequal}\ {\isasymlambda}s{\isachardot}\ P\ {\isacharparenleft}default\ toy{\isacharunderscore}init\ {\isacharparenleft}netlift\ fst\ s{\isacharparenright}{\isacharparenright}}}

\snip{default}{0}{0}{\isa{default\ df\ f\ {\isacharequal}\ {\isacharparenleft}{\isasymlambda}i{\isachardot}\ \textsf{case}\ f\ i\ \textsf{of}\ None\ {\isasymRightarrow}\ df\ i\ {\isacharbar}\ Some\ s\ {\isasymRightarrow}\ s{\isacharparenright}}}
\snip{default_lhs}{0}{0}{\isa{default\ df\ f}}
\snip{default_rhs}{0}{0}{\isa{{\isasymlambda}i{\isachardot}\ \textsf{case}\ f\ i\ \textsf{of}\ None\ {\isasymRightarrow}\ df\ i\ {\isacharbar}\ Some\ s\ {\isasymRightarrow}\ s}}

\snip{netlift1_lhs}{0}{0}{\isa{netlift\ sr\ {\isacharparenleft}\NodeS\ i\ s\ R{\isacharparenright}}}
\snip{netlift1_rhs}{0}{0}{\isa{{\isacharbrackleft}i\ {\isasymmapsto}\ fst\ {\isacharparenleft}sr\ s{\isacharparenright}{\isacharbrackright}}}
\snip{netlift2_lhs}{0}{0}{\isa{netlift\ sr\ {\isacharparenleft}\subnets{s}{t}{\isacharparenright}}}
\snip{netlift2_rhs}{0}{0}{\isa{netlift\ sr\ s\ \listconcat\ netlift\ sr\ t}}

\snip{netliftl1_lhs}{0}{0}{\isa{netliftl\ sr\ {\isacharparenleft}\NodeS\ i\ s\ R{\isacharparenright}}}
\snip{netliftl1_rhs}{0}{0}{\isa{\NodeS\ i\ {{\isacharparenleft}snd\ {\isacharparenleft}sr\ s{\isacharparenright}{\isacharparenright}}\ R}}
\snip{netliftl2_lhs}{0}{0}{\isa{netliftl\ sr\ {\isacharparenleft}\subnets{s}{t}{\isacharparenright}}}
\snip{netliftl2_rhs}{0}{0}{\isa{\subnets{{\isacharparenleft}netliftl\ sr\ s{\isacharparenright}}{{\isacharparenleft}netliftl\ sr\ t{\isacharparenright}}}}

\snip{netgmap1_lhs}{0}{0}{\isa{netgmap\ sr\ {\isacharparenleft}\NodeS\ i\ s\ R{\isacharparenright}}}
\snip{netgmap1_rhs}{0}{0}{\isa{{\isacharparenleft}{\isacharbrackleft}i\ {\isasymmapsto}\ fst\ {\isacharparenleft}sr\ s{\isacharparenright}{\isacharbrackright},\ \NodeS\ i\ {{\isacharparenleft}snd\ {\isacharparenleft}sr\ s{\isacharparenright}{\isacharparenright}}\ R{\isacharparenright}}}
\snip{netgmap2_lhs}{0}{0}{\isa{netgmap\ sr\ {\isacharparenleft}\subnets{s\isactrlsub {\isadigit{1}}}{s\isactrlsub {\isadigit{2}}}{\isacharparenright}}}
\snip{netgmap2_rhs}{0}{0}{\isa{\textsf{let}\ {\isacharparenleft}{\isasymsigma}\isactrlsub {\isadigit{1}},\ ss{\isacharparenright}\ {\isacharequal}\ netgmap\ sr\ s\isactrlsub {\isadigit{1}}{\isacharsemicolon}}\\&&\isa{\phantom{let\ }{\isacharparenleft}{\isasymsigma}\isactrlsub {\isadigit{2}},\ tt{\isacharparenright}\ {\isacharequal}\ netgmap\ sr\ s\isactrlsub {\isadigit{2}}\ }\\&&\isa{\textsf{in}\ {\isacharparenleft}{\isasymsigma}\isactrlsub {\isadigit{1}}\ \listconcat\ {\isasymsigma}\isactrlsub {\isadigit{2}},\ \subnets{ss}{tt}{\isacharparenright}}}
\snip{initmissing_def}{0}{0}{\isa{initmissing\ {\isasymsigma}\ {\isacharequal}\ {\isacharparenleft}{\isasymlambda}i{\isachardot}\ \textsf{case}\ fst\ {\isasymsigma}\ i\ \textsf{of}\ None\ {\isasymRightarrow}\ toy{\isacharunderscore}init\ i\ {\isacharbar}\ Some\ s\ {\isasymRightarrow}\ s,\ snd\ {\isasymsigma}{\isacharparenright}}}
\snip{initmissing_lhs}{0}{0}{\isa{initmissing\ {\isasymsigma}}}
\snip{initmissing_rhs}{0}{0}{\isa{{\isacharparenleft}{\isasymlambda}i{\isachardot}\ \textsf{case}\ fst\ {\isasymsigma}\ i\ \textsf{of}\ None\ {\isasymRightarrow}\ toy{\isacharunderscore}init\ i\ {\isacharbar}\ Some\ s\ {\isasymRightarrow}\ s,\ snd\ {\isasymsigma}{\isacharparenright}}}

\snip{onl}{0}{0}{\isa{onl\ {\isasymGamma}\ P\ {\isacharequal}\ {\isasymlambda}{\isacharparenleft}{\isasymxi},\ p{\isacharparenright}{\isachardot}\ {\isasymforall}l{\isasymin}labels\ {\isasymGamma}\ p{\isachardot}\ P\ {\isacharparenleft}{\isasymxi},\ l{\isacharparenright}}}
\snip{onl_lhs}{0}{0}{\isa{onl\ {\isasymGamma}\ P}}
\snip{onl_rhs}{0}{0}{\isa{{\isasymlambda}{\isacharparenleft}{\isasymxi},\ p{\isacharparenright}{\isachardot}\ {\isasymforall}l{\isasymin}labels\ {\isasymGamma}\ p{\isachardot}\ P\ {\isacharparenleft}{\isasymxi},\ l{\isacharparenright}}}

\snip{cnet_connect}{0}{0}{\isa{\mbox{}\inferrule{\mbox{{\isacharparenleft}s,\ connect{\isacharparenleft}i,\ i{\isacharprime}{\isacharparenright},\ s{\isacharprime}{\isacharparenright}{\isasymin}S}}{\mbox{{\isacharparenleft}s,\ connect{\isacharparenleft}i,\ i{\isacharprime}{\isacharparenright},\ s{\isacharprime}{\isacharparenright}{\isasymin}\cnetsos\ S}}}}
\snip{cnet_disconnect}{0}{0}{\isa{\mbox{}\inferrule{\mbox{{\isacharparenleft}s,\ disconnect{\isacharparenleft}i,\ i{\isacharprime}{\isacharparenright},\ s{\isacharprime}{\isacharparenright}{\isasymin}S}}{\mbox{{\isacharparenleft}s,\ disconnect{\isacharparenleft}i,\ i{\isacharprime}{\isacharparenright},\ s{\isacharprime}{\isacharparenright}{\isasymin}\cnetsos\ S}}}}
\snip{cnet_cast}{0}{0}{\isa{\mbox{}\inferrule{\mbox{{\isacharparenleft}s,\ R{\isacharcolon}{\isacharasterisk}cast{\isacharparenleft}m{\isacharparenright},\ s{\isacharprime}{\isacharparenright}{\isasymin}S}}{\mbox{{\isacharparenleft}s,\ {\isasymtau},\ s{\isacharprime}{\isacharparenright}{\isasymin}\cnetsos\ S}}}}
\snip{cnet_tau}{0}{0}{\isa{\mbox{}\inferrule{\mbox{{\isacharparenleft}s,\ {\isasymtau},\ s{\isacharprime}{\isacharparenright}{\isasymin}S}}{\mbox{{\isacharparenleft}s,\ {\isasymtau},\ s{\isacharprime}{\isacharparenright}{\isasymin}\cnetsos\ S}}}}
\snip{cnet_deliver}{0}{0}{\isa{\mbox{}\inferrule{\mbox{{\isacharparenleft}s,\ i{\isacharcolon}deliver{\isacharparenleft}d{\isacharparenright},\ s{\isacharprime}{\isacharparenright}{\isasymin}S}}{\mbox{{\isacharparenleft}s,\ i{\isacharcolon}deliver{\isacharparenleft}d{\isacharparenright},\ s{\isacharprime}{\isacharparenright}{\isasymin}\cnetsos\ S}}}}
\snip{cnet_newpkt}{0}{0}{\isa{\mbox{}\inferrule{\mbox{{\isacharparenleft}s,\ {\isacharbraceleft}i{\isacharbraceright}{\isasymnot}K{\isacharcolon}arrive{\isacharparenleft}newpkt\ {\isacharparenleft}d,\ di{\isacharparenright}{\isacharparenright},\ s{\isacharprime}{\isacharparenright}{\isasymin}S}}{\mbox{{\isacharparenleft}s,\ i{\isacharcolon}newpkt{\isacharparenleft}d,\ di{\isacharparenright},\ s{\isacharprime}{\isacharparenright}{\isasymin}\cnetsos\ S}}}}

\snip{closed}{0}{0}{\isa{closed\ {\isacharparenleft}pnet\ {\isacharparenleft}{\isasymlambda}i{\isachardot}\ ptoy\ i\ {\isasymlangle}{\isasymlangle}\ qmsg{\isacharparenright}\ any{\isacharparenright}}}
\snip{closed'}{0}{0}{\isa{closed\ A}}
\snip{closed_term}{0}{0}{\isa{A{\isasymlparr}trans\ {\isacharcolon}{\isacharequal}\ \cnetsos\ {\isacharparenleft}trans\ A{\isacharparenright}{\isasymrparr}}}

\snip{closed_type}{0}{0}{\isa{{\isacharparenleft}{\isacharparenleft}{\isacharparenleft}state\ {\isasymtimes}\ {\isacharparenleft}state,\ msg,\ pseqp,\ pseqp\ label{\isacharparenright}\ seqp{\isacharparenright}\ {\isasymtimes}\ msg\ list\ {\isasymtimes}\ {\isacharparenleft}msg\ list,\ msg,\ unit,\ unit\ label{\isacharparenright}\ seqp{\isacharparenright}\ net{\isacharunderscore}state,\ msg\ node{\isacharunderscore}action{\isacharparenright}\ automaton}}
\snip{state_type}{0}{0}{\isa{state}}
\snip{sigma_type}{0}{0}{\isa{ip\ {\isasymRightarrow}\ state}}

\snip{obroadcastT}{0}{0}{\isa{\mbox{}\inferrule{\mbox{{\isasymsigma}{\isacharprime}\ i\ {\isacharequal}\ {\isasymsigma}\ i}}{\mbox{{\isacharparenleft}{\isacharparenleft}{\isasymsigma},\ \gray{{\isacharbraceleft}l{\isacharbraceright}}broadcast{\isacharparenleft}\selmsg{\isacharparenright}\ {\isachardot}\ p{\isacharparenright},\ broadcast\ {\isacharparenleft}\selmsg\ {\isacharparenleft}{\isasymsigma}\ i{\isacharparenright}{\isacharparenright},\ {\isacharparenleft}{\isasymsigma}{\isacharprime},\ p{\isacharparenright}{\isacharparenright}{\isasymin}\oseqpsos\ {\isasymGamma}\ i}}}}
\snip{ogroupcastT}{0}{0}{\isa{\mbox{}\inferrule{\mbox{{\isasymsigma}{\isacharprime}\ i\ {\isacharequal}\ {\isasymsigma}\ i}}{\mbox{{\isacharparenleft}{\isacharparenleft}{\isasymsigma},\ \gray{{\isacharbraceleft}l{\isacharbraceright}}groupcast{\isacharparenleft}\selips,\ \selmsg{\isacharparenright}\ {\isachardot}\ p{\isacharparenright},\ groupcast\ {\isacharparenleft}\selips\ {\isacharparenleft}{\isasymsigma}\ i{\isacharparenright}{\isacharparenright}\ {\isacharparenleft}\selmsg\ {\isacharparenleft}{\isasymsigma}\ i{\isacharparenright}{\isacharparenright},\ {\isacharparenleft}{\isasymsigma}{\isacharprime},\ p{\isacharparenright}{\isacharparenright}{\isasymin}\oseqpsos\ {\isasymGamma}\ i}}}}
\snip{ounicastT}{0}{0}{\isa{\mbox{}\inferrule{\mbox{{\isasymsigma}{\isacharprime}\ i\ {\isacharequal}\ {\isasymsigma}\ i}}{\mbox{{\isacharparenleft}{\isacharparenleft}{\isasymsigma},\ \gray{{\isacharbraceleft}l{\isacharbraceright}}unicast{\isacharparenleft}\selip,\ \hspace{-1.5pt}\selmsg{\isacharparenright}\ {\isachardot}\ p\ {\isasymtriangleright}\ q{\isacharparenright},\ unicast\ {\isacharparenleft}\selip\ \hspace{-1.5pt}{\isacharparenleft}{\isasymsigma}\ \hspace{-1.5pt}i{\isacharparenright}{\isacharparenright}\ {\isacharparenleft}\selmsg\ \hspace{-1.5pt}{\isacharparenleft}{\isasymsigma}\ i{\isacharparenright}{\isacharparenright},\ {\isacharparenleft}{\isasymsigma}{\isacharprime},\ \hspace{-1.5pt}p{\isacharparenright}{\isacharparenright}{\isasymin}\oseqpsos\ {\isasymGamma}\ i}}}}
\snip{onotunicastT}{0}{0}{\isa{\mbox{}\inferrule{\mbox{{\isasymsigma}{\isacharprime}\ i\ {\isacharequal}\ {\isasymsigma}\ i}}{\mbox{{\isacharparenleft}{\isacharparenleft}{\isasymsigma},\ \gray{{\isacharbraceleft}l{\isacharbraceright}}unicast{\isacharparenleft}\selip,\ \hspace{-1.5pt}\selmsg{\isacharparenright}\ {\isachardot}\ p\ {\isasymtriangleright}\ q{\isacharparenright},\ {\isasymnot}unicast\ {\isacharparenleft}\selip\ \hspace{-1.5pt}{\isacharparenleft}{\isasymsigma}\ i{\isacharparenright}{\isacharparenright},\ {\isacharparenleft}{\isasymsigma}{\isacharprime},\ q{\isacharparenright}{\isacharparenright}{\isasymin}\oseqpsos\ {\isasymGamma}\ i}}}}
\snip{osendT}{0}{0}{\isa{\mbox{}\inferrule{\mbox{{\isasymsigma}{\isacharprime}\ i\ {\isacharequal}\ {\isasymsigma}\ i}}{\mbox{{\isacharparenleft}{\isacharparenleft}{\isasymsigma},\ \gray{{\isacharbraceleft}l{\isacharbraceright}}send{\isacharparenleft}\selmsg{\isacharparenright}\ {\isachardot}\ p{\isacharparenright},\ send\ {\isacharparenleft}\selmsg\ {\isacharparenleft}{\isasymsigma}\ i{\isacharparenright}{\isacharparenright},\ {\isacharparenleft}{\isasymsigma}{\isacharprime},\ p{\isacharparenright}{\isacharparenright}{\isasymin}\oseqpsos\ {\isasymGamma}\ i}}}}
\snip{odeliverT}{0}{0}{\isa{\mbox{}\inferrule{\mbox{{\isasymsigma}{\isacharprime}\ i\ {\isacharequal}\ {\isasymsigma}\ i}}{\mbox{{\isacharparenleft}{\isacharparenleft}{\isasymsigma},\ \gray{{\isacharbraceleft}l{\isacharbraceright}}deliver{\isacharparenleft}\seldata{\isacharparenright}\ {\isachardot}\ p{\isacharparenright},\ deliver\ {\isacharparenleft}\seldata\ {\isacharparenleft}{\isasymsigma}\ i{\isacharparenright}{\isacharparenright},\ {\isacharparenleft}{\isasymsigma}{\isacharprime},\ p{\isacharparenright}{\isacharparenright}{\isasymin}\oseqpsos\ {\isasymGamma}\ i}}}}
\snip{oreceiveT}{0}{0}{\isa{\mbox{}\inferrule{\mbox{{\isasymsigma}{\isacharprime}\ i\ {\isacharequal}\ \updmsg\ msg\ {\isacharparenleft}{\isasymsigma}\ i{\isacharparenright}}}{\mbox{{\isacharparenleft}{\isacharparenleft}{\isasymsigma},\ \gray{{\isacharbraceleft}l{\isacharbraceright}}receive{\isacharparenleft}\updmsg{\isacharparenright}\ {\isachardot}\ p{\isacharparenright},\ receive\ msg,\ {\isacharparenleft}{\isasymsigma}{\isacharprime},\ p{\isacharparenright}{\isacharparenright}{\isasymin}\oseqpsos\ {\isasymGamma}\ i}}}}
\snip{oassignT}{0}{0}{\isa{\mbox{}\inferrule{\mbox{{\isasymsigma}{\isacharprime}\ i\ {\isacharequal}\ u\ {\isacharparenleft}{\isasymsigma}\ i{\isacharparenright}}}{\mbox{{\isacharparenleft}{\isacharparenleft}{\isasymsigma},\ \gray{{\isacharbraceleft}l{\isacharbraceright}}{\isasymlbrakk}u{\isasymrbrakk}\ p{\isacharparenright},\ {\isasymtau},\ {\isacharparenleft}{\isasymsigma}{\isacharprime},\ p{\isacharparenright}{\isacharparenright}{\isasymin}\oseqpsos\ {\isasymGamma}\ i}}}}
\snip{ocallT}{0}{0}{\isa{\mbox{}\inferrule{\mbox{{\isacharparenleft}{\isacharparenleft}{\isasymsigma},\ {\isasymGamma}\ pn{\isacharparenright},\ a,\ {\isacharparenleft}{\isasymsigma}{\isacharprime},\ p{\isacharprime}{\isacharparenright}{\isacharparenright}{\isasymin}\oseqpsos\ {\isasymGamma}\ i}}{\mbox{{\isacharparenleft}{\isacharparenleft}{\isasymsigma},\ call{\isacharparenleft}pn{\isacharparenright}{\isacharparenright},\ a,\ {\isacharparenleft}{\isasymsigma}{\isacharprime},\ p{\isacharprime}{\isacharparenright}{\isacharparenright}{\isasymin}\oseqpsos\ {\isasymGamma}\ i}}}}
\snip{ochoiceT1}{0}{0}{\isa{\mbox{}\inferrule{\mbox{{\isacharparenleft}{\isacharparenleft}{\isasymsigma},\ p{\isacharparenright},\ a,\ {\isacharparenleft}{\isasymsigma}{\isacharprime},\ p{\isacharprime}{\isacharparenright}{\isacharparenright}{\isasymin}\oseqpsos\ {\isasymGamma}\ i}}{\mbox{{\isacharparenleft}{\isacharparenleft}{\isasymsigma},\ p\ {\isasymoplus}\ q{\isacharparenright},\ a,\ {\isacharparenleft}{\isasymsigma}{\isacharprime},\ p{\isacharprime}{\isacharparenright}{\isacharparenright}{\isasymin}\oseqpsos\ {\isasymGamma}\ i}}}}
\snip{ochoiceT2}{0}{0}{\isa{\mbox{}\inferrule{\mbox{{\isacharparenleft}{\isacharparenleft}{\isasymsigma},\ q{\isacharparenright},\ a,\ {\isacharparenleft}{\isasymsigma}{\isacharprime},\ q{\isacharprime}{\isacharparenright}{\isacharparenright}{\isasymin}\oseqpsos\ {\isasymGamma}\ i}}{\mbox{{\isacharparenleft}{\isacharparenleft}{\isasymsigma},\ p\ {\isasymoplus}\ q{\isacharparenright},\ a,\ {\isacharparenleft}{\isasymsigma}{\isacharprime},\ q{\isacharprime}{\isacharparenright}{\isacharparenright}{\isasymin}\oseqpsos\ {\isasymGamma}\ i}}}}
\snip{oguardT}{0}{0}{\isa{\mbox{}\inferrule{\mbox{{\isasymsigma}{\isacharprime}\ i{\isasymin}g\ {\isacharparenleft}{\isasymsigma}\ i{\isacharparenright}}}{\mbox{{\isacharparenleft}{\isacharparenleft}{\isasymsigma},\ \gray{{\isacharbraceleft}l{\isacharbraceright}}{\isasymlangle}g{\isasymrangle}\ p{\isacharparenright},\ {\isasymtau},\ {\isacharparenleft}{\isasymsigma}{\isacharprime},\ p{\isacharparenright}{\isacharparenright}{\isasymin}\oseqpsos\ {\isasymGamma}\ i}}}}

\snip{oassignT_prem_1}{0}{0}{\isa{{\isasymsigma}{\isacharprime}\ i\ {\isacharequal}\ u\ {\isacharparenleft}{\isasymsigma}\ i{\isacharparenright}}}
\snip{ounicastT_prem_1}{0}{0}{\isa{{\isasymsigma}{\isacharprime}\ i\ {\isacharequal}\ {\isasymsigma}\ i}}

\snip{oparleft}{0}{0}{\isa{\mbox{}\inferrule{\mbox{{\isacharparenleft}{\isacharparenleft}{\isasymsigma},\ s{\isacharparenright},\ a,\ {\isacharparenleft}{\isasymsigma}{\isacharprime},\ s{\isacharprime}{\isacharparenright}{\isacharparenright}{\isasymin}S}\\\ \mbox{{\isasymAnd}m{\isachardot}\ a\ {\isasymnoteq}\ receive\ m}}{\mbox{{\isacharparenleft}{\isacharparenleft}{\isasymsigma},\ s,\ t{\isacharparenright},\ a,\ {\isasymsigma}{\isacharprime},\ s{\isacharprime},\ t{\isacharparenright}{\isasymin}\oparpsos\ i\ S\ T}}}}
\snip{oparright}{0}{0}{\isa{\mbox{}\inferrule{\mbox{{\isacharparenleft}t,\ {\isacharparenleft}a,\ t{\isacharprime}{\isacharparenright}{\isacharparenright}{\isasymin}T}\\\ \mbox{{\isasymAnd}m{\isachardot}\ a\ {\isasymnoteq}\ send\ m}\\\ \mbox{{\isasymsigma}{\isacharprime}\ i\ {\isacharequal}\ {\isasymsigma}\ i}}{\mbox{{\isacharparenleft}{\isacharparenleft}{\isasymsigma},\ s,\ t{\isacharparenright},\ a,\ {\isasymsigma}{\isacharprime},\ s,\ t{\isacharprime}{\isacharparenright}{\isasymin}\oparpsos\ i\ S\ T}}}}
\snip{oparboth}{0}{0}{\isa{\mbox{}\inferrule{\mbox{{\isacharparenleft}{\isacharparenleft}{\isasymsigma},\ s{\isacharparenright},\ receive\ m,\ {\isacharparenleft}{\isasymsigma}{\isacharprime},\ s{\isacharprime}{\isacharparenright}{\isacharparenright}{\isasymin}S}\\\ \mbox{{\isacharparenleft}t,\ send\ m,\ t{\isacharprime}{\isacharparenright}{\isasymin}T}}{\mbox{{\isacharparenleft}{\isacharparenleft}{\isasymsigma},\ {\isacharparenleft}s,\ t{\isacharparenright}{\isacharparenright},\ {\isasymtau},\ {\isacharparenleft}{\isasymsigma}{\isacharprime},\ {\isacharparenleft}s{\isacharprime},\ t{\isacharprime}{\isacharparenright}{\isacharparenright}{\isacharparenright}{\isasymin}\oparpsos\ i\ S\ T}}}}

\snip{opar_comp}{0}{0}{\mbox{\isa{s\ {\isasymlangle}{\isasymlangle}\isactrlbsub i\isactrlesub \ t\ {\isacharequal}\ {\isasymlparr}}} & \mbox{\isa{init\ {\isacharequal}\ extg\ {\isacharbackquote}\ {\isacharparenleft}init\ s\ {\isasymtimes}\ init\ t{\isacharparenright},}}\\ & \mbox{\isa{trans\ {\isacharequal}\ \oparpsos\ i\ {\isacharparenleft}trans\ s{\isacharparenright}\ {\isacharparenleft}trans\ t{\isacharparenright}{\isasymrparr}}}}
\snip{opar_comp'}{0}{0}{\mbox{\isa{s\ {\isasymlangle}{\isasymlangle}\isactrlbsub i\isactrlesub \ t\ {\isacharequal}\ {\isasymlparr}}} & \mbox{\isa{init\ {\isacharequal}\ {\isacharbraceleft}{\isacharparenleft}{\isasymsigma},\ (s\isactrlsub l,\ t\isactrlsub l){\isacharparenright}\ {\isacharbar}\ {\isacharparenleft}{\isasymsigma},\ s\isactrlsub l{\isacharparenright}{\isasymin}init\ s\ {\isasymand}\ t\isactrlsub l{\isasymin}init\ t{\isacharbraceright},}}\\ & \mbox{\isa{trans\ {\isacharequal}\ \oparpsos\ i\ {\isacharparenleft}trans\ s{\isacharparenright}\ {\isacharparenleft}trans\ t{\isacharparenright}{\isasymrparr}}}}
\snip{opar_comp_term}{0}{0}{\mbox{\isa{s\ {\isasymlangle}{\isasymlangle}\isactrlbsub i\isactrlesub \ t{\isasymColon}{\isacharparenleft}{\isacharparenleft}nat\ {\isasymRightarrow}\ {\isacharprime}a{\isacharparenright}\ {\isasymtimes}\ {\isacharprime}b\ {\isasymtimes}\ {\isacharprime}d,\ {\isacharprime}c\ seq{\isacharunderscore}action{\isacharparenright}\ automaton}}}

\snip{onode_bcast}{0}{0}{\isa{\mbox{}\inferrule{\mbox{{\isacharparenleft}{\isacharparenleft}{\isasymsigma},\ P{\isacharparenright},\ broadcast\ m,\ {\isasymsigma}{\isacharprime},\ P{\isacharprime}{\isacharparenright}{\isasymin}S}}{\mbox{{\isacharparenleft}{\isacharparenleft}{\isasymsigma},\ \NodeS\ i\ P\ R{\isacharparenright},\ R{\isacharcolon}{\isacharasterisk}cast{\isacharparenleft}m{\isacharparenright},\ {\isacharparenleft}{\isasymsigma}{\isacharprime},\ \NodeS\ i\ P{\isacharprime}\ R{\isacharparenright}{\isacharparenright}{\isasymin}\onodesos\ S}}}}
\snip{onode_gcast}{0}{0}{\isa{\mbox{}\inferrule{\mbox{{\isacharparenleft}{\isacharparenleft}{\isasymsigma},\ P{\isacharparenright},\ groupcast\ D\ m,\ {\isasymsigma}{\isacharprime},\ P{\isacharprime}{\isacharparenright}{\isasymin}S}}{\mbox{{\isacharparenleft}{\isacharparenleft}{\isasymsigma},\ \NodeS\ i\ P\ R{\isacharparenright},\ {\isacharparenleft}R\ {\isasyminter}\ D{\isacharparenright}{\isacharcolon}{\isacharasterisk}cast{\isacharparenleft}m{\isacharparenright},\ {\isacharparenleft}{\isasymsigma}{\isacharprime},\ \NodeS\ i\ P{\isacharprime}\ R{\isacharparenright}{\isacharparenright}{\isasymin}\onodesos\ S}}}}
\snip{onode_ucast}{0}{0}{\isa{\mbox{}\inferrule{\mbox{{\isacharparenleft}{\isacharparenleft}{\isasymsigma},\ P{\isacharparenright},\ unicast\ d\ m,\ {\isasymsigma}{\isacharprime},\ P{\isacharprime}{\isacharparenright}{\isasymin}S}\\\ \mbox{d{\isasymin}R}}{\mbox{{\isacharparenleft}{\isacharparenleft}{\isasymsigma},\ \NodeS\ i\ P\ R{\isacharparenright},\ {\isacharbraceleft}d{\isacharbraceright}{\isacharcolon}{\isacharasterisk}cast{\isacharparenleft}m{\isacharparenright},\ {\isacharparenleft}{\isasymsigma}{\isacharprime},\ \NodeS\ i\ P{\isacharprime}\ R{\isacharparenright}{\isacharparenright}{\isasymin}\onodesos\ S}}}}
\snip{onode_notucast}{0}{0}{\isa{\mbox{}\inferrule{\mbox{{\isacharparenleft}{\isacharparenleft}{\isasymsigma},\ P{\isacharparenright},\ {\isasymnot}unicast\ d,\ {\isasymsigma}{\isacharprime},\ P{\isacharprime}{\isacharparenright}{\isasymin}S}\\\ \mbox{d\ {\isasymnotin}\ R}\\\ \mbox{{\isasymforall}j{\isachardot}\ j\ {\isasymnoteq}\ i\ {\isasymlongrightarrow}\ {\isasymsigma}{\isacharprime}\ j\ {\isacharequal}\ {\isasymsigma}\ j}}{\mbox{{\isacharparenleft}{\isacharparenleft}{\isasymsigma},\ \NodeS\ i\ P\ R{\isacharparenright},\ {\isasymtau},\ {\isacharparenleft}{\isasymsigma}{\isacharprime},\ \NodeS\ i\ P{\isacharprime}\ R{\isacharparenright}{\isacharparenright}{\isasymin}\onodesos\ S}}}}
\snip{onode_deliver}{0}{0}{\isa{\mbox{}\inferrule{\mbox{{\isacharparenleft}{\isacharparenleft}{\isasymsigma},\ P{\isacharparenright},\ deliver\ d,\ {\isasymsigma}{\isacharprime},\ P{\isacharprime}{\isacharparenright}{\isasymin}S}\\\ \mbox{{\isasymforall}j{\isachardot}\ j\ {\isasymnoteq}\ i\ {\isasymlongrightarrow}\ {\isasymsigma}{\isacharprime}\ j\ {\isacharequal}\ {\isasymsigma}\ j}}{\mbox{{\isacharparenleft}{\isacharparenleft}{\isasymsigma},\ \NodeS\ i\ P\ R{\isacharparenright},\ i{\isacharcolon}deliver{\isacharparenleft}d{\isacharparenright},\ {\isacharparenleft}{\isasymsigma}{\isacharprime},\ \NodeS\ i\ P{\isacharprime}\ R{\isacharparenright}{\isacharparenright}{\isasymin}\onodesos\ S}}}}

\snip{onode_receive}{0}{0}{\isa{\mbox{}\inferrule{\mbox{{\isacharparenleft}{\isacharparenleft}{\isasymsigma},\ \hspace{-2pt}s{\isacharparenright},\ receive\ m,\ {\isacharparenleft}{\isasymsigma}{\isacharprime}\hspace{-1.5pt},\ \hspace{-2pt}s{\isacharprime}{\isacharparenright}{\isacharparenright}{\isasymin}S}}{\mbox{{\isacharparenleft}{\isacharparenleft}{\isasymsigma},\ \NodeS\ {i}\ \hspace{-2pt}s\ {R}{\isacharparenright},\ \hspace{-1pt}{\isacharbraceleft}i{\isacharbraceright}{\isasymnot}{\isasymemptyset}{\isacharcolon}arrive{\isacharparenleft}m{\isacharparenright},\ \hspace{-1pt}{\isacharparenleft}{\isasymsigma}{\isacharprime}\hspace{-1.5pt},\ \NodeS\ {\hspace{-0.5pt}i}\ \hspace{-2pt}s{\isacharprime}\ {\hspace{-1pt}R}{\isacharparenright}{\isacharparenright}{\isasymin}\onodesos\ \hspace{-1pt}S}}}}

\snip{onode_tau}{0}{0}{\isa{\mbox{}\inferrule{\mbox{{\isacharparenleft}{\isacharparenleft}{\isasymsigma},\ \hspace{-2pt}s{\isacharparenright},\ \hspace{-2pt}{\isasymtau}\hspace{-1pt},\ \hspace{-1pt}{\isacharparenleft}{\isasymsigma}{\isacharprime}\hspace{-1.5pt},\ \hspace{-2pt}s{\isacharprime}{\isacharparenright}{\isacharparenright}{\isasymin}S}\hspace{8pt}\mbox{{\isasymforall}j \hspace{-1pt}{\isasymnoteq}\hspace{-1pt} i{\isachardot}\ \hspace{-0pt}{\isasymsigma}{\isacharprime}\ \hspace{-2pt}j\ \hspace{-1pt}{\isacharequal}\hspace{-1pt}\ {\isasymsigma}\ \hspace{-2pt}j}}{\mbox{{\isacharparenleft}{\isacharparenleft}{\isasymsigma},\ \NodeS\ {i}\ \hspace{-2pt}s\ {R}{\isacharparenright},\ \hspace{-2pt}{\isasymtau}\hspace{-1pt},\ \hspace{-1pt}{\isacharparenleft}{\isasymsigma}{\isacharprime}\hspace{-1.5pt},\ \NodeS\ {\hspace{-0.5pt}i}\ \hspace{-2pt}s{\isacharprime}\ {\hspace{-1.5pt}R}{\isacharparenright}{\isacharparenright}{\isasymin}\onodesos\ \hspace{-1pt}S}}}}

\snip{onode_arrive}{0}{0}{\isa{\mbox{}\inferrule{\mbox{{\isasymsigma}{\isacharprime}\ i\ {\isacharequal}\ {\isasymsigma}\ i}}{\mbox{{\isacharparenleft}{\isacharparenleft}{\isasymsigma},\ \NodeS\ i\ P\ R{\isacharparenright},\ {\isasymemptyset}{\isasymnot}{\isacharbraceleft}i{\isacharbraceright}{\isacharcolon}arrive{\isacharparenleft}m{\isacharparenright},\ {\isacharparenleft}{\isasymsigma}{\isacharprime},\ \NodeS\ i\ P\ R{\isacharparenright}{\isacharparenright}{\isasymin}\onodesos\ S}}}}
\snip{onode_connect1}{0}{0}{\isa{\mbox{}\inferrule{\mbox{{\isasymsigma}{\isacharprime}\ i\ {\isacharequal}\ {\isasymsigma}\ i}}{\mbox{{\isacharparenleft}{\isacharparenleft}{\isasymsigma},\ \NodeS\ i\ P\ R{\isacharparenright},\ connect{\isacharparenleft}i,\ i{\isacharprime}{\isacharparenright},\ {\isacharparenleft}{\isasymsigma}{\isacharprime},\ \NodeS\ i\ P\ {R\ {\isasymunion}\ {\isacharbraceleft}i{\isacharprime}{\isacharbraceright}}{\isacharparenright}{\isacharparenright}{\isasymin}\onodesos\ S}}}}
\snip{onode_connect2}{0}{0}{\isa{\mbox{}\inferrule{\mbox{{\isasymsigma}{\isacharprime}\ i\ {\isacharequal}\ {\isasymsigma}\ i}}{\mbox{{\isacharparenleft}{\isacharparenleft}{\isasymsigma},\ \NodeS\ i\ P\ R{\isacharparenright},\ connect{\isacharparenleft}i{\isacharprime},\ i{\isacharparenright},\ {\isacharparenleft}{\isasymsigma}{\isacharprime},\ \NodeS\ i\ P\ {R\ {\isasymunion}\ {\isacharbraceleft}i{\isacharprime}{\isacharbraceright}}{\isacharparenright}{\isacharparenright}{\isasymin}\onodesos\ S}}}}
\snip{onode_disconnect1}{0}{0}{\isa{\mbox{}\inferrule{\mbox{{\isasymsigma}{\isacharprime}\ i\ {\isacharequal}\ {\isasymsigma}\ i}}{\mbox{{\isacharparenleft}{\isacharparenleft}{\isasymsigma},\ \NodeS\ i\ P\ R{\isacharparenright},\ disconnect{\isacharparenleft}i,\ i{\isacharprime}{\isacharparenright},\ {\isacharparenleft}{\isasymsigma}{\isacharprime},\ \NodeS\ i\ P\ {\isacharparenleft}R\ {\isacharminus}\ {\isacharbraceleft}i{\isacharprime}{\isacharbraceright}{\isacharparenright}{\isacharparenright}{\isacharparenright}{\isasymin}\onodesos\ S}}}}
\snip{onode_disconnect2}{0}{0}{\isa{\mbox{}\inferrule{\mbox{{\isasymsigma}{\isacharprime}\ i\ {\isacharequal}\ {\isasymsigma}\ i}}{\mbox{{\isacharparenleft}{\isacharparenleft}{\isasymsigma},\ \NodeS\ i\ P\ R{\isacharparenright},\ disconnect{\isacharparenleft}i{\isacharprime},\ i{\isacharparenright},\ {\isacharparenleft}{\isasymsigma}{\isacharprime},\ \NodeS\ i\ P\ {\isacharparenleft}R\ {\isacharminus}\ {\isacharbraceleft}i{\isacharprime}{\isacharbraceright}{\isacharparenright}{\isacharparenright}{\isacharparenright}{\isasymin}\onodesos\ S}}}}
\snip{onode_connect_other}{0}{0}{\isa{\mbox{}\inferrule{\mbox{i\ {\isasymnoteq}\ i{\isacharprime}}\\\ \mbox{i\ {\isasymnoteq}\ i{\isacharprime}{\isacharprime}}\\\ \mbox{{\isasymsigma}{\isacharprime}\ i\ {\isacharequal}\ {\isasymsigma}\ i}}{\mbox{{\isacharparenleft}{\isacharparenleft}{\isasymsigma},\ \NodeS\ i\ P\ R{\isacharparenright},\ connect{\isacharparenleft}i{\isacharprime},\ i{\isacharprime}{\isacharprime}{\isacharparenright},\ {\isacharparenleft}{\isasymsigma}{\isacharprime},\ \NodeS\ i\ P\ R{\isacharparenright}{\isacharparenright}{\isasymin}\onodesos\ S}}}}
\snip{onode_disconnect_other}{0}{0}{\isa{\mbox{}\inferrule{\mbox{i\ {\isasymnoteq}\ i{\isacharprime}}\\\ \mbox{i\ {\isasymnoteq}\ i{\isacharprime}{\isacharprime}}\\\ \mbox{{\isasymsigma}{\isacharprime}\ i\ {\isacharequal}\ {\isasymsigma}\ i}}{\mbox{{\isacharparenleft}{\isacharparenleft}{\isasymsigma},\ \NodeS\ i\ P\ R{\isacharparenright},\ disconnect{\isacharparenleft}i{\isacharprime},\ i{\isacharprime}{\isacharprime}{\isacharparenright},\ {\isacharparenleft}{\isasymsigma}{\isacharprime},\ \NodeS\ i\ P\ R{\isacharparenright}{\isacharparenright}{\isasymin}\onodesos\ S}}}}

\snip{onode_tau_prem_2}{0}{0}{\isa{{\isasymforall}j {\isasymnoteq} i{\isachardot}\ {\isasymsigma}{\isacharprime}\ j\ {\isacharequal}\ {\isasymsigma}\ j}}

\snip{onode_comp}{0}{0}{\isa{{\isasymlangle}i\ {\isacharcolon}\ onp\ {\isacharcolon}\ \Ri{\isasymrangle}\isactrlsub o\ {\isacharequal}\ {\isasymlparr}init\ {\isacharequal}\ {\isacharbraceleft}{\isacharparenleft}{\isasymsigma},\ \NodeS\ i\ s\ \Ri{\isacharparenright}\ {\isacharbar}\ {\isacharparenleft}{\isasymsigma},\ s{\isacharparenright}{\isasymin}init\ onp{\isacharbraceright},\ trans\ {\isacharequal}\ \onodesos\ {\isacharparenleft}trans\ onp{\isacharparenright}{\isasymrparr}}}
\snip{onode_comp_term}{0}{0}{\isa{{\isasymlangle}i\ {\isacharcolon}\ S\ {\isacharcolon}\ R{\isasymrangle}\isactrlsub o{\isasymColon}{\isacharparenleft}{\isacharparenleft}nat\ {\isasymRightarrow}\ {\isacharprime}a{\isacharparenright}\ {\isasymtimes}\ {\isacharprime}b\ net{\isacharunderscore}state,\ {\isacharprime}c\ node{\isacharunderscore}action{\isacharparenright}\ automaton}}

\snip{opnet_cast1}{0}{0}{\isa{\mbox{}\inferrule{\mbox{{\isacharparenleft}{\isacharparenleft}{\isasymsigma},\ s{\isacharparenright},\ R{\isacharcolon}{\isacharasterisk}cast{\isacharparenleft}m{\isacharparenright},\ {\isasymsigma}{\isacharprime},\ s{\isacharprime}{\isacharparenright}{\isasymin}S}\\\ \mbox{{\isacharparenleft}{\isacharparenleft}{\isasymsigma},\ t{\isacharparenright},\ H{\isasymnot}K{\isacharcolon}arrive{\isacharparenleft}m{\isacharparenright},\ {\isasymsigma}{\isacharprime},\ t{\isacharprime}{\isacharparenright}{\isasymin}T}\\\ \mbox{H\ {\isasymsubseteq}\ R}\\\ \mbox{K\ {\isasyminter}\ R\ {\isacharequal}\ {\isasymemptyset}}}{\mbox{{\isacharparenleft}{\isacharparenleft}{\isasymsigma},\ \subnets{s}{t}{\isacharparenright},\ R{\isacharcolon}{\isacharasterisk}cast{\isacharparenleft}m{\isacharparenright},\ {\isacharparenleft}{\isasymsigma}{\isacharprime},\ \subnets{s{\isacharprime}}{t{\isacharprime}}{\isacharparenright}{\isacharparenright}{\isasymin}\opnetsos\ S\ T}}}}
\snip{opnet_cast2}{0}{0}{\isa{\mbox{}\inferrule{\mbox{{\isacharparenleft}{\isacharparenleft}{\isasymsigma},\ s{\isacharparenright},\ H{\isasymnot}K{\isacharcolon}arrive{\isacharparenleft}m{\isacharparenright},\ {\isasymsigma}{\isacharprime},\ s{\isacharprime}{\isacharparenright}{\isasymin}S}\\\ \mbox{{\isacharparenleft}{\isacharparenleft}{\isasymsigma},\ t{\isacharparenright},\ R{\isacharcolon}{\isacharasterisk}cast{\isacharparenleft}m{\isacharparenright},\ {\isasymsigma}{\isacharprime},\ t{\isacharprime}{\isacharparenright}{\isasymin}T}\\\ \mbox{H\ {\isasymsubseteq}\ R}\\\ \mbox{K\ {\isasyminter}\ R\ {\isacharequal}\ {\isasymemptyset}}}{\mbox{{\isacharparenleft}{\isacharparenleft}{\isasymsigma},\ \subnets{s}{t}{\isacharparenright},\ R{\isacharcolon}{\isacharasterisk}cast{\isacharparenleft}m{\isacharparenright},\ {\isacharparenleft}{\isasymsigma}{\isacharprime},\ \subnets{s{\isacharprime}}{t{\isacharprime}}{\isacharparenright}{\isacharparenright}{\isasymin}\opnetsos\ S\ T}}}}
\snip{opnet_arrive}{0}{0}{\isa{\mbox{}\inferrule{\mbox{{\isacharparenleft}{\isacharparenleft}{\isasymsigma},\ s{\isacharparenright},\ H{\isasymnot}K{\isacharcolon}arrive{\isacharparenleft}m{\isacharparenright},\ {\isacharparenleft}{\isasymsigma}{\isacharprime},\ s{\isacharprime}{\isacharparenright}{\isacharparenright}{\isasymin}S}\\\ \mbox{{\isacharparenleft}{\isacharparenleft}{\isasymsigma},\ t{\isacharparenright},\ H{\isacharprime}{\isasymnot}K{\isacharprime}{\isacharcolon}arrive{\isacharparenleft}m{\isacharparenright},\ {\isacharparenleft}{\isasymsigma}{\isacharprime},\ t{\isacharprime}{\isacharparenright}{\isacharparenright}{\isasymin}T}}{\mbox{{\isacharparenleft}{\isacharparenleft}{\isasymsigma},\ \subnets{s}{t}{\isacharparenright},\ {\isacharparenleft}H\ {\isasymunion}\ H{\isacharprime}{\isacharparenright}{\isasymnot}{\isacharparenleft}K\ {\isasymunion}\ K{\isacharprime}{\isacharparenright}{\isacharcolon}arrive{\isacharparenleft}m{\isacharparenright},\ {\isacharparenleft}{\isasymsigma}{\isacharprime},\ \subnetsadj{s{\isacharprime}}{t{\isacharprime}}{\isacharparenright}{\isacharparenright}{\isasymin}\opnetsos\ S\ T}}}}
\snip{opnet_deliver1}{0}{0}{\isa{\mbox{}\inferrule{\mbox{{\isacharparenleft}{\isacharparenleft}{\isasymsigma},\ s{\isacharparenright},\ i{\isacharcolon}deliver{\isacharparenleft}d{\isacharparenright},\ {\isasymsigma}{\isacharprime},\ s{\isacharprime}{\isacharparenright}{\isasymin}S}}{\mbox{{\isacharparenleft}{\isacharparenleft}{\isasymsigma},\ \subnets{s}{t}{\isacharparenright},\ i{\isacharcolon}deliver{\isacharparenleft}d{\isacharparenright},\ {\isacharparenleft}{\isasymsigma}{\isacharprime},\ \subnets{s{\isacharprime}}{t}{\isacharparenright}{\isacharparenright}{\isasymin}\opnetsos\ S\ T}}}}
\snip{opnet_deliver2}{0}{0}{\isa{\mbox{}\inferrule{\mbox{{\isacharparenleft}{\isacharparenleft}{\isasymsigma},\ t{\isacharparenright},\ i{\isacharcolon}deliver{\isacharparenleft}d{\isacharparenright},\ {\isasymsigma}{\isacharprime},\ t{\isacharprime}{\isacharparenright}{\isasymin}T}}{\mbox{{\isacharparenleft}{\isacharparenleft}{\isasymsigma},\ \subnets{s}{t}{\isacharparenright},\ i{\isacharcolon}deliver{\isacharparenleft}d{\isacharparenright},\ {\isacharparenleft}{\isasymsigma}{\isacharprime},\ \subnets{s}{t{\isacharprime}}{\isacharparenright}{\isacharparenright}{\isasymin}\opnetsos\ S\ T}}}}
\snip{opnet_tau1}{0}{0}{\isa{\mbox{}\inferrule{\mbox{{\isacharparenleft}{\isacharparenleft}{\isasymsigma},\ s{\isacharparenright},\ {\isasymtau},\ {\isasymsigma}{\isacharprime},\ s{\isacharprime}{\isacharparenright}{\isasymin}S}}{\mbox{{\isacharparenleft}{\isacharparenleft}{\isasymsigma},\ \subnets{s}{t}{\isacharparenright},\ {\isasymtau},\ {\isacharparenleft}{\isasymsigma}{\isacharprime},\ \subnets{s{\isacharprime}}{t}{\isacharparenright}{\isacharparenright}{\isasymin}\opnetsos\ S\ T}}}}
\snip{opnet_tau2}{0}{0}{\isa{\mbox{}\inferrule{\mbox{{\isacharparenleft}{\isacharparenleft}{\isasymsigma},\ t{\isacharparenright},\ {\isasymtau},\ {\isasymsigma}{\isacharprime},\ t{\isacharprime}{\isacharparenright}{\isasymin}T}}{\mbox{{\isacharparenleft}{\isacharparenleft}{\isasymsigma},\ \subnets{s}{t}{\isacharparenright},\ {\isasymtau},\ {\isacharparenleft}{\isasymsigma}{\isacharprime},\ \subnets{s}{t{\isacharprime}}{\isacharparenright}{\isacharparenright}{\isasymin}\opnetsos\ S\ T}}}}
\snip{opnet_connect}{0}{0}{\isa{\mbox{}\inferrule{\mbox{{\isacharparenleft}{\isacharparenleft}{\isasymsigma},\ s{\isacharparenright},\ connect{\isacharparenleft}i,\ i{\isacharprime}{\isacharparenright},\ {\isasymsigma}{\isacharprime},\ s{\isacharprime}{\isacharparenright}{\isasymin}S}\\\ \mbox{{\isacharparenleft}{\isacharparenleft}{\isasymsigma},\ t{\isacharparenright},\ connect{\isacharparenleft}i,\ i{\isacharprime}{\isacharparenright},\ {\isasymsigma}{\isacharprime},\ t{\isacharprime}{\isacharparenright}{\isasymin}T}}{\mbox{{\isacharparenleft}{\isacharparenleft}{\isasymsigma},\ \subnets{s}{t}{\isacharparenright},\ connect{\isacharparenleft}i,\ i{\isacharprime}{\isacharparenright},\ {\isacharparenleft}{\isasymsigma}{\isacharprime},\ \subnets{s{\isacharprime}}{t{\isacharprime}}{\isacharparenright}{\isacharparenright}{\isasymin}\opnetsos\ S\ T}}}}
\snip{opnet_disconnect}{0}{0}{\isa{\mbox{}\inferrule{\mbox{{\isacharparenleft}{\isacharparenleft}{\isasymsigma},\ s{\isacharparenright},\ disconnect{\isacharparenleft}i,\ i{\isacharprime}{\isacharparenright},\ {\isasymsigma}{\isacharprime},\ s{\isacharprime}{\isacharparenright}{\isasymin}S}\\\ \mbox{{\isacharparenleft}{\isacharparenleft}{\isasymsigma},\ t{\isacharparenright},\ disconnect{\isacharparenleft}i,\ i{\isacharprime}{\isacharparenright},\ {\isasymsigma}{\isacharprime},\ t{\isacharprime}{\isacharparenright}{\isasymin}T}}{\mbox{{\isacharparenleft}{\isacharparenleft}{\isasymsigma},\ \subnets{s}{t}{\isacharparenright},\ disconnect{\isacharparenleft}i,\ i{\isacharprime}{\isacharparenright},\ {\isacharparenleft}{\isasymsigma}{\isacharprime},\ \subnets{s{\isacharprime}}{t{\isacharprime}}{\isacharparenright}{\isacharparenright}{\isasymin}\opnetsos\ S\ T}}}}

\snip{opnet}{0}{0}{\isa{opnet\ onp\ {\isasymlangle}i{\isacharsemicolon}\ \Ri{\isasymrangle}\ {\isacharequal}\ {\isasymlangle}i\ {\isacharcolon}\ onp\ i\ {\isacharcolon}\ \Ri{\isasymrangle}\isactrlsub o\isasep\isanewline%
opnet\ onp\ {\isacharparenleft}p\isactrlsub {\isadigit{1}}\parallelcomp{}p\isactrlsub {\isadigit{2}}{\isacharparenright}\ {\isacharequal}\ {\isasymlparr}init\ {\isacharequal}\ {\isacharbraceleft}{\isacharparenleft}{\isasymsigma},\ \subnets{s\isactrlsub {\isadigit{1}}}{s\isactrlsub {\isadigit{2}}}{\isacharparenright}\ {\isacharbar}\ {\isacharparenleft}{\isasymsigma},\ s\isactrlsub {\isadigit{1}}{\isacharparenright}{\isasymin}init\ {\isacharparenleft}opnet\ onp\ p\isactrlsub {\isadigit{1}}{\isacharparenright}\ {\isasymand}\ {\isacharparenleft}{\isasymsigma},\ s\isactrlsub {\isadigit{2}}{\isacharparenright}{\isasymin}init\ {\isacharparenleft}opnet\ onp\ p\isactrlsub {\isadigit{2}}{\isacharparenright}\ {\isasymand}\ net{\isacharunderscore}ips\ s\isactrlsub {\isadigit{1}}\ {\isasyminter}\ net{\isacharunderscore}ips\ s\isactrlsub {\isadigit{2}}\ {\isacharequal}\ {\isasymemptyset}{\isacharbraceright},\ trans\ {\isacharequal}\ \opnetsos\ {\isacharparenleft}trans\ {\isacharparenleft}opnet\ onp\ p\isactrlsub {\isadigit{1}}{\isacharparenright}{\isacharparenright}\ {\isacharparenleft}trans\ {\isacharparenleft}opnet\ onp\ p\isactrlsub {\isadigit{2}}{\isacharparenright}{\isacharparenright}{\isasymrparr}}}

\snip{ocnet_connect}{0}{0}{\isa{\mbox{}\inferrule{\mbox{{\isacharparenleft}{\isacharparenleft}{\isasymsigma},\ s{\isacharparenright},\ connect{\isacharparenleft}i,\ i{\isacharprime}{\isacharparenright},\ {\isasymsigma}{\isacharprime},\ s{\isacharprime}{\isacharparenright}{\isasymin}S}\\\ \mbox{{\isasymforall}j{\isachardot}\ j\ {\isasymnotin}\ net{\isacharunderscore}ips\ s\ {\isasymlongrightarrow}\ {\isasymsigma}{\isacharprime}\ j\ {\isacharequal}\ {\isasymsigma}\ j}}{\mbox{{\isacharparenleft}{\isacharparenleft}{\isasymsigma},\ s{\isacharparenright},\ connect{\isacharparenleft}i,\ i{\isacharprime}{\isacharparenright},\ {\isasymsigma}{\isacharprime},\ s{\isacharprime}{\isacharparenright}{\isasymin}\ocnetsos\ S}}}}
\snip{ocnet_disconnect}{0}{0}{\isa{\mbox{}\inferrule{\mbox{{\isacharparenleft}{\isacharparenleft}{\isasymsigma},\ s{\isacharparenright},\ disconnect{\isacharparenleft}i,\ i{\isacharprime}{\isacharparenright},\ {\isasymsigma}{\isacharprime},\ s{\isacharprime}{\isacharparenright}{\isasymin}S}\\\ \mbox{{\isasymforall}j{\isachardot}\ j\ {\isasymnotin}\ net{\isacharunderscore}ips\ s\ {\isasymlongrightarrow}\ {\isasymsigma}{\isacharprime}\ j\ {\isacharequal}\ {\isasymsigma}\ j}}{\mbox{{\isacharparenleft}{\isacharparenleft}{\isasymsigma},\ s{\isacharparenright},\ disconnect{\isacharparenleft}i,\ i{\isacharprime}{\isacharparenright},\ {\isasymsigma}{\isacharprime},\ s{\isacharprime}{\isacharparenright}{\isasymin}\ocnetsos\ S}}}}
\snip{ocnet_cast}{0}{0}{\isa{\mbox{}\inferrule{\mbox{{\isacharparenleft}{\isacharparenleft}{\isasymsigma},\ s{\isacharparenright},\ R{\isacharcolon}{\isacharasterisk}cast{\isacharparenleft}m{\isacharparenright},\ {\isasymsigma}{\isacharprime},\ s{\isacharprime}{\isacharparenright}{\isasymin}S}\\\ \mbox{{\isasymforall}j{\isachardot}\ j\ {\isasymnotin}\ net{\isacharunderscore}ips\ s\ {\isasymlongrightarrow}\ {\isasymsigma}{\isacharprime}\ j\ {\isacharequal}\ {\isasymsigma}\ j}}{\mbox{{\isacharparenleft}{\isacharparenleft}{\isasymsigma},\ s{\isacharparenright},\ {\isasymtau},\ {\isasymsigma}{\isacharprime},\ s{\isacharprime}{\isacharparenright}{\isasymin}\ocnetsos\ S}}}}
\snip{ocnet_tau}{0}{0}{\isa{\mbox{}\inferrule{\mbox{{\isacharparenleft}{\isacharparenleft}{\isasymsigma},\ s{\isacharparenright},\ {\isasymtau},\ {\isasymsigma}{\isacharprime},\ s{\isacharprime}{\isacharparenright}{\isasymin}S}\\\ \mbox{{\isasymforall}j{\isachardot}\ j\ {\isasymnotin}\ net{\isacharunderscore}ips\ s\ {\isasymlongrightarrow}\ {\isasymsigma}{\isacharprime}\ j\ {\isacharequal}\ {\isasymsigma}\ j}}{\mbox{{\isacharparenleft}{\isacharparenleft}{\isasymsigma},\ s{\isacharparenright},\ {\isasymtau},\ {\isasymsigma}{\isacharprime},\ s{\isacharprime}{\isacharparenright}{\isasymin}\ocnetsos\ S}}}}
\snip{ocnet_deliver}{0}{0}{\isa{\mbox{}\inferrule{\mbox{{\isacharparenleft}{\isacharparenleft}{\isasymsigma},\ s{\isacharparenright},\ i{\isacharcolon}deliver{\isacharparenleft}d{\isacharparenright},\ {\isasymsigma}{\isacharprime},\ s{\isacharprime}{\isacharparenright}{\isasymin}S}\\\ \mbox{{\isasymforall}j{\isachardot}\ j\ {\isasymnotin}\ net{\isacharunderscore}ips\ s\ {\isasymlongrightarrow}\ {\isasymsigma}{\isacharprime}\ j\ {\isacharequal}\ {\isasymsigma}\ j}}{\mbox{{\isacharparenleft}{\isacharparenleft}{\isasymsigma},\ s{\isacharparenright},\ i{\isacharcolon}deliver{\isacharparenleft}d{\isacharparenright},\ {\isasymsigma}{\isacharprime},\ s{\isacharprime}{\isacharparenright}{\isasymin}\ocnetsos\ S}}}}
\snip{ocnet_newpkt}{0}{0}{\isa{\mbox{}\inferrule{\mbox{{\isacharparenleft}{\isacharparenleft}{\isasymsigma},\ s{\isacharparenright},\ {\isacharbraceleft}i{\isacharbraceright}{\isasymnot}K{\isacharcolon}arrive{\isacharparenleft}newpkt\ {\isacharparenleft}d,\ di{\isacharparenright}{\isacharparenright},\ {\isasymsigma}{\isacharprime},\ s{\isacharprime}{\isacharparenright}{\isasymin}S}\\\ \mbox{{\isasymforall}j{\isachardot}\ j\ {\isasymnotin}\ net{\isacharunderscore}ips\ s\ {\isasymlongrightarrow}\ {\isasymsigma}{\isacharprime}\ j\ {\isacharequal}\ {\isasymsigma}\ j}}{\mbox{{\isacharparenleft}{\isacharparenleft}{\isasymsigma},\ s{\isacharparenright},\ i{\isacharcolon}newpkt{\isacharparenleft}d,\ di{\isacharparenright},\ {\isasymsigma}{\isacharprime},\ s{\isacharprime}{\isacharparenright}{\isasymin}\ocnetsos\ S}}}}

\snip{oclosed}{0}{0}{\isa{oclosed\ {\isacharparenleft}opnet\ {\isacharparenleft}{\isasymlambda}i{\isachardot}\ optoy\ i\ {\isasymlangle}{\isasymlangle}\isactrlbsub i\isactrlesub \ qmsg{\isacharparenright}\ n{\isacharparenright}}}
\snip{oclosed_withtype}{0}{0}{\isa{oclosed\ {\isacharparenleft}opnet\ {\isacharparenleft}{\isasymlambda}i{\isachardot}\ optoy\ i\ {\isasymlangle}{\isasymlangle}\isactrlbsub i\isactrlesub \ qmsg{\isacharparenright}\ n{\isacharparenright}{\isasymColon}{\isacharparenleft}{\isacharparenleft}nat\ {\isasymRightarrow}\ state{\isacharparenright}\ {\isasymtimes}\ {\isacharparenleft}{\isacharparenleft}state,\ msg,\ pseqp,\ pseqp\ label{\isacharparenright}\ seqp\ {\isasymtimes}\ msg\ list\ {\isasymtimes}\ {\isacharparenleft}msg\ list,\ msg,\ unit,\ unit\ label{\isacharparenright}\ seqp{\isacharparenright}\ net{\isacharunderscore}state,\ msg\ node{\isacharunderscore}action{\isacharparenright}\ automaton}}
\snip{microstep_choiceI1}{0}{0}{\isa{{\isacharparenleft}p\isactrlsub {\isadigit{1}}\ {\isasymoplus}\ p\isactrlsub {\isadigit{2}}{\isacharparenright}\ {\isasymleadsto}\isactrlbsub {\isasymGamma}\isactrlesub \ p\isactrlsub {\isadigit{1}}}}
\snip{microstep_choiceI2}{0}{0}{\isa{{\isacharparenleft}p\isactrlsub {\isadigit{1}}\ {\isasymoplus}\ p\isactrlsub {\isadigit{2}}{\isacharparenright}\ {\isasymleadsto}\isactrlbsub {\isasymGamma}\isactrlesub \ p\isactrlsub {\isadigit{2}}}}
\snip{microstep_callI}{0}{0}{\isa{{\isacharparenleft}call{\isacharparenleft}pn{\isacharparenright}{\isacharparenright}\ {\isasymleadsto}\isactrlbsub {\isasymGamma}\isactrlesub \ {\isasymGamma}\ pn}}

\snip{wellformed}{0}{0}{\isa{wellformed\ {\isasymGamma}\ {\isacharequal}\ wf\ {\isacharbraceleft}{\isacharparenleft}q,\ p{\isacharparenright}\ {\isacharbar}\ p\ {\isasymleadsto}\isactrlbsub {\isasymGamma}\isactrlesub \ q{\isacharbraceright}}}

\snip{sterms_termination}{0}{0}{\isa{wellformed\ {\isasymGamma}\ {\isasymLongrightarrow}\ sterms{\isacharunderscore}dom\ {\isacharparenleft}{\isasymGamma},\ p{\isacharparenright}}}
\snip{sterms_termination_concl}{0}{0}{\isa{sterms{\isacharunderscore}dom\ {\isacharparenleft}{\isasymGamma},\ p{\isacharparenright}}}

\snip{sterms_choice}{0}{0}{\isa{\mbox{}\inferrule{\mbox{wellformed\ {\isasymGamma}}}{\mbox{sterms\ {\isasymGamma}\ {\isacharparenleft}p\isactrlsub {\isadigit{1}}\ {\isasymoplus}\ p\isactrlsub {\isadigit{2}}{\isacharparenright}\ {\isacharequal}\ sterms\ {\isasymGamma}\ p\isactrlsub {\isadigit{1}}\ {\isasymunion}\ sterms\ {\isasymGamma}\ p\isactrlsub {\isadigit{2}}}}}}

\snip{sterms_call}{0}{0}{\isa{\mbox{}\inferrule{\mbox{wellformed\ {\isasymGamma}}}{\mbox{sterms\ {\isasymGamma}\ {\isacharparenleft}call{\isacharparenleft}pn{\isacharparenright}{\isacharparenright}\ {\isacharequal}\ sterms\ {\isasymGamma}\ {\isacharparenleft}{\isasymGamma}\ pn{\isacharparenright}}}}}
\snip{sterms_other1}{0}{0}{\isa{\mbox{}\inferrule{\mbox{wellformed\ {\isasymGamma}}}{\mbox{sterms\ {\isasymGamma}\ {\isacharparenleft}{\isacharbraceleft}v{\isacharbraceright}{\isasymlangle}va{\isasymrangle}\ vb{\isacharparenright}\ {\isacharequal}\ {\isacharbraceleft}{\isacharbraceleft}v{\isacharbraceright}{\isasymlangle}va{\isasymrangle}\ vb{\isacharbraceright}}}}}
\snip{sterms_other2}{0}{0}{\isa{\mbox{}\inferrule{\mbox{wellformed\ {\isasymGamma}}}{\mbox{sterms\ {\isasymGamma}\ {\isacharparenleft}{\isacharbraceleft}v{\isacharbraceright}{\isasymlbrakk}va{\isasymrbrakk}\ vb{\isacharparenright}\ {\isacharequal}\ {\isacharbraceleft}{\isacharbraceleft}v{\isacharbraceright}{\isasymlbrakk}va{\isasymrbrakk}\ vb{\isacharbraceright}}}}}

\snip{sterms_choice_concl}{0}{0}{\isa{sterms\ {\isasymGamma}\ {\isacharparenleft}p\isactrlsub {\isadigit{1}}\ {\isasymoplus}\ p\isactrlsub {\isadigit{2}}{\isacharparenright}\ {\isacharequal}\ sterms\ {\isasymGamma}\ p\isactrlsub {\isadigit{1}}\ {\isasymunion}\ sterms\ {\isasymGamma}\ p\isactrlsub {\isadigit{2}}}}
\snip{sterms_choice_concl_lhs}{0}{0}{\isa{sterms\ {\isasymGamma}\ {\isacharparenleft}p\isactrlsub {\isadigit{1}}\ {\isasymoplus}\ p\isactrlsub {\isadigit{2}}{\isacharparenright}}}
\snip{sterms_choice_concl_rhs}{0}{0}{\isa{sterms\ {\isasymGamma}\ p\isactrlsub {\isadigit{1}}\ {\isasymunion}\ sterms\ {\isasymGamma}\ p\isactrlsub {\isadigit{2}}}}

\snip{sterms_other}{0}{0}{\isa{sterms\ {\isasymGamma}\ p\ {\isacharequal}\ {\isacharbraceleft}p{\isacharbraceright}}}
\snip{sterms_other_lhs}{0}{0}{\isa{sterms\ {\isasymGamma}\ p}}
\snip{sterms_other_rhs}{0}{0}{\isa{{\isacharbraceleft}p{\isacharbraceright}}}

\snip{sterms_call_concl}{0}{0}{\isa{sterms\ {\isasymGamma}\ {\isacharparenleft}call{\isacharparenleft}pn{\isacharparenright}{\isacharparenright}\ {\isacharequal}\ sterms\ {\isasymGamma}\ {\isacharparenleft}{\isasymGamma}\ pn{\isacharparenright}}}
\snip{sterms_call_concl_lhs}{0}{0}{\isa{sterms\ {\isasymGamma}\ {\isacharparenleft}call{\isacharparenleft}pn{\isacharparenright}{\isacharparenright}}}
\snip{sterms_call_concl_rhs}{0}{0}{\isa{sterms\ {\isasymGamma}\ {\isacharparenleft}{\isasymGamma}\ pn{\isacharparenright}}}

\snip{sterms_other1_concl}{0}{0}{\isa{sterms\ {\isasymGamma}\ {\isacharparenleft}{\isacharbraceleft}v{\isacharbraceright}{\isasymlangle}va{\isasymrangle}\ vb{\isacharparenright}\ {\isacharequal}\ {\isacharbraceleft}{\isacharbraceleft}v{\isacharbraceright}{\isasymlangle}va{\isasymrangle}\ vb{\isacharbraceright}}}
\snip{sterms_other2_concl}{0}{0}{\isa{sterms\ {\isasymGamma}\ {\isacharparenleft}{\isacharbraceleft}v{\isacharbraceright}{\isasymlbrakk}va{\isasymrbrakk}\ vb{\isacharparenright}\ {\isacharequal}\ {\isacharbraceleft}{\isacharbraceleft}v{\isacharbraceright}{\isasymlbrakk}va{\isasymrbrakk}\ vb{\isacharbraceright}}}

\snip{sterms_maximal_microstep}{0}{0}{\isa{\mbox{}\inferrule{\mbox{wellformed\ {\isasymGamma}}}{\mbox{sterms\ {\isasymGamma}\ p\ {\isacharequal}\ {\isacharbraceleft}q\ {\isacharbar}\ p\ {\isasymleadsto}\isactrlbsub {\isasymGamma}\isactrlesub \isactrlsup {\isacharasterisk}\ q\ {\isasymand}\ {\isacharparenleft}{\isasymnexists}q{\isacharprime}{\isachardot}\ q\ {\isasymleadsto}\isactrlbsub {\isasymGamma}\isactrlesub \ q{\isacharprime}{\isacharparenright}{\isacharbraceright}}}}}
\snip{sterms_maximal_microstep_rhs}{0}{0}{\isa{{\isacharbraceleft}q\ {\isacharbar}\ p\ {\isasymleadsto}\isactrlbsub {\isasymGamma}\isactrlesub \isactrlsup {\isacharasterisk}\ q\ {\isasymand}\ {\isacharparenleft}{\isasymnexists}q{\isacharprime}{\isachardot}\ q\ {\isasymleadsto}\isactrlbsub {\isasymGamma}\isactrlesub \ q{\isacharprime}{\isacharparenright}{\isacharbraceright}}}

\snip{dterms_choice}{0}{0}{\isa{\mbox{}\inferrule{\mbox{wellformed\ {\isasymGamma}}}{\mbox{dterms\ {\isasymGamma}\ {\isacharparenleft}p{\isadigit{1}}{\isachardot}{\isadigit{0}}\ {\isasymoplus}\ p{\isadigit{2}}{\isachardot}{\isadigit{0}}{\isacharparenright}\ {\isacharequal}\ dterms\ {\isasymGamma}\ p{\isadigit{1}}{\isachardot}{\isadigit{0}}\ {\isasymunion}\ dterms\ {\isasymGamma}\ p{\isadigit{2}}{\isachardot}{\isadigit{0}}}}}}
\snip{dterms_call}{0}{0}{\isa{\mbox{}\inferrule{\mbox{wellformed\ {\isasymGamma}}}{\mbox{dterms\ {\isasymGamma}\ {\isacharparenleft}call{\isacharparenleft}pn{\isacharparenright}{\isacharparenright}\ {\isacharequal}\ dterms\ {\isasymGamma}\ {\isacharparenleft}{\isasymGamma}\ pn{\isacharparenright}}}}}
\snip{dterms_other1}{0}{0}{\isa{\mbox{}\inferrule{\mbox{wellformed\ {\isasymGamma}}}{\mbox{dterms\ {\isasymGamma}\ {\isacharparenleft}\gray{{\isacharbraceleft}l{\isacharbraceright}}{\isasymlangle}g{\isasymrangle}\ p{\isacharparenright}\ {\isacharequal}\ sterms\ {\isasymGamma}\ p}}}}
\snip{dterms_other2}{0}{0}{\isa{\mbox{}\inferrule{\mbox{wellformed\ {\isasymGamma}}}{\mbox{dterms\ {\isasymGamma}\ {\isacharparenleft}\gray{{\isacharbraceleft}l{\isacharbraceright}}{\isasymlbrakk}u{\isasymrbrakk}\ p{\isacharparenright}\ {\isacharequal}\ sterms\ {\isasymGamma}\ p}}}}

\snip{dterms_choice_concl}{0}{0}{\isa{dterms\ {\isasymGamma}\ {\isacharparenleft}p\isactrlsub {\isadigit{1}}\ {\isasymoplus}\ p\isactrlsub {\isadigit{2}}{\isacharparenright}\ {\isacharequal}\ dterms\ {\isasymGamma}\ p\isactrlsub {\isadigit{1}}\ {\isasymunion}\ dterms\ {\isasymGamma}\ p\isactrlsub {\isadigit{2}}}}
\snip{dterms_choice_concl_lhs}{0}{0}{\isa{dterms\ {\isasymGamma}\ {\isacharparenleft}p\isactrlsub {\isadigit{1}}\ {\isasymoplus}\ p\isactrlsub {\isadigit{2}}{\isacharparenright}}}
\snip{dterms_choice_concl_rhs}{0}{0}{\isa{dterms\ {\isasymGamma}\ p\isactrlsub {\isadigit{1}}\ {\isasymunion}\ dterms\ {\isasymGamma}\ p\isactrlsub {\isadigit{2}}}}

\snip{dterms_call_concl}{0}{0}{\isa{dterms\ {\isasymGamma}\ {\isacharparenleft}call{\isacharparenleft}pn{\isacharparenright}{\isacharparenright}\ {\isacharequal}\ dterms\ {\isasymGamma}\ {\isacharparenleft}{\isasymGamma}\ pn{\isacharparenright}}}
\snip{dterms_call_concl_lhs}{0}{0}{\isa{dterms\ {\isasymGamma}\ {\isacharparenleft}call{\isacharparenleft}pn{\isacharparenright}{\isacharparenright}}}
\snip{dterms_call_concl_rhs}{0}{0}{\isa{dterms\ {\isasymGamma}\ {\isacharparenleft}{\isasymGamma}\ pn{\isacharparenright}}}

\snip{dterms_other1_concl}{0}{0}{\isa{dterms\ {\isasymGamma}\ {\isacharparenleft}\gray{{\isacharbraceleft}l{\isacharbraceright}}{\isasymlangle}g{\isasymrangle}\ p{\isacharparenright}\ {\isacharequal}\ sterms\ {\isasymGamma}\ p}}

\snip{dterms_other2_concl}{0}{0}{\isa{dterms\ {\isasymGamma}\ {\isacharparenleft}\gray{{\isacharbraceleft}l{\isacharbraceright}}{\isasymlbrakk}u{\isasymrbrakk}\ p{\isacharparenright}\ {\isacharequal}\ sterms\ {\isasymGamma}\ p}}
\snip{dterms_other2_concl_lhs}{0}{0}{\isa{dterms\ {\isasymGamma}\ {\isacharparenleft}\gray{{\isacharbraceleft}l{\isacharbraceright}}{\isasymlbrakk}u{\isasymrbrakk}\ p{\isacharparenright}}}
\snip{dterms_other2_concl_rhs}{0}{0}{\isa{sterms\ {\isasymGamma}\ p}}

\snip{dterms_unicast_concl}{0}{0}{\isa{dterms\ {\isasymGamma}\ {\isacharparenleft}\gray{{\isacharbraceleft}l{\isacharbraceright}}unicast{\isacharparenleft}\selip,\ \selmsg{\isacharparenright}\ {\isachardot}\ p\ {\isasymtriangleright}\ q{\isacharparenright}\ {\isacharequal}\ sterms\ {\isasymGamma}\ p\ {\isasymunion}\ sterms\ {\isasymGamma}\ q}}
\snip{dterms_unicast_concl_lhs}{0}{0}{\isa{dterms\ {\isasymGamma}\ {\isacharparenleft}\gray{{\isacharbraceleft}l{\isacharbraceright}}unicast{\isacharparenleft}\selip,\ \selmsg{\isacharparenright}\ {\isachardot}\ p\ {\isasymtriangleright}\ q{\isacharparenright}}}
\snip{dterms_unicast_concl_rhs}{0}{0}{\isa{sterms\ {\isasymGamma}\ p\ {\isasymunion}\ sterms\ {\isasymGamma}\ q}}

\snip{dterms}{0}{0}{\isa{wellformed\ {\isasymGamma}\ {\isasymLongrightarrow}\ dterms\ {\isasymGamma}\ {\isacharparenleft}\gray{{\isacharbraceleft}l{\isacharbraceright}}{\isasymlangle}g{\isasymrangle}\ p{\isacharparenright}\ {\isacharequal}\ sterms\ {\isasymGamma}\ p\isasep\isanewline%
wellformed\ {\isasymGamma}\ {\isasymLongrightarrow}\ dterms\ {\isasymGamma}\ {\isacharparenleft}\gray{{\isacharbraceleft}l{\isacharbraceright}}{\isasymlbrakk}u{\isasymrbrakk}\ p{\isacharparenright}\ {\isacharequal}\ sterms\ {\isasymGamma}\ p\isasep\isanewline%
wellformed\ {\isasymGamma}\ {\isasymLongrightarrow}\ dterms\ {\isasymGamma}\ {\isacharparenleft}p{\isadigit{1}}{\isachardot}{\isadigit{0}}\ {\isasymoplus}\ p{\isadigit{2}}{\isachardot}{\isadigit{0}}{\isacharparenright}\ {\isacharequal}\ dterms\ {\isasymGamma}\ p{\isadigit{1}}{\isachardot}{\isadigit{0}}\ {\isasymunion}\ dterms\ {\isasymGamma}\ p{\isadigit{2}}{\isachardot}{\isadigit{0}}\isasep\isanewline%
wellformed\ {\isasymGamma}\ {\isasymLongrightarrow}\ dterms\ {\isasymGamma}\ {\isacharparenleft}\gray{{\isacharbraceleft}l{\isacharbraceright}}unicast{\isacharparenleft}\selip,\ \selmsg{\isacharparenright}\ {\isachardot}\ p\ {\isasymtriangleright}\ q{\isacharparenright}\ {\isacharequal}\ sterms\ {\isasymGamma}\ p\ {\isasymunion}\ sterms\ {\isasymGamma}\ q\isasep\isanewline%
wellformed\ {\isasymGamma}\ {\isasymLongrightarrow}\ dterms\ {\isasymGamma}\ {\isacharparenleft}\gray{{\isacharbraceleft}l{\isacharbraceright}}broadcast{\isacharparenleft}\selmsg{\isacharparenright}\ {\isachardot}\ p{\isacharparenright}\ {\isacharequal}\ sterms\ {\isasymGamma}\ p\isasep\isanewline%
wellformed\ {\isasymGamma}\ {\isasymLongrightarrow}\ dterms\ {\isasymGamma}\ {\isacharparenleft}\gray{{\isacharbraceleft}l{\isacharbraceright}}groupcast{\isacharparenleft}\selips,\ \selmsg{\isacharparenright}\ {\isachardot}\ p{\isacharparenright}\ {\isacharequal}\ sterms\ {\isasymGamma}\ p\isasep\isanewline%
wellformed\ {\isasymGamma}\ {\isasymLongrightarrow}\ dterms\ {\isasymGamma}\ {\isacharparenleft}\gray{{\isacharbraceleft}l{\isacharbraceright}}send{\isacharparenleft}\selmsg{\isacharparenright}\ {\isachardot}\ p{\isacharparenright}\ {\isacharequal}\ sterms\ {\isasymGamma}\ p\isasep\isanewline%
wellformed\ {\isasymGamma}\ {\isasymLongrightarrow}\ dterms\ {\isasymGamma}\ {\isacharparenleft}\gray{{\isacharbraceleft}l{\isacharbraceright}}deliver{\isacharparenleft}\seldata{\isacharparenright}\ {\isachardot}\ p{\isacharparenright}\ {\isacharequal}\ sterms\ {\isasymGamma}\ p\isasep\isanewline%
wellformed\ {\isasymGamma}\ {\isasymLongrightarrow}\ dterms\ {\isasymGamma}\ {\isacharparenleft}\gray{{\isacharbraceleft}l{\isacharbraceright}}receive{\isacharparenleft}\updmsg{\isacharparenright}\ {\isachardot}\ p{\isacharparenright}\ {\isacharequal}\ sterms\ {\isasymGamma}\ p\isasep\isanewline%
wellformed\ {\isasymGamma}\ {\isasymLongrightarrow}\ dterms\ {\isasymGamma}\ {\isacharparenleft}call{\isacharparenleft}pn{\isacharparenright}{\isacharparenright}\ {\isacharequal}\ dterms\ {\isasymGamma}\ {\isacharparenleft}{\isasymGamma}\ pn{\isacharparenright}}}

\snip{wf_no_direct_calls}{0}{0}{\isa{\mbox{}\inferrule{\mbox{{\isasymAnd}pn{\isachardot}\ {\isasymforall}pn{\isacharprime}{\isachardot}\ call{\isacharparenleft}pn{\isacharprime}{\isacharparenright}\ {\isasymnotin}\ stermsl\ {\isacharparenleft}{\isasymGamma}\ pn{\isacharparenright}}}{\mbox{wellformed\ {\isasymGamma}}}}}
\snip{wf_no_direct_calls_prem_1}{0}{0}{\isa{call{\isacharparenleft}pn{\isacharprime}{\isacharparenright}\ {\isasymnotin}\ stermsl\ {\isacharparenleft}{\isasymGamma}\ pn{\isacharparenright}}}

\snip{cterms_SI}{0}{0}{\isa{\mbox{}\inferrule{\mbox{p{\isasymin}sterms\ {\isasymGamma}\ {\isacharparenleft}{\isasymGamma}\ pn{\isacharparenright}}}{\mbox{p{\isasymin}cterms\ {\isasymGamma}}}}}
\snip{cterms_DI}{0}{0}{\isa{\mbox{}\inferrule{\mbox{pp{\isasymin}cterms\ {\isasymGamma}}\\\ \mbox{p{\isasymin}dterms\ {\isasymGamma}\ pp}}{\mbox{p{\isasymin}cterms\ {\isasymGamma}}}}}

\snip{cterms_def'}{0}{0}{\isa{\mbox{}\inferrule{\mbox{wellformed\ {\isasymGamma}}}{\mbox{cterms\ {\isasymGamma}\ {\isacharequal}\ {\isacharbraceleft}p\ {\isacharbar}\ {\isasymexists}pn{\isachardot}\ p{\isasymin}ctermsl\ {\isacharparenleft}{\isasymGamma}\ pn{\isacharparenright}\ {\isasymand}\ not{\isacharunderscore}call\ p{\isacharbraceright}}}}}
\snip{cterms_def'_concl}{0}{0}{\isa{cterms\ {\isasymGamma}\ {\isacharequal}\ {\isacharbraceleft}p\ {\isacharbar}\ {\isasymexists}pn{\isachardot}\ p{\isasymin}ctermsl\ {\isacharparenleft}{\isasymGamma}\ pn{\isacharparenright}\ {\isasymand}\ not{\isacharunderscore}call\ p{\isacharbraceright}}}

\snip{cterms_subterms}{0}{0}{\isa{\mbox{}\inferrule{\mbox{wellformed\ {\isasymGamma}}}{\mbox{cterms\ {\isasymGamma}\ {\isacharequal}\ {\isacharbraceleft}p\ {\isacharbar}\ {\isasymexists}pn{\isachardot}\ p{\isasymin}subterms\ {\isacharparenleft}{\isasymGamma}\ pn{\isacharparenright}\ {\isasymand}\ not{\isacharunderscore}call\ p\ {\isasymand}\ not{\isacharunderscore}choice\ p{\isacharbraceright}}}}}  
\snip{cterms_subterms_concl}{0}{0}{\isa{cterms\ {\isasymGamma}\ {\isacharequal}\ {\isacharbraceleft}p\ {\isacharbar}\ {\isasymexists}pn{\isachardot}\ p{\isasymin}subterms\ {\isacharparenleft}{\isasymGamma}\ pn{\isacharparenright}\ {\isasymand}\ not{\isacharunderscore}call\ p\ {\isasymand}\ not{\isacharunderscore}choice\ p{\isacharbraceright}}}

\snip{ctermsl_choice}{0}{0}{\isa{ctermsl\ {\isacharparenleft}p\isactrlsub {\isadigit{1}}\ {\isasymoplus}\ p\isactrlsub {\isadigit{2}}{\isacharparenright}\ {\isacharequal}\ ctermsl\ p\isactrlsub {\isadigit{1}}\ {\isasymunion}\ ctermsl\ p\isactrlsub {\isadigit{2}}}}
\snip{ctermsl_choice_lhs}{0}{0}{\isa{ctermsl\ {\isacharparenleft}p\isactrlsub {\isadigit{1}}\ {\isasymoplus}\ p\isactrlsub {\isadigit{2}}{\isacharparenright}}}
\snip{ctermsl_choice_rhs}{0}{0}{\isa{ctermsl\ p\isactrlsub {\isadigit{1}}\ {\isasymunion}\ ctermsl\ p\isactrlsub {\isadigit{2}}}}

\snip{ctermsl_call}{0}{0}{\isa{ctermsl\ {\isacharparenleft}call{\isacharparenleft}pn{\isacharparenright}{\isacharparenright}\ {\isacharequal}\ {\isacharbraceleft}call{\isacharparenleft}pn{\isacharparenright}{\isacharbraceright}}}
\snip{ctermsl_call_lhs}{0}{0}{\isa{ctermsl\ {\isacharparenleft}call{\isacharparenleft}pn{\isacharparenright}{\isacharparenright}}}
\snip{ctermsl_call_rhs}{0}{0}{\isa{{\isacharbraceleft}call{\isacharparenleft}pn{\isacharparenright}{\isacharbraceright}}}

\snip{ctermsl_other2}{0}{0}{\isa{ctermsl\ {\isacharparenleft}\gray{{\isacharbraceleft}l{\isacharbraceright}}{\isasymlbrakk}u{\isasymrbrakk}\ p{\isacharparenright}\ {\isacharequal}\ {\isacharbraceleft}{\isacharbraceleft}l{\isacharbraceright}{\isasymlbrakk}u{\isasymrbrakk}\ p{\isacharbraceright}\ {\isasymunion}\ ctermsl\ p}}
\snip{ctermsl_other2_lhs}{0}{0}{\isa{ctermsl\ {\isacharparenleft}\gray{{\isacharbraceleft}l{\isacharbraceright}}{\isasymlbrakk}u{\isasymrbrakk}\ p{\isacharparenright}}}
\snip{ctermsl_other2_rhs}{0}{0}{\isa{{\isacharbraceleft}\gray{{\isacharbraceleft}l{\isacharbraceright}}{\isasymlbrakk}u{\isasymrbrakk}\ p{\isacharbraceright}\ {\isasymunion}\ ctermsl\ p}}

\snip{ctermsl_unicast}{0}{0}{\isa{ctermsl\ {\isacharparenleft}\gray{{\isacharbraceleft}l{\isacharbraceright}}broadcast{\isacharparenleft}\selmsg{\isacharparenright}\ {\isachardot}\ p{\isacharparenright}\ {\isacharequal}\ {\isacharbraceleft}{\isacharbraceleft}l{\isacharbraceright}broadcast{\isacharparenleft}\selmsg{\isacharparenright}\ {\isachardot}\ p{\isacharbraceright}\ {\isasymunion}\ ctermsl\ p}}

\snip{stermsl_ctermsl}{0}{0}{\isa{\mbox{}\inferrule{\mbox{q{\isasymin}stermsl\ p}}{\mbox{q{\isasymin}ctermsl\ p}}}}
\snip{stermsl_ctermsl_prem}{0}{0}{\isa{q{\isasymin}stermsl\ p}}
\snip{stermsl_ctermsl_concl}{0}{0}{\isa{q{\isasymin}ctermsl\ p}}

\snip{stermsl}{0}{0}{\isa{stermsl}}
\snip{stermsl_choice}{0}{0}{\isa{stermsl\ {\isacharparenleft}p\isactrlsub {\isadigit{1}}\ {\isasymoplus}\ p\isactrlsub {\isadigit{2}}{\isacharparenright}\ {\isacharequal}\ stermsl\ p\isactrlsub {\isadigit{1}}\ {\isasymunion}\ stermsl\ p\isactrlsub {\isadigit{2}}}}
\snip{stermsl_other}{0}{0}{\isa{stermsl\ p\ {\isacharequal}\ {\isacharbraceleft}p{\isacharbraceright}}}

\snip{seq_reachable_in_cterms_prem_1}{0}{0}{\isa{wellformed\ {\isasymGamma}}}
\snip{seq_reachable_in_cterms_prem_2}{0}{0}{\isa{control{\isacharunderscore}within\ {\isasymGamma}\ {\isacharparenleft}init\ A{\isacharparenright}}}
\snip{seq_reachable_in_cterms_prem_3}{0}{0}{\isa{trans\ A\ {\isacharequal}\ \seqpsos\ {\isasymGamma}}}
\snip{seq_reachable_in_cterms_prem_4}{0}{0}{\isa{{\isacharparenleft}{\isasymxi},\ p{\isacharparenright}{\isasymin}reachable\ A\ I}}
\snip{seq_reachable_in_cterms_prem_5}{0}{0}{\isa{q{\isasymin}sterms\ {\isasymGamma}\ p}}
\snip{seq_reachable_in_cterms_concl}{0}{0}{\isa{q{\isasymin}cterms\ {\isasymGamma}}}

\snip{control_within}{0}{0}{\isa{control{\isacharunderscore}within\ {\isasymGamma}\ {\isasymsigma}\ {\isacharequal}\ {\isasymforall}{\isacharparenleft}{\isasymxi},\ p{\isacharparenright}{\isasymin}{\isasymsigma}{\isachardot}\ {\isasymexists}pn{\isachardot}\ p{\isasymin}subterms\ {\isacharparenleft}{\isasymGamma}\ pn{\isacharparenright}}}
\snip{simple_labels}{0}{0}{\isa{simple{\isacharunderscore}labels\ {\isasymGamma}\ {\isacharequal}\ {\isasymforall}pn{\isachardot}\ {\isasymforall}p{\isasymin}subterms\ {\isacharparenleft}{\isasymGamma}\ pn{\isacharparenright}{\isachardot}\ {\isasymexists}\hspace{-.2em}{\isacharbang}\hspace{.15em}l{\isachardot}\ labels\ {\isasymGamma}\ p\ {\isacharequal}\ \gray{{\isacharbraceleft}l{\isacharbraceright}}}}

\snip{seq_invariant_ctermsI_prem_1}{0}{0}{\isa{wellformed\ {\isasymGamma}}}
\snip{seq_invariant_ctermsI_prem_2}{0}{0}{\isa{control{\isacharunderscore}within\ {\isasymGamma}\ {\isacharparenleft}init\ A{\isacharparenright}}}
\snip{seq_invariant_ctermsI_prem_3}{0}{0}{\isa{simple{\isacharunderscore}labels\ {\isasymGamma}}}
\snip{seq_invariant_ctermsI_prem_4}{0}{0}{\isa{trans\ A\ {\isacharequal}\ \seqpsos\ {\isasymGamma}}}

\snip{seq_invariant_ctermsI_init}{0}{0}{\isa{{\isasymAnd}{\isasymxi}\ p\ l{\isachardot}\ \mbox{}\inferrule{\mbox{{\isacharparenleft}{\isasymxi},\ p{\isacharparenright}{\isasymin}init\ A}\\\ \mbox{l{\isasymin}labels\ {\isasymGamma}\ p}}{\mbox{P\ {\isacharparenleft}{\isasymxi},\ l{\isacharparenright}}}}}
\snip{seq_invariant_ctermsI_init_1}{0}{0}{\isa{{\isacharparenleft}{\isasymxi},\ p{\isacharparenright}{\isasymin}init\ A}}
\snip{seq_invariant_ctermsI_init_2}{0}{0}{\isa{l{\isasymin}labels\ {\isasymGamma}\ p}}
\snip{seq_invariant_ctermsI_init_3}{0}{0}{\isa{P\ {\isacharparenleft}{\isasymxi},\ l{\isacharparenright}}}

\snip{seq_invariant_ctermsI_trans_1}{0}{0}{\isa{p{\isasymin}ctermsl\ {\isacharparenleft}{\isasymGamma}\ pn{\isacharparenright}}}
\snip{seq_invariant_ctermsI_trans_2}{0}{0}{\isa{not{\isacharunderscore}call\ p}}
\snip{seq_invariant_ctermsI_trans_3}{0}{0}{\isa{l{\isasymin}labels\ {\isasymGamma}\ p}}
\snip{seq_invariant_ctermsI_trans_4}{0}{0}{\isa{P\ {\isacharparenleft}{\isasymxi},\ l{\isacharparenright}}}
\snip{seq_invariant_ctermsI_trans_5}{0}{0}{\isa{{\isacharparenleft}{\isacharparenleft}{\isasymxi},\ p{\isacharparenright},\ a,\ {\isacharparenleft}{\isasymxi}{\isacharprime},\ q{\isacharparenright}{\isacharparenright}{\isasymin}\seqpsos\ {\isasymGamma}}}
\snip{seq_invariant_ctermsI_trans_6}{0}{0}{\isa{l{\isacharprime}{\isasymin}labels\ {\isasymGamma}\ q}}
\snip{seq_invariant_ctermsI_trans_7}{0}{0}{\isa{{\isacharparenleft}{\isasymxi},\ pp{\isacharparenright}{\isasymin}reachable\ A\ I}}
\snip{seq_invariant_ctermsI_trans_8}{0}{0}{\isa{p{\isasymin}sterms\ {\isasymGamma}\ pp}}
\snip{seq_invariant_ctermsI_trans_9}{0}{0}{\isa{I\ a}}
\snip{seq_invariant_ctermsI_trans_10}{0}{0}{\isa{P\ {\isacharparenleft}{\isasymxi}{\isacharprime},\ l{\isacharprime}{\isacharparenright}}}

\snip{seq_invariant_ctermsI_concl}{0}{0}{\isa{A\ {\isasymTTurnstile}\ {\isacharparenleft}I\ {\isasymrightarrow}{\isacharparenright}\ onl\ {\isasymGamma}\ P}}

\snip{oseq_invariant_ctermsI_trans_1}{0}{0}{\isa{{\isacharparenleft}{\isasymsigma},\ p{\isacharparenright}{\isasymin}oreachable\ A\ S\ U}}
\snip{oseq_invariant_ctermsI_trans_2}{0}{0}{\isa{l{\isasymin}labels\ {\isasymGamma}\ p}}
\snip{oseq_invariant_ctermsI_trans_3}{0}{0}{\isa{P\ {\isacharparenleft}{\isasymsigma},\ l{\isacharparenright}}}
\snip{oseq_invariant_ctermsI_trans_4}{0}{0}{\isa{U\ {\isasymsigma}\ {\isasymsigma}{\isacharprime}}}
\snip{oseq_invariant_ctermsI_trans_5}{0}{0}{\isa{P\ {\isacharparenleft}{\isasymsigma}{\isacharprime},\ l{\isacharparenright}}}

\snip{oseq_invariant_ctermsI_concl}{0}{0}{\isa{A\ {\isasymTurnstile}\ {\isacharparenleft}S,\ U\ {\isasymrightarrow}{\isacharparenright}\ onl\ {\isasymGamma}\ P}}
\snip{seq_invariant_ctermI}{0}{0}{\begin{isabelle}%
{\isasymlbrakk}wellformed\ {\isasymGamma}{\isacharsemicolon}\ control{\isacharunderscore}within\ {\isasymGamma}\ {\isacharparenleft}init\ A{\isacharparenright}{\isacharsemicolon}\ simple{\isacharunderscore}labels\ {\isasymGamma}{\isacharsemicolon}\isanewline
\isaindent{\ }trans\ A\ {\isacharequal}\ \seqpsos\ {\isasymGamma}{\isacharsemicolon}\isanewline
\isaindent{\ }{\isasymAnd}{\isasymxi}\ p\ l{\isachardot}\ {\isasymlbrakk}{\isacharparenleft}{\isasymxi},\ p{\isacharparenright}{\isasymin}init\ A{\isacharsemicolon}\ l{\isasymin}labels\ {\isasymGamma}\ p{\isasymrbrakk}\ {\isasymLongrightarrow}\ P\ {\isacharparenleft}{\isasymxi},\ l{\isacharparenright}{\isacharsemicolon}\isanewline
\isaindent{\ }{\isasymAnd}p\ l\ {\isasymxi}\ a\ q\ l{\isacharprime}\ {\isasymxi}{\isacharprime}\ pp{\isachardot}\isanewline
\isaindent{\ \ \ \ }{\isasymlbrakk}p{\isasymin}cterms\ {\isasymGamma}{\isacharsemicolon}\ l{\isasymin}labels\ {\isasymGamma}\ p{\isacharsemicolon}\ P\ {\isacharparenleft}{\isasymxi},\ l{\isacharparenright}{\isacharsemicolon}\isanewline
\isaindent{\ \ \ \ \ }{\isacharparenleft}{\isacharparenleft}{\isasymxi},\ p{\isacharparenright},\ a,\ {\isacharparenleft}{\isasymxi}{\isacharprime},\ q{\isacharparenright}{\isacharparenright}{\isasymin}\seqpsos\ {\isasymGamma}{\isacharsemicolon}\isanewline
\isaindent{\ \ \ \ \ }{\isacharparenleft}{\isacharparenleft}{\isasymxi},\ p{\isacharparenright},\ a,\ {\isasymxi}{\isacharprime},\ q{\isacharparenright}{\isasymin}trans\ A{\isacharsemicolon}\ l{\isacharprime}{\isasymin}labels\ {\isasymGamma}\ q{\isacharsemicolon}\isanewline
\isaindent{\ \ \ \ \ }{\isacharparenleft}{\isasymxi},\ pp{\isacharparenright}{\isasymin}reachable\ A\ I{\isacharsemicolon}\ p{\isasymin}sterms\ {\isasymGamma}\ pp{\isacharsemicolon}\ {\isacharparenleft}{\isasymxi}{\isacharprime},\ q{\isacharparenright}{\isasymin}reachable\ A\ I{\isacharsemicolon}\ I\ a{\isasymrbrakk}\isanewline
\isaindent{\ \ \ \ }{\isasymLongrightarrow}\ P\ {\isacharparenleft}{\isasymxi}{\isacharprime},\ l{\isacharprime}{\isacharparenright}{\isasymrbrakk}\isanewline
{\isasymLongrightarrow}\ A\ {\isasymTTurnstile}\ {\isacharparenleft}I\ {\isasymrightarrow}{\isacharparenright}\ onl\ {\isasymGamma}\ P%
\end{isabelle}}
\snip{seq_invariant_ctermI_wf}{0}{0}{\isa{wellformed\ {\isasymGamma}}}
\snip{seq_invariant_ctermI_cw}{0}{0}{\isa{control{\isacharunderscore}within\ {\isasymGamma}\ {\isacharparenleft}init\ A{\isacharparenright}}}
\snip{seq_invariant_ctermI_sl}{0}{0}{\isa{simple{\isacharunderscore}labels\ {\isasymGamma}}}
\snip{seq_invariant_ctermI_init}{0}{0}{\begin{isabelle}%
trans\ A\ {\isacharequal}\ \seqpsos\ {\isasymGamma}%
\end{isabelle}}
\snip{seq_invariant_ctermI_step}{0}{0}{\begin{isabelle}%
{\isasymAnd}{\isasymxi}\ p\ l{\isachardot}\ {\isasymlbrakk}{\isacharparenleft}{\isasymxi},\ p{\isacharparenright}{\isasymin}init\ A{\isacharsemicolon}\ l{\isasymin}labels\ {\isasymGamma}\ p{\isasymrbrakk}\ {\isasymLongrightarrow}\ P\ {\isacharparenleft}{\isasymxi},\ l{\isacharparenright}%
\end{isabelle}}
\snip{seq_invariant_ctermI_concl}{0}{0}{\isa{A\ {\isasymTTurnstile}\ {\isacharparenleft}I\ {\isasymrightarrow}{\isacharparenright}\ onl\ {\isasymGamma}\ P}}

\snip{seq_reachable_in_cterms}{0}{0}{\isa{\mbox{}\inferrule{\mbox{wellformed\ {\isasymGamma}}\\\ \mbox{control{\isacharunderscore}within\ {\isasymGamma}\ {\isacharparenleft}init\ A{\isacharparenright}}\\\ \mbox{trans\ A\ {\isacharequal}\ \seqpsos\ {\isasymGamma}}\\\ \mbox{{\isacharparenleft}{\isasymxi},\ p{\isacharparenright}{\isasymin}reachable\ A\ I}\\\ \mbox{p{\isacharprime}{\isasymin}sterms\ {\isasymGamma}\ p}}{\mbox{p{\isacharprime}{\isasymin}cterms\ {\isasymGamma}}}}}
\snip{open_seq_step_invariant}{0}{0}{\isa{\mbox{}\inferrule{\mbox{A\ \stepinv\ {\isacharparenleft}I\ {\isasymrightarrow}{\isacharparenright}\ P}\\\ \mbox{initiali\ i\ {\isacharparenleft}init\ OA{\isacharparenright}\ {\isacharparenleft}init\ A{\isacharparenright}}\\\ \mbox{trans\ OA\ {\isacharequal}\ \oseqpsos\ {\isasymGamma}\ i}\\\ \mbox{trans\ A\ {\isacharequal}\ \seqpsos\ {\isasymGamma}}}{\mbox{OA\ \ostepinv\ {\isacharparenleft}act\ I,\ other\ ANY\ {\isacharbraceleft}i{\isacharbraceright}\ {\isasymrightarrow}{\isacharparenright}\ seqll\ i\ P}}}}

\snip{open_seq_invariant}{0}{0}{\isa{\mbox{}\inferrule{\mbox{A\ {\isasymTTurnstile}\ {\isacharparenleft}I\ {\isasymrightarrow}{\isacharparenright}\ P}\\\ \mbox{initiali\ i\ {\isacharparenleft}init\ OA{\isacharparenright}\ {\isacharparenleft}init\ A{\isacharparenright}}\\\ \mbox{trans\ OA\ {\isacharequal}\ \oseqpsos\ {\isasymGamma}\ i}\\\ \mbox{trans\ A\ {\isacharequal}\ \seqpsos\ {\isasymGamma}}}{\mbox{OA\ {\isasymTurnstile}\ {\isacharparenleft}act\ I,\ other\ ANY\ {\isacharbraceleft}i{\isacharbraceright}\ {\isasymrightarrow}{\isacharparenright}\ seql\ i\ P}}}}
\snip{open_seq_invariant_prem_1}{0}{0}{\isa{A\ {\isasymTTurnstile}\ {\isacharparenleft}I\ {\isasymrightarrow}{\isacharparenright}\ P}}
\snip{open_seq_invariant_prem_2}{0}{0}{\isa{initiali\ i\ {\isacharparenleft}init\ A{\isacharprime}{\isacharparenright}\ {\isacharparenleft}init\ A{\isacharparenright}}}
\snip{open_seq_invariant_prem_2'}{0}{0}{\isa{init\ A\ {\isacharequal}\ {\isacharbraceleft}{\isacharparenleft}{\isasymsigma}\ i,\ p{\isacharparenright}\ {\isacharbar}\ {\isacharparenleft}{\isasymsigma},\ p{\isacharparenright}{\isasymin}init\ A{\isacharprime}{\isacharbraceright}}}
\snip{open_seq_invariant_prem_3}{0}{0}{\isa{trans\ A{\isacharprime}\ {\isacharequal}\ \oseqpsos\ {\isasymGamma}\ i}}
\snip{open_seq_invariant_prem_4}{0}{0}{\isa{trans\ A\ {\isacharequal}\ \seqpsos\ {\isasymGamma}}}
\snip{open_seq_invariant_concl}{0}{0}{\isa{A{\isacharprime}\ {\isasymTurnstile}\ {\isacharparenleft}act\ I,\ other\ F\ {\isacharbraceleft}i{\isacharbraceright}\ {\isasymrightarrow}{\isacharparenright}\ seql\ i\ P}}
\snip{open_seq_invariant_concl'}{0}{0}{\isa{A{\isacharprime}\ {\isasymTurnstile}\ {\isacharparenleft}{\isasymlambda}{\isacharunderscore}\ {\isacharunderscore}{\isachardot}\ I,\ other\ F\ {\isacharbraceleft}i{\isacharbraceright}\ {\isasymrightarrow}{\isacharparenright}\ {\isacharparenleft}{\isasymlambda}{\isacharparenleft}{\isasymsigma},\ p{\isacharparenright}{\isachardot}\ P\ {\isacharparenleft}{\isasymsigma}\ i,\ p{\isacharparenright}{\isacharparenright}}}

\snip{oinvariant}{0}{0}{\begin{isabelle}%
{\isacharparenleft}A\ {\isasymTurnstile}\ {\isacharparenleft}S,\ U\ {\isasymrightarrow}{\isacharparenright}\ P{\isacharparenright}\ {\isacharequal}\ {\isacharparenleft}{\isasymforall}s{\isasymin}oreachable\ A\ S\ U{\isachardot}\ P\ s{\isacharparenright}%
\end{isabelle}}
\snip{oinvariant_lhs}{0}{0}{\isa{A\ {\isasymTurnstile}\ {\isacharparenleft}S,\ U\ {\isasymrightarrow}{\isacharparenright}\ P}}
\snip{oinvariant_rhs}{0}{0}{\isa{{\isasymforall}s{\isasymin}oreachable\ A\ S\ U{\isachardot}\ P\ s}}
\snip{ostep_invariant}{0}{0}{\begin{isabelle}%
{\isacharparenleft}A\ \ostepinv\ {\isacharparenleft}S,\ U\ {\isasymrightarrow}{\isacharparenright}\ P{\isacharparenright}\ {\isacharequal}\isanewline
{\isacharparenleft}{\isasymforall}s{\isasymin}oreachable\ A\ S\ U{\isachardot}\isanewline
\isaindent{{\isacharparenleft}\ \ \ }{\isasymforall}a\ s{\isacharprime}{\isachardot}\isanewline
\isaindent{{\isacharparenleft}\ \ \ \ \ \ }{\isacharparenleft}s,\ a,\ s{\isacharprime}{\isacharparenright}{\isasymin}trans\ A\ {\isasymand}\ S\ {\isacharparenleft}fst\ s{\isacharparenright}\ {\isacharparenleft}fst\ s{\isacharprime}{\isacharparenright}\ a\ {\isasymlongrightarrow}\isanewline
\isaindent{{\isacharparenleft}\ \ \ \ \ \ }P\ {\isacharparenleft}s,\ a,\ s{\isacharprime}{\isacharparenright}{\isacharparenright}%
\end{isabelle}}
\snip{ostep_invariant_lhs}{0}{0}{\isa{A\ \ostepinv\ {\isacharparenleft}S,\ U\ {\isasymrightarrow}{\isacharparenright}\ P}}
\snip{ostep_invariant_rhs}{0}{0}{\isa{{\isasymforall}s{\isasymin}oreachable\ A\ S\ U{\isachardot}\ {\isasymforall}a\ s{\isacharprime}{\isachardot}\ {\isacharparenleft}s,\ a,\ s{\isacharprime}{\isacharparenright}{\isasymin}trans\ A\ {\isasymand}\ S\ {\isacharparenleft}fst\ s{\isacharparenright}\ {\isacharparenleft}fst\ s{\isacharprime}{\isacharparenright}\ a\ {\isasymlongrightarrow}\ P\ {\isacharparenleft}s,\ a,\ s{\isacharprime}{\isacharparenright}}}

\snip{invariant}{0}{0}{\begin{isabelle}%
{\isacharparenleft}A\ {\isasymTTurnstile}\ {\isacharparenleft}I\ {\isasymrightarrow}{\isacharparenright}\ P{\isacharparenright}\ {\isacharequal}\ {\isacharparenleft}{\isasymforall}s{\isasymin}reachable\ A\ I{\isachardot}\ P\ s{\isacharparenright}%
\end{isabelle}}
\snip{invariant_lhs}{0}{0}{\isa{A\ {\isasymTTurnstile}\ {\isacharparenleft}I\ {\isasymrightarrow}{\isacharparenright}\ P}}
\snip{invariant_rhs}{0}{0}{\isa{{\isasymforall}s{\isasymin}reachable\ A\ I{\isachardot}\ P\ s}}
\snip{invariant_TT_lhs}{0}{0}{\isa{A\ {\isasymTTurnstile}\ P}}

\snip{step_invariant}{0}{0}{\begin{isabelle}%
{\isacharparenleft}A\ \stepinv\ {\isacharparenleft}I\ {\isasymrightarrow}{\isacharparenright}\ P{\isacharparenright}\ {\isacharequal}\isanewline
{\isacharparenleft}{\isasymforall}a{\isachardot}\ I\ a\ {\isasymlongrightarrow}\isanewline
\isaindent{{\isacharparenleft}{\isasymforall}a{\isachardot}\ }{\isacharparenleft}{\isasymforall}s{\isasymin}reachable\ A\ I{\isachardot}\isanewline
\isaindent{{\isacharparenleft}{\isasymforall}a{\isachardot}\ {\isacharparenleft}\ \ \ }{\isasymforall}s{\isacharprime}{\isachardot}\ {\isacharparenleft}s,\ a,\ s{\isacharprime}{\isacharparenright}{\isasymin}trans\ A\ {\isasymlongrightarrow}\ P\ {\isacharparenleft}s,\ a,\ s{\isacharprime}{\isacharparenright}{\isacharparenright}{\isacharparenright}%
\end{isabelle}}
\snip{step_invariant_lhs}{0}{0}{\isa{A\ \stepinv\ {\isacharparenleft}I\ {\isasymrightarrow}{\isacharparenright}\ P}}
\snip{step_invariant_rhs}{0}{0}{\isa{{\isasymforall}a{\isachardot}\ I\ a\ {\isasymlongrightarrow}\ {\isacharparenleft}{\isasymforall}s{\isasymin}reachable\ A\ I{\isachardot}\ {\isasymforall}s{\isacharprime}{\isachardot}\ {\isacharparenleft}s,\ a,\ s{\isacharprime}{\isacharparenright}{\isasymin}trans\ A\ {\isasymlongrightarrow}\ P\ {\isacharparenleft}s,\ a,\ s{\isacharprime}{\isacharparenright}{\isacharparenright}}}

\snip{TT}{0}{0}{\isa{TT\ {\isacharequal}\ {\isacharparenleft}{\isasymlambda}{\isacharunderscore}{\isachardot}\ True{\isacharparenright}}}
\snip{TT_rhs}{0}{0}{\isa{{\isasymlambda}{\isacharunderscore}{\isachardot}\ True}}

\snip{open_closed_invariant}{0}{0}{\isa{\mbox{}\inferrule{\mbox{A\ {\isasymTTurnstile}\ {\isacharparenleft}I\ {\isasymrightarrow}{\isacharparenright}\ P}\\\ \mbox{subreachable\ A\ U\ J}\\\ \mbox{{\isasymAnd}{\isasymsigma}\ {\isasymsigma}{\isacharprime}\ s{\isachardot}\ \mbox{}\inferrule{\mbox{{\isasymforall}j{\isasymin}J{\isachardot}\ {\isasymsigma}{\isacharprime}\ j\ {\isacharequal}\ {\isasymsigma}\ j}\\\ \mbox{P\ {\isacharparenleft}{\isasymsigma}{\isacharprime},\ s{\isacharparenright}}}{\mbox{P\ {\isacharparenleft}{\isasymsigma},\ s{\isacharparenright}}}}}{\mbox{A\ {\isasymTurnstile}\ {\isacharparenleft}act\ I,\ U\ {\isasymrightarrow}{\isacharparenright}\ P}}}}  

\snip{oseq_invariant_ctermI}{0}{0}{\begin{isabelle}%
{\isasymlbrakk}wellformed\ {\isasymGamma}{\isacharsemicolon}\ control{\isacharunderscore}within\ {\isasymGamma}\ {\isacharparenleft}init\ A{\isacharparenright}{\isacharsemicolon}\ simple{\isacharunderscore}labels\ {\isasymGamma}{\isacharsemicolon}\isanewline
\isaindent{\ }trans\ A\ {\isacharequal}\ \oseqpsos\ {\isasymGamma}\ i{\isacharsemicolon}\isanewline
\isaindent{\ }{\isasymAnd}{\isasymsigma}\ p\ l{\isachardot}\ {\isasymlbrakk}{\isacharparenleft}{\isasymsigma},\ p{\isacharparenright}{\isasymin}init\ A{\isacharsemicolon}\ l{\isasymin}labels\ {\isasymGamma}\ p{\isasymrbrakk}\ {\isasymLongrightarrow}\ P\ {\isacharparenleft}{\isasymsigma},\ l{\isacharparenright}{\isacharsemicolon}\isanewline
\isaindent{\ }{\isasymAnd}{\isasymsigma}\ {\isasymsigma}{\isacharprime}\ p\ l{\isachardot}\isanewline
\isaindent{\ \ \ \ }{\isasymlbrakk}{\isacharparenleft}{\isasymsigma},\ p{\isacharparenright}{\isasymin}oreachable\ A\ S\ U{\isacharsemicolon}\ l{\isasymin}labels\ {\isasymGamma}\ p{\isacharsemicolon}\ P\ {\isacharparenleft}{\isasymsigma},\ l{\isacharparenright}{\isacharsemicolon}\ U\ {\isasymsigma}\ {\isasymsigma}{\isacharprime}{\isasymrbrakk}\isanewline
\isaindent{\ \ \ \ }{\isasymLongrightarrow}\ P\ {\isacharparenleft}{\isasymsigma}{\isacharprime},\ l{\isacharparenright}{\isacharsemicolon}\isanewline
\isaindent{\ }{\isasymAnd}p\ l\ {\isasymsigma}\ a\ q\ l{\isacharprime}\ {\isasymsigma}{\isacharprime}\ pp{\isachardot}\isanewline
\isaindent{\ \ \ \ }{\isasymlbrakk}p{\isasymin}cterms\ {\isasymGamma}{\isacharsemicolon}\ l{\isasymin}labels\ {\isasymGamma}\ p{\isacharsemicolon}\ P\ {\isacharparenleft}{\isasymsigma},\ l{\isacharparenright}{\isacharsemicolon}\isanewline
\isaindent{\ \ \ \ \ }{\isacharparenleft}{\isacharparenleft}{\isasymsigma},\ p{\isacharparenright},\ a,\ {\isacharparenleft}{\isasymsigma}{\isacharprime},\ q{\isacharparenright}{\isacharparenright}{\isasymin}\oseqpsos\ {\isasymGamma}\ i{\isacharsemicolon}\isanewline
\isaindent{\ \ \ \ \ }{\isacharparenleft}{\isacharparenleft}{\isasymsigma},\ p{\isacharparenright},\ a,\ {\isasymsigma}{\isacharprime},\ q{\isacharparenright}{\isasymin}trans\ A{\isacharsemicolon}\ l{\isacharprime}{\isasymin}labels\ {\isasymGamma}\ q{\isacharsemicolon}\isanewline
\isaindent{\ \ \ \ \ }{\isacharparenleft}{\isasymsigma},\ pp{\isacharparenright}{\isasymin}oreachable\ A\ S\ U{\isacharsemicolon}\ p{\isasymin}sterms\ {\isasymGamma}\ pp{\isacharsemicolon}\isanewline
\isaindent{\ \ \ \ \ }{\isacharparenleft}{\isasymsigma}{\isacharprime},\ q{\isacharparenright}{\isasymin}oreachable\ A\ S\ U{\isacharsemicolon}\ S\ {\isasymsigma}\ {\isasymsigma}{\isacharprime}\ a{\isasymrbrakk}\isanewline
\isaindent{\ \ \ \ }{\isasymLongrightarrow}\ P\ {\isacharparenleft}{\isasymsigma}{\isacharprime},\ l{\isacharprime}{\isacharparenright}{\isasymrbrakk}\isanewline
{\isasymLongrightarrow}\ A\ {\isasymTurnstile}\ {\isacharparenleft}S,\ U\ {\isasymrightarrow}{\isacharparenright}\ onl\ {\isasymGamma}\ P%
\end{isabelle}}
\snip{oseq_invariant_ctermI_wf}{0}{0}{\isa{wellformed\ {\isasymGamma}}}
\snip{oseq_invariant_ctermI_cw}{0}{0}{\isa{control{\isacharunderscore}within\ {\isasymGamma}\ {\isacharparenleft}init\ A{\isacharparenright}}}
\snip{oseq_invariant_ctermI_sl}{0}{0}{\isa{simple{\isacharunderscore}labels\ {\isasymGamma}}}
\snip{oseq_invariant_ctermI_init}{0}{0}{\begin{isabelle}%
trans\ A\ {\isacharequal}\ \oseqpsos\ {\isasymGamma}\ i%
\end{isabelle}}
\snip{oseq_invariant_ctermI_other}{0}{0}{\isa{{\isasymAnd}{\isasymsigma}\ p\ l{\isachardot}\ {\isasymlbrakk}{\isacharparenleft}{\isasymsigma},\ p{\isacharparenright}{\isasymin}init\ A{\isacharsemicolon}\ l{\isasymin}labels\ {\isasymGamma}\ p{\isasymrbrakk}\ {\isasymLongrightarrow}\ P\ {\isacharparenleft}{\isasymsigma},\ l{\isacharparenright}}}
\snip{oseq_invariant_ctermI_step}{0}{0}{\begin{isabelle}%
{\isasymAnd}{\isasymsigma}\ {\isasymsigma}{\isacharprime}\ p\ l{\isachardot}\isanewline
\isaindent{\ \ \ }{\isasymlbrakk}{\isacharparenleft}{\isasymsigma},\ p{\isacharparenright}{\isasymin}oreachable\ A\ S\ U{\isacharsemicolon}\ l{\isasymin}labels\ {\isasymGamma}\ p{\isacharsemicolon}\ P\ {\isacharparenleft}{\isasymsigma},\ l{\isacharparenright}{\isacharsemicolon}\ U\ {\isasymsigma}\ {\isasymsigma}{\isacharprime}{\isasymrbrakk}\isanewline
\isaindent{\ \ \ }{\isasymLongrightarrow}\ P\ {\isacharparenleft}{\isasymsigma}{\isacharprime},\ l{\isacharparenright}%
\end{isabelle}}
\snip{oseq_invariant_ctermI_concl}{0}{0}{\isa{A\ {\isasymTurnstile}\ {\isacharparenleft}S,\ U\ {\isasymrightarrow}{\isacharparenright}\ onl\ {\isasymGamma}\ P}}

\snip{oseq_step_invariant_ctermI}{0}{0}{\begin{isabelle}%
{\isasymlbrakk}wellformed\ {\isasymGamma}{\isacharsemicolon}\ control{\isacharunderscore}within\ {\isasymGamma}\ {\isacharparenleft}init\ A{\isacharparenright}{\isacharsemicolon}\ simple{\isacharunderscore}labels\ {\isasymGamma}{\isacharsemicolon}\isanewline
\isaindent{\ }trans\ A\ {\isacharequal}\ \oseqpsos\ {\isasymGamma}\ i{\isacharsemicolon}\isanewline
\isaindent{\ }{\isasymAnd}p\ l\ {\isasymsigma}\ a\ q\ l{\isacharprime}\ {\isasymsigma}{\isacharprime}\ pp{\isachardot}\isanewline
\isaindent{\ \ \ \ }{\isasymlbrakk}p{\isasymin}cterms\ {\isasymGamma}{\isacharsemicolon}\ l{\isasymin}labels\ {\isasymGamma}\ p{\isacharsemicolon}\ {\isacharparenleft}{\isacharparenleft}{\isasymsigma},\ p{\isacharparenright},\ a,\ {\isacharparenleft}{\isasymsigma}{\isacharprime},\ q{\isacharparenright}{\isacharparenright}{\isasymin}\oseqpsos\ {\isasymGamma}\ i{\isacharsemicolon}\isanewline
\isaindent{\ \ \ \ \ }{\isacharparenleft}{\isacharparenleft}{\isasymsigma},\ p{\isacharparenright},\ a,\ {\isasymsigma}{\isacharprime},\ q{\isacharparenright}{\isasymin}trans\ A{\isacharsemicolon}\ l{\isacharprime}{\isasymin}labels\ {\isasymGamma}\ q{\isacharsemicolon}\isanewline
\isaindent{\ \ \ \ \ }{\isacharparenleft}{\isasymsigma},\ pp{\isacharparenright}{\isasymin}oreachable\ A\ S\ U{\isacharsemicolon}\ p{\isasymin}sterms\ {\isasymGamma}\ pp{\isacharsemicolon}\isanewline
\isaindent{\ \ \ \ \ }{\isacharparenleft}{\isasymsigma}{\isacharprime},\ q{\isacharparenright}{\isasymin}oreachable\ A\ S\ U{\isacharsemicolon}\ S\ {\isasymsigma}\ {\isasymsigma}{\isacharprime}\ a{\isasymrbrakk}\isanewline
\isaindent{\ \ \ \ }{\isasymLongrightarrow}\ P\ {\isacharparenleft}{\isacharparenleft}{\isasymsigma},\ l{\isacharparenright},\ a,\ {\isasymsigma}{\isacharprime},\ l{\isacharprime}{\isacharparenright}{\isasymrbrakk}\isanewline
{\isasymLongrightarrow}\ A\ \ostepinv\ {\isacharparenleft}S,\ U\ {\isasymrightarrow}{\isacharparenright}\ onll\ {\isasymGamma}\ P%
\end{isabelle}}

\snip{reachable_init}{0}{0}{\isa{\mbox{}\inferrule{\mbox{s{\isasymin}init\ A}}{\mbox{s{\isasymin}reachable\ A\ I}}}}
\snip{reachable_step}{0}{0}{\isa{\mbox{}\inferrule{\mbox{s{\isasymin}reachable\ A\ I}\\\ \mbox{{\isacharparenleft}s,\ a,\ s{\isacharprime}{\isacharparenright}{\isasymin}trans\ A}\\\ \mbox{I\ a}}{\mbox{s{\isacharprime}{\isasymin}reachable\ A\ I}}}}

\snip{oreachable_init}{0}{0}{\isa{\mbox{}\inferrule{\mbox{{\isacharparenleft}{\isasymsigma},\ p{\isacharparenright}{\isasymin}init\ A}}{\mbox{{\isacharparenleft}{\isasymsigma},\ p{\isacharparenright}{\isasymin}oreachable\ A\ S\ U}}}}
\snip{oreachable_local}{0}{0}{\isa{\mbox{}\inferrule{\mbox{s{\isasymin}oreachable\ A\ S\ U}\\\ \mbox{{\isacharparenleft}s,\ {\isacharparenleft}a,\ s{\isacharprime}{\isacharparenright}{\isacharparenright}{\isasymin}trans\ A}\\\ \mbox{S\ {\isacharparenleft}fst\ s{\isacharparenright}\ {\isacharparenleft}fst\ s{\isacharprime}{\isacharparenright}\ a}}{\mbox{s{\isacharprime}{\isasymin}oreachable\ A\ S\ U}}}}
\snip{oreachable_other}{0}{0}{\isa{\mbox{}\inferrule{\mbox{s{\isasymin}oreachable\ A\ S\ U}\\\ \mbox{U\ {\isacharparenleft}fst\ s{\isacharparenright}\ {\isasymsigma}{\isacharprime}}}{\mbox{{\isacharparenleft}{\isasymsigma}{\isacharprime},\ snd\ s{\isacharparenright}{\isasymin}oreachable\ A\ S\ U}}}}
\snip{oreachable_local'}{0}{0}{\isa{\mbox{}\inferrule{\mbox{{\isacharparenleft}{\isasymsigma},\ p{\isacharparenright}{\isasymin}oreachable\ A\ S\ U}\\\ \mbox{{\isacharparenleft}{\isacharparenleft}{\isasymsigma},\ p{\isacharparenright},\ a,\ {\isacharparenleft}{\isasymsigma}{\isacharprime},\ p{\isacharprime}{\isacharparenright}{\isacharparenright}{\isasymin}trans\ A}\\\ \mbox{S\ {\isasymsigma}\ {\isasymsigma}{\isacharprime}\ a}}{\mbox{{\isacharparenleft}{\isasymsigma}{\isacharprime},\ p{\isacharprime}{\isacharparenright}{\isasymin}oreachable\ A\ S\ U}}}}
\snip{oreachable_other'}{0}{0}{\isa{\mbox{}\inferrule{\mbox{{\isacharparenleft}{\isasymsigma},\ p{\isacharparenright}{\isasymin}oreachable\ A\ S\ U}\\\ \mbox{U\ {\isasymsigma}\ {\isasymsigma}{\isacharprime}}}{\mbox{{\isacharparenleft}{\isasymsigma}{\isacharprime},\ p{\isacharparenright}{\isasymin}oreachable\ A\ S\ U}}}}

\snip{otherwith}{0}{0}{\isa{otherwith\ {\isacharparenleft}op\ {\isacharequal}{\isacharparenright}\ {\isacharbraceleft}i{\isacharbraceright}\ {\isacharparenleft}orecvmsg\ P{\isacharparenright}}}
\snip{other}{0}{0}{\isa{other\ quality{\isacharunderscore}increases\ {\isacharbraceleft}i{\isacharbraceright}}}
\snip{otherwith_def}{0}{0}{\isa{otherwith\ E\ N\ I\ {\isasymsigma}\ {\isasymsigma}{\isacharprime}\ a\ {\isacharequal}\ {\isacharparenleft}{\isasymforall}i{\isachardot}\ i\ {\isasymnotin}\ N\ {\isasymlongrightarrow}\ E\ {\isacharparenleft}{\isasymsigma}\ i{\isacharparenright}\ {\isacharparenleft}{\isasymsigma}{\isacharprime}\ i{\isacharparenright}{\isacharparenright}\ {\isasymand}\ I\ {\isasymsigma}\ a}}
\snip{other_def}{0}{0}{\isa{other\ F\ N\ {\isasymsigma}\ {\isasymsigma}{\isacharprime}\ {\isacharequal}\ {\isasymforall}i{\isachardot}\ \textsf{if}\ i{\isasymin}N\ \textsf{then}\ {\isasymsigma}{\isacharprime}\ i\ {\isacharequal}\ {\isasymsigma}\ i\ \textsf{else}\ F\ {\isacharparenleft}{\isasymsigma}\ i{\isacharparenright}\ {\isacharparenleft}{\isasymsigma}{\isacharprime}\ i{\isacharparenright}}}
\snip{qmsg_msgs_not_empty}{0}{0}{\begin{isabelle}%
qmsg\ {\isasymTTurnstile}\ onl\ \gammaqmsg\ {\isacharparenleft}{\isasymlambda}{\isacharparenleft}msgs,\ l{\isacharparenright}{\isachardot}\ l\ {\isacharequal}\ {\isacharparenleft}{\isacharparenright}{\isacharminus}{\isacharcolon}{\isadigit{1}}\ {\isasymlongrightarrow}\ msgs\ {\isasymnoteq}\ {\isacharbrackleft}{\isacharbrackright}{\isacharparenright}%
\end{isabelle}}
\snip{qmsg_send_from_queue}{0}{0}{\begin{isabelle}%
qmsg\ \stepinv\ {\isacharparenleft}{\isasymlambda}{\isacharparenleft}{\isacharparenleft}msg,\ q{\isacharparenright},\ a,\ {\isasymzeta}{\isacharprime}{\isacharparenright}{\isachardot}\ sendmsg\ {\isacharparenleft}{\isasymlambda}m{\isachardot}\ m{\isasymin}set\ msg{\isacharparenright}\ a{\isacharparenright}%
\end{isabelle}}
\snip{qmsg_send_receive_or_tau}{0}{0}{\begin{isabelle}%
qmsg\ \stepinv\ {\isacharparenleft}{\isasymlambda}{\isacharparenleft}uu{\isacharunderscore},\ a,\ {\isacharunderscore}{\isacharparenright}{\isachardot}\ {\isasymexists}m{\isachardot}\ a\ {\isacharequal}\ send\ m\ {\isasymor}\ a\ {\isacharequal}\ receive\ m\ {\isasymor}\ a\ {\isacharequal}\ {\isasymtau}{\isacharparenright}%
\end{isabelle}}

\snip{par_qmsg_oreachable}{0}{0}{\begin{isabelle}%
{\isasymlbrakk}{\isacharparenleft}{\isasymsigma},\ {\isasymzeta}{\isacharparenright}\isanewline
\isaindent{{\isasymlbrakk}}{\isasymin}\ oreachable\ {\isacharparenleft}A\ {\isasymlangle}{\isasymlangle}\isactrlbsub i\isactrlesub \ qmsg{\isacharparenright}\ {\isacharparenleft}otherwith\ S\ {\isacharbraceleft}i{\isacharbraceright}\ {\isacharparenleft}orecvmsg\ R{\isacharparenright}{\isacharparenright}\ {\isacharparenleft}other\ U\ {\isacharbraceleft}i{\isacharbraceright}{\isacharparenright}{\isacharsemicolon}\isanewline
\isaindent{\ }A\ \ostepinv\ {\isacharparenleft}otherwith\ S\ {\isacharbraceleft}i{\isacharbraceright}\ {\isacharparenleft}orecvmsg\ R{\isacharparenright},\ other\ U\ {\isacharbraceleft}i{\isacharbraceright}\ {\isasymrightarrow}{\isacharparenright}\isanewline
\isaindent{\ A\ \ostepinv\ \ }globala\ {\isacharparenleft}{\isasymlambda}{\isacharparenleft}{\isasymsigma},\ {\isacharunderscore},\ {\isasymsigma}{\isacharprime}{\isacharparenright}{\isachardot}\ U\ {\isacharparenleft}{\isasymsigma}\ i{\isacharparenright}\ {\isacharparenleft}{\isasymsigma}{\isacharprime}\ i{\isacharparenright}{\isacharparenright}{\isacharsemicolon}\isanewline
\isaindent{\ }{\isasymAnd}{\isasymxi}{\isachardot}\ U\ {\isasymxi}\ {\isasymxi}{\isacharsemicolon}\ {\isasymAnd}{\isasymxi}\ {\isasymxi}{\isacharprime}{\isachardot}\ S\ {\isasymxi}\ {\isasymxi}{\isacharprime}\ {\isasymLongrightarrow}\ U\ {\isasymxi}\ {\isasymxi}{\isacharprime}{\isacharsemicolon}\isanewline
\isaindent{\ }{\isasymAnd}{\isasymsigma}\ {\isasymsigma}{\isacharprime}\ m{\isachardot}\ {\isasymlbrakk}{\isasymforall}j{\isachardot}\ U\ {\isacharparenleft}{\isasymsigma}\ j{\isacharparenright}\ {\isacharparenleft}{\isasymsigma}{\isacharprime}\ j{\isacharparenright}{\isacharsemicolon}\ R\ {\isasymsigma}\ m{\isasymrbrakk}\ {\isasymLongrightarrow}\ R\ {\isasymsigma}{\isacharprime}\ m{\isasymrbrakk}\isanewline
{\isasymLongrightarrow}\ {\isacharparenleft}{\isasymsigma},\ fst\ {\isasymzeta}{\isacharparenright}{\isasymin}oreachable\ A\ {\isacharparenleft}otherwith\ S\ {\isacharbraceleft}i{\isacharbraceright}\ {\isacharparenleft}orecvmsg\ R{\isacharparenright}{\isacharparenright}\ {\isacharparenleft}other\ U\ {\isacharbraceleft}i{\isacharbraceright}{\isacharparenright}\ {\isasymand}\isanewline
\isaindent{{\isasymLongrightarrow}\ }snd\ {\isasymzeta}{\isasymin}reachable\ qmsg\ {\isacharparenleft}recvmsg\ {\isacharparenleft}R\ {\isasymsigma}{\isacharparenright}{\isacharparenright}\ {\isasymand}\ {\isacharparenleft}{\isasymforall}m{\isasymin}set\ {\isacharparenleft}fst\ {\isacharparenleft}snd\ {\isasymzeta}{\isacharparenright}{\isacharparenright}{\isachardot}\ R\ {\isasymsigma}\ m{\isacharparenright}%
\end{isabelle}}
\snip{par_qmsg_oreachable_upreservesq}{0}{0}{\begin{isabelle}%
{\isasymAnd}{\isasymsigma}\ {\isasymsigma}{\isacharprime}\ m{\isachardot}\ {\isasymlbrakk}{\isasymforall}j{\isachardot}\ U\ {\isacharparenleft}{\isasymsigma}\ j{\isacharparenright}\ {\isacharparenleft}{\isasymsigma}{\isacharprime}\ j{\isacharparenright}{\isacharsemicolon}\ R\ {\isasymsigma}\ m{\isasymrbrakk}\ {\isasymLongrightarrow}\ R\ {\isasymsigma}{\isacharprime}\ m%
\end{isabelle}}
\snip{par_qmsg_oreachable_concl3}{0}{0}{\begin{isabelle}%
{\isasymforall}m{\isasymin}set\ {\isacharparenleft}fst\ {\isacharparenleft}snd\ {\isasymzeta}{\isacharparenright}{\isacharparenright}{\isachardot}\ R\ {\isasymsigma}\ m%
\end{isabelle}}

\snip{lift_into_qmsg}{0}{0}{\begin{isabelle}%
{\isasymlbrakk}A\ {\isasymTurnstile}\ {\isacharparenleft}otherwith\ S\ {\isacharbraceleft}i{\isacharbraceright}\ {\isacharparenleft}orecvmsg\ R{\isacharparenright},\ other\ U\ {\isacharbraceleft}i{\isacharbraceright}\ {\isasymrightarrow}{\isacharparenright}\ global\ P{\isacharsemicolon}\ {\isasymAnd}{\isasymxi}{\isachardot}\ U\ {\isasymxi}\ {\isasymxi}{\isacharsemicolon}\isanewline
\isaindent{\ }{\isasymAnd}{\isasymxi}\ {\isasymxi}{\isacharprime}{\isachardot}\ S\ {\isasymxi}\ {\isasymxi}{\isacharprime}\ {\isasymLongrightarrow}\ U\ {\isasymxi}\ {\isasymxi}{\isacharprime}{\isacharsemicolon}\ {\isasymAnd}{\isasymsigma}\ {\isasymsigma}{\isacharprime}\ m{\isachardot}\ {\isasymlbrakk}{\isasymforall}j{\isachardot}\ U\ {\isacharparenleft}{\isasymsigma}\ j{\isacharparenright}\ {\isacharparenleft}{\isasymsigma}{\isacharprime}\ j{\isacharparenright}{\isacharsemicolon}\ R\ {\isasymsigma}\ m{\isasymrbrakk}\ {\isasymLongrightarrow}\ R\ {\isasymsigma}{\isacharprime}\ m{\isacharsemicolon}\isanewline
\isaindent{\ }A\ \ostepinv\ {\isacharparenleft}otherwith\ S\ {\isacharbraceleft}i{\isacharbraceright}\ {\isacharparenleft}orecvmsg\ R{\isacharparenright},\ other\ U\ {\isacharbraceleft}i{\isacharbraceright}\ {\isasymrightarrow}{\isacharparenright}\isanewline
\isaindent{\ A\ \ostepinv\ \ }globala\ {\isacharparenleft}{\isasymlambda}{\isacharparenleft}{\isasymsigma},\ {\isacharunderscore},\ {\isasymsigma}{\isacharprime}{\isacharparenright}{\isachardot}\ U\ {\isacharparenleft}{\isasymsigma}\ i{\isacharparenright}\ {\isacharparenleft}{\isasymsigma}{\isacharprime}\ i{\isacharparenright}{\isacharparenright}{\isasymrbrakk}\isanewline
{\isasymLongrightarrow}\ A\ {\isasymlangle}{\isasymlangle}\isactrlbsub i\isactrlesub \ qmsg\ {\isasymTurnstile}\ {\isacharparenleft}otherwith\ S\ {\isacharbraceleft}i{\isacharbraceright}\ {\isacharparenleft}orecvmsg\ R{\isacharparenright},\ other\ U\ {\isacharbraceleft}i{\isacharbraceright}\ {\isasymrightarrow}{\isacharparenright}\ global\ P%
\end{isabelle}}
\snip{lift_step_into_qmsg}{0}{0}{\begin{isabelle}%
{\isasymlbrakk}A\ \ostepinv\ {\isacharparenleft}otherwith\ S\ {\isacharbraceleft}i{\isacharbraceright}\ {\isacharparenleft}orecvmsg\ R{\isacharparenright},\ other\ U\ {\isacharbraceleft}i{\isacharbraceright}\ {\isasymrightarrow}{\isacharparenright}\ globala\ P{\isacharsemicolon}\ {\isasymAnd}{\isasymxi}{\isachardot}\ U\ {\isasymxi}\ {\isasymxi}{\isacharsemicolon}\isanewline
\isaindent{\ }{\isasymAnd}{\isasymxi}\ {\isasymxi}{\isacharprime}{\isachardot}\ S\ {\isasymxi}\ {\isasymxi}{\isacharprime}\ {\isasymLongrightarrow}\ U\ {\isasymxi}\ {\isasymxi}{\isacharprime}{\isacharsemicolon}\ {\isasymAnd}{\isasymsigma}\ {\isasymsigma}{\isacharprime}\ m{\isachardot}\ {\isasymlbrakk}{\isasymforall}j{\isachardot}\ U\ {\isacharparenleft}{\isasymsigma}\ j{\isacharparenright}\ {\isacharparenleft}{\isasymsigma}{\isacharprime}\ j{\isacharparenright}{\isacharsemicolon}\ R\ {\isasymsigma}\ m{\isasymrbrakk}\ {\isasymLongrightarrow}\ R\ {\isasymsigma}{\isacharprime}\ m{\isacharsemicolon}\isanewline
\isaindent{\ }A\ \ostepinv\ {\isacharparenleft}otherwith\ S\ {\isacharbraceleft}i{\isacharbraceright}\ {\isacharparenleft}orecvmsg\ R{\isacharparenright},\ other\ U\ {\isacharbraceleft}i{\isacharbraceright}\ {\isasymrightarrow}{\isacharparenright}\isanewline
\isaindent{\ A\ \ostepinv\ \ }globala\ {\isacharparenleft}{\isasymlambda}{\isacharparenleft}{\isasymsigma},\ {\isacharunderscore},\ {\isasymsigma}{\isacharprime}{\isacharparenright}{\isachardot}\ U\ {\isacharparenleft}{\isasymsigma}\ i{\isacharparenright}\ {\isacharparenleft}{\isasymsigma}{\isacharprime}\ i{\isacharparenright}{\isacharparenright}{\isacharsemicolon}\isanewline
\isaindent{\ }{\isasymAnd}{\isasymsigma}\ {\isasymsigma}{\isacharprime}\ m{\isachardot}\ {\isasymlbrakk}{\isasymforall}j{\isachardot}\ U\ {\isacharparenleft}{\isasymsigma}\ j{\isacharparenright}\ {\isacharparenleft}{\isasymsigma}{\isacharprime}\ j{\isacharparenright}{\isacharsemicolon}\ {\isasymsigma}{\isacharprime}\ i\ {\isacharequal}\ {\isasymsigma}\ i{\isasymrbrakk}\ {\isasymLongrightarrow}\ P\ {\isacharparenleft}{\isasymsigma},\ receive\ m,\ {\isasymsigma}{\isacharprime}{\isacharparenright}{\isacharsemicolon}\isanewline
\isaindent{\ }{\isasymAnd}{\isasymsigma}\ {\isasymsigma}{\isacharprime}\ m{\isachardot}\ P\ {\isacharparenleft}{\isasymsigma},\ receive\ m,\ {\isasymsigma}{\isacharprime}{\isacharparenright}\ {\isasymLongrightarrow}\ P\ {\isacharparenleft}{\isasymsigma},\ {\isasymtau},\ {\isasymsigma}{\isacharprime}{\isacharparenright}{\isasymrbrakk}\isanewline
{\isasymLongrightarrow}\ A\ {\isasymlangle}{\isasymlangle}\isactrlbsub i\isactrlesub \ qmsg\ \ostepinv\ {\isacharparenleft}otherwith\ S\ {\isacharbraceleft}i{\isacharbraceright}\ {\isacharparenleft}orecvmsg\ R{\isacharparenright},\ other\ U\ {\isacharbraceleft}i{\isacharbraceright}\ {\isasymrightarrow}{\isacharparenright}\ globala\ P%
\end{isabelle}}

\snip{node_lift}{0}{0}{\begin{isabelle}%
{\isasymlbrakk}T\ {\isasymTurnstile}\ {\isacharparenleft}otherwith\ S\ {\isacharbraceleft}ii{\isacharbraceright}\ {\isacharparenleft}orecvmsg\ I{\isacharparenright},\ other\ U\ {\isacharbraceleft}ii{\isacharbraceright}\ {\isasymrightarrow}{\isacharparenright}\ global\ P{\isacharsemicolon}\isanewline
\isaindent{\ }{\isasymAnd}{\isasymxi}\ {\isasymxi}{\isacharprime}{\isachardot}\ S\ {\isasymxi}\ {\isasymxi}{\isacharprime}\ {\isasymLongrightarrow}\ U\ {\isasymxi}\ {\isasymxi}{\isacharprime}{\isasymrbrakk}\isanewline
{\isasymLongrightarrow}\ {\isasymlangle}ii\ {\isacharcolon}\ T\ {\isacharcolon}\ \Ri{\isasymrangle}\isactrlsub o\ {\isasymTurnstile}\ {\isacharparenleft}otherwith\ S\ {\isacharbraceleft}ii{\isacharbraceright}\ {\isacharparenleft}oarrivemsg\ I{\isacharparenright},\ other\ U\ {\isacharbraceleft}ii{\isacharbraceright}\ {\isasymrightarrow}{\isacharparenright}\isanewline
\isaindent{{\isasymLongrightarrow}\ {\isasymlangle}ii\ {\isacharcolon}\ T\ {\isacharcolon}\ \Ri{\isasymrangle}\isactrlsub o\ {\isasymTurnstile}\ \ }global\ P%
\end{isabelle}}
\snip{node_lift_step}{0}{0}{\begin{isabelle}%
{\isasymlbrakk}T\ \ostepinv\ {\isacharparenleft}otherwith\ S\ {\isacharbraceleft}i{\isacharbraceright}\ {\isacharparenleft}orecvmsg\ I{\isacharparenright},\ other\ U\ {\isacharbraceleft}i{\isacharbraceright}\ {\isasymrightarrow}{\isacharparenright}\isanewline
\isaindent{{\isasymlbrakk}T\ \ostepinv\ \ }globala\ {\isacharparenleft}{\isasymlambda}{\isacharparenleft}{\isasymsigma},\ {\isacharunderscore},\ {\isasymsigma}{\isacharprime}{\isacharparenright}{\isachardot}\ Q\ {\isasymsigma}\ {\isasymsigma}{\isacharprime}{\isacharparenright}{\isacharsemicolon}\isanewline
\isaindent{\ }{\isasymAnd}{\isasymsigma}\ {\isasymsigma}{\isacharprime}{\isachardot}\ other\ U\ {\isacharbraceleft}i{\isacharbraceright}\ {\isasymsigma}\ {\isasymsigma}{\isacharprime}\ {\isasymLongrightarrow}\ Q\ {\isasymsigma}\ {\isasymsigma}{\isacharprime}{\isacharsemicolon}\ {\isasymAnd}{\isasymxi}\ {\isasymxi}{\isacharprime}{\isachardot}\ S\ {\isasymxi}\ {\isasymxi}{\isacharprime}\ {\isasymLongrightarrow}\ U\ {\isasymxi}\ {\isasymxi}{\isacharprime}{\isasymrbrakk}\isanewline
{\isasymLongrightarrow}\ {\isasymlangle}i\ {\isacharcolon}\ T\ {\isacharcolon}\ \Ri{\isasymrangle}\isactrlsub o\ \ostepinv\ {\isacharparenleft}otherwith\ S\ {\isacharbraceleft}i{\isacharbraceright}\ {\isacharparenleft}oarrivemsg\ I{\isacharparenright},\ other\ U\ {\isacharbraceleft}i{\isacharbraceright}\ {\isasymrightarrow}{\isacharparenright}\isanewline
\isaindent{{\isasymLongrightarrow}\ {\isasymlangle}i\ {\isacharcolon}\ T\ {\isacharcolon}\ \Ri{\isasymrangle}\isactrlsub o\ \ostepinv\ \ }globala\ {\isacharparenleft}{\isasymlambda}{\isacharparenleft}{\isasymsigma},\ {\isacharunderscore},\ {\isasymsigma}{\isacharprime}{\isacharparenright}{\isachardot}\ Q\ {\isasymsigma}\ {\isasymsigma}{\isacharprime}{\isacharparenright}%
\end{isabelle}}

\snip{optoy_qmsg_term}{0}{0}{\isa{optoy\ i\ {\isasymlangle}{\isasymlangle}\isactrlbsub i\isactrlesub \ qmsg}}

\snip{otherwith_r_term}{0}{0}{\isa{otherwith\ S\ {\isacharbraceleft}i{\isacharbraceright}\ {\isacharparenleft}orecvmsg\ R{\isacharparenright}}}
\snip{stmt_guard}{0}{0}{\isa{{\isasymlangle}{\isasymlambda}{\isasymxi}{\isachardot}\ \textsf{if}\ g\ {\isasymxi}\ \textsf{then}\ {\isacharbraceleft}{\isasymxi}{\isacharbraceright}\ \textsf{else}\ {\isasymemptyset}{\isasymrangle}\ p}}
\snip{stmt_bind}{0}{0}{\isa{{\isasymlangle}{\isasymlambda}{\isasymxi}{\isachardot}\ {\isacharbraceleft}{\isasymxi}{\isasymlparr}no\ {\isacharcolon}{\isacharequal}\ n{\isasymrparr}\ {\isacharbar}\ n\ {\isacharless}\ {\isadigit{5}}{\isacharbraceright}{\isasymrangle}\ p}}

\snip{act_unicast}{0}{0}{\isa{unicast\ i\ m}}
\snip{act_not_unicast}{0}{0}{\isa{{\isasymnot}unicast\ i}}

\snip{eg1_pnet}{0}{0}{\isa{pnet\ {\isacharparenleft}{\isasymlambda}i{\isachardot}\ ptoy\ i\ {\isasymlangle}{\isasymlangle}\ qmsg{\isacharparenright}\ {\isacharparenleft}{\isasymlangle}A{\isacharsemicolon}\ {\isacharbraceleft}B{\isacharbraceright}{\isasymrangle}\parallelcomp{}{\isasymlangle}B{\isacharsemicolon}\ {\isacharbraceleft}A,\ C{\isacharbraceright}{\isasymrangle}\parallelcomp{}{\isasymlangle}C{\isacharsemicolon}\ {\isacharbraceleft}B{\isacharbraceright}{\isasymrangle}{\isacharparenright}}}

\snip{eg1_all}{0}{0}{\isa{closed\ {\isacharparenleft}pnet\ {\isacharparenleft}{\isasymlambda}i{\isachardot}\ ptoy\ i\ {\isasymlangle}{\isasymlangle}\ qmsg{\isacharparenright}\ {\isacharparenleft}{\isasymlangle}A{\isacharsemicolon}\ {\isacharbraceleft}B,\ D{\isacharbraceright}{\isasymrangle}\parallelcomp{}{\isacharparenleft}{\isasymlangle}B{\isacharsemicolon}\ {\isacharbraceleft}A,\ C{\isacharbraceright}{\isasymrangle}\parallelcomp{}{\isacharparenleft}{\isasymlangle}C{\isacharsemicolon}\ {\isacharbraceleft}B{\isacharbraceright}{\isasymrangle}\parallelcomp{}{\isasymlangle}D{\isacharsemicolon}\ {\isacharbraceleft}A{\isacharbraceright}{\isasymrangle}{\isacharparenright}{\isacharparenright}{\isacharparenright}{\isacharparenright}}}

\snip{eg1_node_a'}{0}{0}{\isa{{\isasymlangle}A\ {\isacharcolon}\ ptoy\ A\ {\isasymlangle}{\isasymlangle}\ qmsg\ {\isacharcolon}\ {\isacharbraceleft}B,\ D{\isacharbraceright}{\isasymrangle}}}
\snip{eg1_node_b'}{0}{0}{\isa{{\isasymlangle}B\ {\isacharcolon}\ ptoy\ B\ {\isasymlangle}{\isasymlangle}\ qmsg\ {\isacharcolon}\ {\isacharbraceleft}A,\ C{\isacharbraceright}{\isasymrangle}}}
\snip{eg1_node_c'}{0}{0}{\isa{{\isasymlangle}C\ {\isacharcolon}\ ptoy\ C\ {\isasymlangle}{\isasymlangle}\ qmsg\ {\isacharcolon}\ {\isacharbraceleft}B{\isacharbraceright}{\isasymrangle}}}
\snip{eg1_node_d'}{0}{0}{\isa{{\isasymlangle}D\ {\isacharcolon}\ ptoy\ D\ {\isasymlangle}{\isasymlangle}\ qmsg\ {\isacharcolon}\ {\isacharbraceleft}A{\isacharbraceright}{\isasymrangle}}}

\snip{eg1_node_a}{0}{0}{\isa{{\isasymlangle}A{\isacharsemicolon}\ {\isacharbraceleft}B,\ D{\isacharbraceright}{\isasymrangle}}}
\snip{eg1_node_b}{0}{0}{\isa{{\isasymlangle}B{\isacharsemicolon}\ {\isacharbraceleft}A,\ C{\isacharbraceright}{\isasymrangle}}}
\snip{eg1_node_c}{0}{0}{\isa{{\isasymlangle}C{\isacharsemicolon}\ {\isacharbraceleft}B{\isacharbraceright}{\isasymrangle}}}
\snip{eg1_node_d}{0}{0}{\isa{{\isasymlangle}D{\isacharsemicolon}\ {\isacharbraceleft}A{\isacharbraceright}{\isasymrangle}}}

\snip{dummy_par}{0}{0}{\isa{\_\parallelcomp{}\_}}

\snip{eg_sterms_foo}{0}{0}{\isa{sterms\ {\isasymGamma}\ {\isacharparenleft}{\isasymGamma}\ foo{\isacharparenright}}}
\snip{eg_sterms_groupcast}{0}{0}{\isa{dterms\ {\isasymGamma}\ {\isacharparenleft}groupcast{\isacharparenleft}\_,\ \_{\isacharparenright}\ {\isachardot}\ p{\isacharparenright}}}
\snip{eg_sterms_p}{0}{0}{\isa{sterms\ {\isasymGamma}\ p}}
\snip{gammap}{0}{0}{\isa{\gammatoy}}

\snip{gammap_ptoy}{0}{0}{\begin{isabelle}%
\gammatoy\ PToy\ {\isacharequal}\ {\isacharbraceleft}PToy{\isacharminus}{\isacharcolon}{\isadigit{0}}{\isacharbraceright}receive{\isacharparenleft}{\isasymlambda}msg{\isacharprime}{\isachardot}\ msg{\isacharunderscore}update\ {\isacharparenleft}{\isasymlambda}{\isacharunderscore}{\isachardot}\ msg{\isacharprime}{\isacharparenright}{\isacharparenright}\ {\isachardot}\isanewline
{\isacharbraceleft}PToy{\isacharminus}{\isacharcolon}{\isadigit{1}}{\isacharbraceright}{\isasymlbrakk}{\isasymlambda}{\isasymxi}{\isachardot}\ {\isasymxi}{\isasymlparr}nhip\ {\isacharcolon}{\isacharequal}\ ip\ {\isasymxi}{\isasymrparr}{\isasymrbrakk}\isanewline
{\isacharparenleft}{\isacharbraceleft}PToy{\isacharminus}{\isacharcolon}{\isadigit{2}}{\isacharbraceright}{\isasymlangle}is{\isacharunderscore}newpkt{\isasymrangle}\isanewline
\isaindent{{\isacharparenleft}}{\isacharbraceleft}PToy{\isacharminus}{\isacharcolon}{\isadigit{3}}{\isacharbraceright}{\isasymlbrakk}{\isasymlambda}{\isasymxi}{\isachardot}\ {\isasymxi}{\isasymlparr}no\ {\isacharcolon}{\isacharequal}\ max\ {\isacharparenleft}no\ {\isasymxi}{\isacharparenright}\ {\isacharparenleft}num\ {\isasymxi}{\isacharparenright}{\isasymrparr}{\isasymrbrakk}\isanewline
\isaindent{{\isacharparenleft}}{\isacharbraceleft}PToy{\isacharminus}{\isacharcolon}{\isadigit{4}}{\isacharbraceright}broadcast{\isacharparenleft}{\isasymlambda}{\isasymxi}{\isachardot}\ Pkt\ {\isacharparenleft}no\ {\isasymxi}{\isacharparenright}\ {\isacharparenleft}ip\ {\isasymxi}{\isacharparenright}{\isacharparenright}\ {\isachardot}\isanewline
\isaindent{{\isacharparenleft}}{\isacharbraceleft}PToy{\isacharminus}{\isacharcolon}{\isadigit{5}}{\isacharbraceright}{\isasymlbrakk}clear{\isacharunderscore}locals{\isasymrbrakk}\isanewline
\isaindent{{\isacharparenleft}}call{\isacharparenleft}PToy{\isacharparenright}\isanewline
\isaindent{{\isacharparenleft}}{\isasymoplus}\isanewline
\isaindent{{\isacharparenleft}}{\isacharbraceleft}PToy{\isacharminus}{\isacharcolon}{\isadigit{2}}{\isacharbraceright}{\isasymlangle}is{\isacharunderscore}pkt{\isasymrangle}\isanewline
\isaindent{{\isacharparenleft}}{\isacharparenleft}{\isacharbraceleft}PToy{\isacharminus}{\isacharcolon}{\isadigit{6}}{\isacharbraceright}{\isasymlangle}{\isasymlambda}{\isasymxi}{\isachardot}\ \textsf{if}\ no\ {\isasymxi}\ {\isasymle}\ num\ {\isasymxi}\ \textsf{then}\ {\isacharbraceleft}{\isasymxi}{\isacharbraceright}\ \textsf{else}\ {\isasymemptyset}{\isasymrangle}\isanewline
\isaindent{{\isacharparenleft}{\isacharparenleft}}{\isacharbraceleft}PToy{\isacharminus}{\isacharcolon}{\isadigit{7}}{\isacharbraceright}{\isasymlbrakk}{\isasymlambda}{\isasymxi}{\isachardot}\ {\isasymxi}{\isasymlparr}no\ {\isacharcolon}{\isacharequal}\ num\ {\isasymxi}{\isasymrparr}{\isasymrbrakk}\isanewline
\isaindent{{\isacharparenleft}{\isacharparenleft}}{\isacharbraceleft}PToy{\isacharminus}{\isacharcolon}{\isadigit{8}}{\isacharbraceright}{\isasymlbrakk}{\isasymlambda}{\isasymxi}{\isachardot}\ {\isasymxi}{\isasymlparr}nhip\ {\isacharcolon}{\isacharequal}\ sip\ {\isasymxi}{\isasymrparr}{\isasymrbrakk}\isanewline
\isaindent{{\isacharparenleft}{\isacharparenleft}}{\isacharbraceleft}PToy{\isacharminus}{\isacharcolon}{\isadigit{9}}{\isacharbraceright}broadcast{\isacharparenleft}{\isasymlambda}{\isasymxi}{\isachardot}\ Pkt\ {\isacharparenleft}no\ {\isasymxi}{\isacharparenright}\ {\isacharparenleft}ip\ {\isasymxi}{\isacharparenright}{\isacharparenright}\ {\isachardot}\isanewline
\isaindent{{\isacharparenleft}{\isacharparenleft}}{\isacharbraceleft}PToy{\isacharminus}{\isacharcolon}{\isadigit{1}}{\isadigit{0}}{\isacharbraceright}{\isasymlbrakk}clear{\isacharunderscore}locals{\isasymrbrakk}\isanewline
\isaindent{{\isacharparenleft}{\isacharparenleft}}call{\isacharparenleft}PToy{\isacharparenright}\isanewline
\isaindent{{\isacharparenleft}{\isacharparenleft}}{\isasymoplus}\isanewline
\isaindent{{\isacharparenleft}{\isacharparenleft}}{\isacharbraceleft}PToy{\isacharminus}{\isacharcolon}{\isadigit{6}}{\isacharbraceright}{\isasymlangle}{\isasymlambda}{\isasymxi}{\isachardot}\ \textsf{if}\ num\ {\isasymxi}\ {\isacharless}\ no\ {\isasymxi}\ \textsf{then}\ {\isacharbraceleft}{\isasymxi}{\isacharbraceright}\ \textsf{else}\ {\isasymemptyset}{\isasymrangle}\isanewline
\isaindent{{\isacharparenleft}{\isacharparenleft}}{\isacharbraceleft}PToy{\isacharminus}{\isacharcolon}{\isadigit{1}}{\isadigit{1}}{\isacharbraceright}{\isasymlbrakk}clear{\isacharunderscore}locals{\isasymrbrakk}\isanewline
\isaindent{{\isacharparenleft}{\isacharparenleft}}call{\isacharparenleft}PToy{\isacharparenright}{\isacharparenright}{\isacharparenright}%
\end{isabelle}}

\snip{sigmap}{0}{0}{\isa{\sigmatoy\ i\ {\isacharequal}\ {\isacharbraceleft}{\isacharparenleft}toy{\isacharunderscore}init\ i,\ \gammatoy\ PToy{\isacharparenright}{\isacharbraceright}}}
\snip{ptoy}{0}{0}{\isa{ptoy}}
\snip{ptoy_term}{0}{0}{\isa{{\isasymlparr}init\ {\isacharequal}\ {\isacharbraceleft}{\isacharparenleft}toy{\isacharunderscore}init\ i,\ \gammatoy\ PToy{\isacharparenright}{\isacharbraceright},\ trans\ {\isacharequal}\ \seqpsos\ \gammatoy{\isasymrparr}}}
\snip{ptoy_term'}{0}{0}{\isa{{\isasymlparr}init\ {\isacharequal}\ \sigmatoy\ i,\ trans\ {\isacharequal}\ \seqpsos\ \gammatoy{\isasymrparr}}}
\snip{ptoy_type}{0}{0}{\isa{nat\ {\isasymRightarrow}\ {\isacharparenleft}state\ {\isasymtimes}\ {\isacharparenleft}state,\ msg,\ pseqp,\ pseqp\ label{\isacharparenright}\ seqp,\ msg\ seq{\isacharunderscore}action{\isacharparenright}\ automaton}}

\snip{sigmap'}{0}{0}{\isa{{\isasymsigma}\isactrlsub O\isactrlsub T\isactrlsub O\isactrlsub Y\ {\isacharequal}\ {\isacharbraceleft}{\isacharparenleft}toy{\isacharunderscore}init,\ \gammatoy\ PToy{\isacharparenright}{\isacharbraceright}}}
\snip{optoy}{0}{0}{\isa{optoy}}
\snip{optoy_term}{0}{0}{\isa{{\isasymlparr}init\ {\isacharequal}\ {\isacharbraceleft}{\isacharparenleft}toy{\isacharunderscore}init,\ \gammatoy\ PToy{\isacharparenright}{\isacharbraceright},\ trans\ {\isacharequal}\ \oseqpsos\ \gammatoy\ i{\isasymrparr}}}

\snip{gammaq}{0}{0}{\isa{\gammaqmsg\ {\isacharparenleft}{\isacharparenright}}}
\snip{gammaq_unit}{0}{0}{\begin{isabelle}%
\gammaqmsg\ {\isacharparenleft}{\isacharparenright}\ {\isacharequal}\isanewline
{\isacharbraceleft}{\isacharparenleft}{\isacharparenright}{\isacharminus}{\isacharcolon}{\isadigit{0}}{\isacharbraceright}receive{\isacharparenleft}{\isasymlambda}msg\ msgs{\isachardot}\ msgs\ {\isacharat}\ {\isacharbrackleft}msg{\isacharbrackright}{\isacharparenright}\ {\isachardot}\isanewline
call{\isacharparenleft}{\isacharparenleft}{\isacharparenright}{\isacharparenright}\isanewline
{\isasymoplus}\isanewline
{\isacharbraceleft}{\isacharparenleft}{\isacharparenright}{\isacharminus}{\isacharcolon}{\isadigit{0}}{\isacharbraceright}{\isasymlangle}{\isasymlambda}msgs{\isachardot}\ \textsf{if}\ msgs\ {\isasymnoteq}\ {\isacharbrackleft}{\isacharbrackright}\ \textsf{then}\ {\isacharbraceleft}msgs{\isacharbraceright}\ \textsf{else}\ {\isasymemptyset}{\isasymrangle}\isanewline
{\isacharparenleft}{\isacharbraceleft}{\isacharparenleft}{\isacharparenright}{\isacharminus}{\isacharcolon}{\isadigit{1}}{\isacharbraceright}send{\isacharparenleft}hd{\isacharparenright}\ {\isachardot}\isanewline
\isaindent{{\isacharparenleft}}{\isacharbraceleft}{\isacharparenleft}{\isacharparenright}{\isacharminus}{\isacharcolon}{\isadigit{2}}{\isacharbraceright}{\isasymlbrakk}tl{\isasymrbrakk}\isanewline
\isaindent{{\isacharparenleft}}call{\isacharparenleft}{\isacharparenleft}{\isacharparenright}{\isacharparenright}\isanewline
\isaindent{{\isacharparenleft}}{\isasymoplus}\isanewline
\isaindent{{\isacharparenleft}}{\isacharbraceleft}{\isacharparenleft}{\isacharparenright}{\isacharminus}{\isacharcolon}{\isadigit{1}}{\isacharbraceright}receive{\isacharparenleft}{\isasymlambda}msg\ msgs{\isachardot}\ msgs\ {\isacharat}\ {\isacharbrackleft}msg{\isacharbrackright}{\isacharparenright}\ {\isachardot}\isanewline
\isaindent{{\isacharparenleft}}call{\isacharparenleft}{\isacharparenleft}{\isacharparenright}{\isacharparenright}{\isacharparenright}%
\end{isabelle}}
\snip{sigmaq}{0}{0}{\isa{\gammaqmsg\ {\isacharequal}\ {\isacharbraceleft}{\isacharparenleft}{\isacharbrackleft}{\isacharbrackright},\ \gammaqmsg\ {\isacharparenleft}{\isacharparenright}{\isacharparenright}{\isacharbraceright}}}
\snip{qmsg}{0}{0}{\isa{qmsg}}
\snip{qmsg_term}{0}{0}{\isa{{\isasymlparr}init\ {\isacharequal}\ \gammaqmsg,\ trans\ {\isacharequal}\ \seqpsos\ \gammaqmsg{\isasymrparr}}}

\snip{clear_locals_sip}{0}{0}{\isa{{\isasymxi}{\isasymlparr}sip\ {\isacharcolon}{\isacharequal}\ SOME\ x{\isachardot}\ x\ {\isasymnoteq}\ ip\ {\isasymxi}{\isasymrparr}}}

\snip{ptoy_qmsg_term}{0}{0}{\isa{ptoy\ i\ {\isasymlangle}{\isasymlangle}\ qmsg}}
\snip{opnet_p}{0}{0}{\isa{opnet\ p}}
\snip{opnet_node}{0}{0}{\isa{opnet\ {\isacharparenleft}{\isasymlambda}i{\isachardot}\ optoy\ i\ {\isasymlangle}{\isasymlangle}\isactrlbsub i\isactrlesub \ qmsg{\isacharparenright}\ {\isasymlangle}i{\isacharsemicolon}\ R{\isasymrangle}}}
\snip{net_tree_ips_node}{0}{0}{\isa{net{\isacharunderscore}tree{\isacharunderscore}ips\ {\isasymlangle}i{\isacharsemicolon}\ R{\isasymrangle}\ {\isacharequal}\ {\isacharbraceleft}i{\isacharbraceright}}}

\snip{opnet_p1_par_p2}{0}{0}{\isa{opnet\ {\isacharparenleft}{\isasymlambda}i{\isachardot}\ optoy\ i\ {\isasymlangle}{\isasymlangle}\isactrlbsub i\isactrlesub \ qmsg{\isacharparenright}\ {\isacharparenleft}p\isactrlsub {\isadigit{1}}\parallelcomp{}p\isactrlsub {\isadigit{2}}{\isacharparenright}}}
\snip{net_tree_ips_par}{0}{0}{\isa{net{\isacharunderscore}tree{\isacharunderscore}ips\ {\isacharparenleft}p{\isadigit{1}}{\isachardot}{\isadigit{0}}\parallelcomp{}p{\isadigit{2}}{\isachardot}{\isadigit{0}}{\isacharparenright}\ {\isacharequal}\ net{\isacharunderscore}tree{\isacharunderscore}ips\ p{\isadigit{1}}{\isachardot}{\isadigit{0}}\ {\isasymunion}\ net{\isacharunderscore}tree{\isacharunderscore}ips\ p{\isadigit{2}}{\isachardot}{\isadigit{0}}}}

\snip{pnet_lift_prem1}{0}{0}{\isa{opnet\ onp\ {\isasymlangle}i{\isacharsemicolon}\ R{\isasymrangle}\ {\isasymTurnstile}\ {\isacharparenleft}otherwith\ S\ {\isacharbraceleft}i{\isacharbraceright}\ {\isacharparenleft}oarrivemsg\ I{\isacharparenright},\ other\ U\ {\isacharbraceleft}i{\isacharbraceright}\ {\isasymrightarrow}{\isacharparenright}\ global\ {\isacharparenleft}P\ i{\isacharparenright}}}

\snip{pnet_lift_assume_s}{0}{0}{\isa{otherwith\ S\ {\isacharparenleft}net{\isacharunderscore}tree{\isacharunderscore}ips\ p{\isacharparenright}\ {\isacharparenleft}oarrivemsg\ I{\isacharparenright}}}
\snip{pnet_lift_assume_u}{0}{0}{\isa{other\ U\ {\isacharparenleft}net{\isacharunderscore}tree{\isacharunderscore}ips\ p{\isacharparenright}}}
\snip{pnet_lift_shows}{0}{0}{\isa{global\ {\isacharparenleft}{\isasymlambda}{\isasymsigma}{\isachardot}\ {\isasymforall}i{\isasymin}net{\isacharunderscore}tree{\isacharunderscore}ips\ p{\isachardot}\ P\ i\ {\isasymsigma}{\isacharparenright}}}

\snip{pnet_lift_castmsg_pred}{0}{0}{\isa{castmsg\ {\isacharparenleft}I\ {\isasymsigma}{\isacharparenright}}}
\snip{pnet_lift_arrivemsg_pred}{0}{0}{\isa{oarrivemsg\ I\ {\isasymsigma}}}

\snip{pnet_lift_castmsg}{0}{0}{\isa{opnet\ onp\ {\isasymlangle}i{\isacharsemicolon}\ R{\isasymrangle}\ \ostepinv\ {\isacharparenleft}{\isasymlambda}{\isasymsigma}\ {\isacharunderscore}{\isachardot}\ oarrivemsg\ I\ {\isasymsigma},\ other\ U\ {\isacharbraceleft}i{\isacharbraceright}\ {\isasymrightarrow}{\isacharparenright}\ globala\ {\isacharparenleft}{\isasymlambda}{\isacharparenleft}{\isasymsigma},\ a,\ {\isasymsigma}{\isacharprime}{\isacharparenright}{\isachardot}\ castmsg\ {\isacharparenleft}I\ {\isasymsigma}{\isacharparenright}\ a{\isacharparenright}}}
\snip{pnet_lift_s}{0}{0}{\isa{opnet\ onp\ {\isasymlangle}i{\isacharsemicolon}\ R{\isasymrangle}\ \ostepinv\ {\isacharparenleft}{\isasymlambda}{\isasymsigma}\ {\isacharunderscore}{\isachardot}\ oarrivemsg\ I\ {\isasymsigma},\ other\ U\ {\isacharbraceleft}i{\isacharbraceright}\ {\isasymrightarrow}{\isacharparenright}\ globala\ {\isacharparenleft}{\isasymlambda}{\isacharparenleft}{\isasymsigma},\ a,\ {\isasymsigma}{\isacharprime}{\isacharparenright}{\isachardot}\ S\ {\isacharparenleft}{\isasymsigma}\ i{\isacharparenright}\ {\isacharparenleft}{\isasymsigma}{\isacharprime}\ i{\isacharparenright}{\isacharparenright}}}
\snip{pnet_lift_u}{0}{0}{\isa{opnet\ onp\ {\isasymlangle}i{\isacharsemicolon}\ R{\isasymrangle}\ \ostepinv\ {\isacharparenleft}{\isasymlambda}{\isasymsigma}\ {\isacharunderscore}{\isachardot}\ oarrivemsg\ I\ {\isasymsigma},\ other\ U\ {\isacharbraceleft}i{\isacharbraceright}\ {\isasymrightarrow}{\isacharparenright}\ globala\ {\isacharparenleft}{\isasymlambda}{\isacharparenleft}{\isasymsigma},\ a,\ {\isasymsigma}{\isacharprime}{\isacharparenright}{\isachardot}\ U\ {\isacharparenleft}{\isasymsigma}\ i{\isacharparenright}\ {\isacharparenleft}{\isasymsigma}{\isacharprime}\ i{\isacharparenright}{\isacharparenright}}}

\snip{pnet_lift_concl}{0}{0}{\isa{opnet\ onp\ p\ {\isasymTurnstile}\ {\isacharparenleft}otherwith\ S\ {\isacharparenleft}net{\isacharunderscore}tree{\isacharunderscore}ips\ p{\isacharparenright}\ {\isacharparenleft}oarrivemsg\ I{\isacharparenright},\ other\ U\ {\isacharparenleft}net{\isacharunderscore}tree{\isacharunderscore}ips\ p{\isacharparenright}\ {\isasymrightarrow}{\isacharparenright}\ global\ {\isacharparenleft}{\isasymlambda}{\isasymsigma}{\isachardot}\ {\isasymforall}i{\isasymin}net{\isacharunderscore}tree{\isacharunderscore}ips\ p{\isachardot}\ P\ i\ {\isasymsigma}{\isacharparenright}}}
\snip{other_net_tree_ips_par_left}{0}{0}{\isa{\mbox{}\inferrule{\mbox{other\ U\ {\isacharparenleft}net{\isacharunderscore}tree{\isacharunderscore}ips\ {\isacharparenleft}p\isactrlsub {\isadigit{1}}\parallelcomp{}p\isactrlsub {\isadigit{2}}{\isacharparenright}{\isacharparenright}\ {\isasymsigma}\ {\isasymsigma}{\isacharprime}}\\\ \mbox{{\isasymAnd}{\isasymxi}{\isachardot}\ U\ {\isasymxi}\ {\isasymxi}}}{\mbox{other\ U\ {\isacharparenleft}net{\isacharunderscore}tree{\isacharunderscore}ips\ p\isactrlsub {\isadigit{1}}{\isacharparenright}\ {\isasymsigma}\ {\isasymsigma}{\isacharprime}}}}}
\snip{net_tree_ips_p1_par_p2}{0}{0}{\isa{i{\isasymin}net{\isacharunderscore}tree{\isacharunderscore}ips\ {\isacharparenleft}p\isactrlsub {\isadigit{1}}\parallelcomp{}p\isactrlsub {\isadigit{2}}{\isacharparenright}}}
\snip{net_tree_ips_not_p1_par_p2}{0}{0}{\isa{i\ {\isasymnotin}\ net{\isacharunderscore}tree{\isacharunderscore}ips\ {\isacharparenleft}p\isactrlsub {\isadigit{1}}\parallelcomp{}p\isactrlsub {\isadigit{2}}{\isacharparenright}}}
\snip{net_tree_ips_p1}{0}{0}{\isa{i{\isasymin}net{\isacharunderscore}tree{\isacharunderscore}ips\ p\isactrlsub {\isadigit{1}}}}
\snip{net_tree_ips_p2}{0}{0}{\isa{i{\isasymin}net{\isacharunderscore}tree{\isacharunderscore}ips\ p\isactrlsub {\isadigit{2}}}}
\snip{net_tree_ips_not_p1_sigma}{0}{0}{\isa{{\isasymforall}j{\isachardot}\ j\ {\isasymnotin}\ net{\isacharunderscore}tree{\isacharunderscore}ips\ p\isactrlsub {\isadigit{1}}\ {\isasymlongrightarrow}\ {\isasymsigma}{\isacharprime}\ j\ {\isacharequal}\ {\isasymsigma}\ j}}
\snip{node_proc_reachable_prem_1}{0}{0}{\isa{{\isacharparenleft}{\isasymsigma},\ \NodeS\ i\ s\ R{\isacharparenright}{\isasymin}oreachable\ {\isacharparenleft}\hspace{.1em}{\isasymlangle}i\ {\isacharcolon}\ A\ {\isacharcolon}\ \Ri{\isasymrangle}\isactrlsub o{\isacharparenright}\ {\isacharparenleft}otherwith\ E\ {\isacharbraceleft}i{\isacharbraceright}\ {\isacharparenleft}oarrivemsg\ I{\isacharparenright}{\isacharparenright}\ {\isacharparenleft}other\ F\ {\isacharbraceleft}i{\isacharbraceright}{\isacharparenright}}}

\snip{node_proc_reachable_prem_2}{0}{0}{\isa{{\isasymAnd}{\isasymxi}\ {\isasymxi}{\isacharprime}{\isachardot}\ E\ {\isasymxi}\ {\isasymxi}{\isacharprime}\ {\isasymLongrightarrow}\ F\ {\isasymxi}\ {\isasymxi}{\isacharprime}}}
\snip{node_proc_reachable_prem_2_prem}{0}{0}{\isa{E\ {\isasymxi}\ {\isasymxi}{\isacharprime}}}
\snip{node_proc_reachable_prem_2_concl}{0}{0}{\isa{F\ {\isasymxi}\ {\isasymxi}{\isacharprime}}}
\snip{node_proc_reachable_concl}{0}{0}{\isa{{\isacharparenleft}{\isasymsigma},\ s{\isacharparenright}{\isasymin}oreachable\ A\ {\isacharparenleft}otherwith\ E\ {\isacharbraceleft}i{\isacharbraceright}\ {\isacharparenleft}orecvmsg\ I{\isacharparenright}{\isacharparenright}\ {\isacharparenleft}other\ F\ {\isacharbraceleft}i{\isacharbraceright}{\isacharparenright}}}

\snip{subnet_oreachable_prem_1}{0}{0}{\isa{{\isacharparenleft}{\isasymsigma},\ \subnetsit{s}{t}{\isacharparenright}{\isasymin}oreachable\ {\isacharparenleft}opnet\ onp\ {\isacharparenleft}p\isactrlsub {\isadigit{1}}\parallelcomp{}p\isactrlsub {\isadigit{2}}{\isacharparenright}{\isacharparenright}\ S\ U}}
\snip{subnet_oreachable_prem_1_L}{0}{0}{\isa{S\ {\isacharequal}\ otherwith\ E\ {\isacharparenleft}net{\isacharunderscore}tree{\isacharunderscore}ips\ {\isacharparenleft}p\isactrlsub {\isadigit{1}}\parallelcomp{}p\isactrlsub {\isadigit{2}}{\isacharparenright}{\isacharparenright}\ {\isacharparenleft}oarrivemsg\ I{\isacharparenright}}}
\snip{subnet_oreachable_prem_1_E}{0}{0}{\isa{U\ {\isacharequal}\ other\ F\ {\isacharparenleft}net{\isacharunderscore}tree{\isacharunderscore}ips\ {\isacharparenleft}p\isactrlsub {\isadigit{1}}\parallelcomp{}p\isactrlsub {\isadigit{2}}{\isacharparenright}{\isacharparenright}}}

\snip{subnet_oreachable_prem_2}{0}{0}{\isa{{\isasymAnd}{\isasymxi}{\isachardot}\ E\ {\isasymxi}\ {\isasymxi}}}
\snip{subnet_oreachable_prem_3}{0}{0}{\isa{{\isasymAnd}{\isasymxi}{\isachardot}\ F\ {\isasymxi}\ {\isasymxi}}}
\snip{subnet_oreachable_prem_4}{0}{0}{\isa{{\isasymAnd}i\ R{\isachardot}\ {\isasymlangle}i\ {\isacharcolon}\ onp\ i\ {\isacharcolon}\ R{\isasymrangle}\isactrlsub o\ \ostepinv\ {\isacharparenleft}{\isasymlambda}{\isasymsigma}\ {\isacharunderscore}{\isachardot}\ oarrivemsg\ I\ {\isasymsigma},\ other\ F\ {\isacharbraceleft}i{\isacharbraceright}\ {\isasymrightarrow}{\isacharparenright}\ globala\ {\isacharparenleft}{\isasymlambda}{\isacharparenleft}{\isasymsigma},\ a,\ {\isacharunderscore}{\isacharparenright}{\isachardot}\ castmsg\ {\isacharparenleft}I\ {\isasymsigma}{\isacharparenright}\ a{\isacharparenright}}}

\snip{subnet_oreachable_prem_4_p}{0}{0}{\isa{p\ i\ \ostepinv\ {\isacharparenleft}{\isasymlambda}{\isasymsigma}\ {\isacharunderscore}{\isachardot}\ oarrivemsg\ I\ {\isasymsigma},\ other\ F\ {\isacharbraceleft}i{\isacharbraceright}\ {\isasymrightarrow}{\isacharparenright}\ {\isacharparenleft}{\isasymlambda}{\isacharparenleft}{\isacharparenleft}{\isasymsigma},\ {\isacharunderscore}{\isacharparenright},\ a,\ ({\isasymsigma}{\isacharprime},\ {\isacharunderscore}){\isacharparenright}{\isachardot}\ castmsg\ {\isacharparenleft}I\ {\isasymsigma}{\isacharparenright}\ a{\isacharparenright}}}
\snip{subnet_oreachable_prem_4_p'}{0}{0}{\isa{{\isasymlangle}i\ {\isacharcolon}\ onp\ i\ {\isacharcolon}\ R{\isasymrangle}\isactrlsub o}}

\snip{subnet_oreachable_prem_5}{0}{0}{\isa{{\isasymAnd}i\ R{\isachardot}\ {\isasymlangle}i\ {\isacharcolon}\ onp\ i\ {\isacharcolon}\ R{\isasymrangle}\isactrlsub o\ \ostepinv\ {\isacharparenleft}{\isasymlambda}{\isasymsigma}\ {\isacharunderscore}{\isachardot}\ oarrivemsg\ I\ {\isasymsigma},\ other\ F\ {\isacharbraceleft}i{\isacharbraceright}\ {\isasymrightarrow}{\isacharparenright}\ globala\ {\isacharparenleft}{\isasymlambda}{\isacharparenleft}{\isasymsigma},\ a,\ {\isasymsigma}{\isacharprime}{\isacharparenright}{\isachardot}\ a\ {\isasymnoteq}\ {\isasymtau}\ {\isasymand}\ {\isacharparenleft}{\isasymforall}i\ d{\isachardot}\ a\ {\isasymnoteq}\ i{\isacharcolon}deliver{\isacharparenleft}d{\isacharparenright}{\isacharparenright}\ {\isasymlongrightarrow}\ E\ {\isacharparenleft}{\isasymsigma}\ i{\isacharparenright}\ {\isacharparenleft}{\isasymsigma}{\isacharprime}\ i{\isacharparenright}{\isacharparenright}}}
\snip{subnet_oreachable_prem_5_pred}{0}{0}{\isa{E\ {\isacharparenleft}{\isasymsigma}\ i{\isacharparenright}\ {\isacharparenleft}{\isasymsigma}{\isacharprime}\ i{\isacharparenright}}}

\snip{subnet_oreachable_prem_6}{0}{0}{\isa{{\isasymAnd}i\ R{\isachardot}\ {\isasymlangle}i\ {\isacharcolon}\ onp\ i\ {\isacharcolon}\ R{\isasymrangle}\isactrlsub o\ \ostepinv\ {\isacharparenleft}{\isasymlambda}{\isasymsigma}\ {\isacharunderscore}{\isachardot}\ oarrivemsg\ I\ {\isasymsigma},\ other\ F\ {\isacharbraceleft}i{\isacharbraceright}\ {\isasymrightarrow}{\isacharparenright}\ globala\ {\isacharparenleft}{\isasymlambda}{\isacharparenleft}{\isasymsigma},\ a,\ {\isasymsigma}{\isacharprime}{\isacharparenright}{\isachardot}\ a\ {\isacharequal}\ {\isasymtau}\ {\isasymor}\ {\isacharparenleft}{\isasymexists}d{\isachardot}\ a\ {\isacharequal}\ i{\isacharcolon}deliver{\isacharparenleft}d{\isacharparenright}{\isacharparenright}\ {\isasymlongrightarrow}\ F\ {\isacharparenleft}{\isasymsigma}\ i{\isacharparenright}\ {\isacharparenleft}{\isasymsigma}{\isacharprime}\ i{\isacharparenright}{\isacharparenright}}}
\snip{subnet_oreachable_prem_6_pred}{0}{0}{\isa{F\ {\isacharparenleft}{\isasymsigma}\ i{\isacharparenright}\ {\isacharparenleft}{\isasymsigma}{\isacharprime}\ i{\isacharparenright}}}

\snip{subnet_oreachable_concl_1}{0}{0}{\isa{{\isacharparenleft}{\isasymsigma},\ s{\isacharparenright}{\isasymin}oreachable\ {\isacharparenleft}opnet\ onp\ p\isactrlsub {\isadigit{1}}{\isacharparenright}\ S\isactrlsub {\isadigit{1}}\ U\isactrlsub {\isadigit{1}}}}
\snip{subnet_oreachable_concl_2}{0}{0}{\isa{{\isacharparenleft}{\isasymsigma},\ t{\isacharparenright}{\isasymin}oreachable\ {\isacharparenleft}opnet\ onp\ p\isactrlsub {\isadigit{2}}{\isacharparenright}\ S\isactrlsub {\isadigit{2}}\ U\isactrlsub {\isadigit{2}}}}
\snip{subnet_oreachable_concl_3}{0}{0}{\isa{net{\isacharunderscore}tree{\isacharunderscore}ips\ p\isactrlsub {\isadigit{1}}\ {\isasyminter}\ net{\isacharunderscore}tree{\isacharunderscore}ips\ p\isactrlsub {\isadigit{2}}\ {\isacharequal}\ {\isasymemptyset}}}

\snip{S1}{0}{0}{\isa{S\isactrlsub {\isadigit{1}}}}
\snip{U1}{0}{0}{\isa{U\isactrlsub {\isadigit{1}}}}
\snip{S2}{0}{0}{\isa{S\isactrlsub {\isadigit{2}}}}
\snip{U2}{0}{0}{\isa{U\isactrlsub {\isadigit{2}}}}

\snip{subnet_oreachable_proof_step}{0}{0}{\isa{{\isacharparenleft}{\isacharparenleft}{\isasymsigma},\ \subnets{s}{t}{\isacharparenright},\ a,\ {\isacharparenleft}{\isasymsigma}{\isacharprime},\ \subnetsadj{s{\isacharprime}}{t{\isacharprime}}{\isacharparenright}{\isacharparenright}{\isasymin}trans\ {\isacharparenleft}opnet\ onp\ {\isacharparenleft}p\isactrlsub {\isadigit{1}}\parallelcomp{}p\isactrlsub {\isadigit{2}}{\isacharparenright}{\isacharparenright}}}
\snip{subnet_oreachable_proof_jL}{0}{0}{\isa{j\ {\isasymnotin}\ net{\isacharunderscore}tree{\isacharunderscore}ips\ {\isacharparenleft}p\isactrlsub {\isadigit{1}}\parallelcomp{}p\isactrlsub {\isadigit{2}}{\isacharparenright}}}
\snip{subnet_oreachable_proof_oarrivemsg}{0}{0}{\isa{oarrivemsg\ I\ {\isasymsigma}}}

\snip{closed_reachable_oreachable'_prem_2}{0}{0}{\isa{s{\isasymin}reachable\ {\isacharparenleft}closed\ {\isacharparenleft}pnet\ {\isacharparenleft}{\isasymlambda}i{\isachardot}\ ptoy\ i\ {\isasymlangle}{\isasymlangle}\ qmsg{\isacharparenright}\ any{\isacharparenright}{\isacharparenright}\ {\isacharparenleft}{\isasymlambda}{\isacharunderscore}{\isachardot}\ True{\isacharparenright}}}
\snip{closed_reachable_oreachable'_concl}{0}{0}{\isa{netgmap\ {\isacharparenleft}{\isasymlambda}{\isacharparenleft}p,\ q{\isacharparenright}{\isachardot}\ {\isacharparenleft}fst\ {\isacharparenleft}id\ p{\isacharparenright},\ snd\ {\isacharparenleft}id\ p{\isacharparenright},\ q{\isacharparenright}{\isacharparenright}\ s{\isasymin}netmask\ {\isacharparenleft}net{\isacharunderscore}tree{\isacharunderscore}ips\ any{\isacharparenright}\ {\isacharbackquote}\ oreachable\ {\isacharparenleft}oclosed\ {\isacharparenleft}opnet\ {\isacharparenleft}{\isasymlambda}i{\isachardot}\ optoy\ i\ {\isasymlangle}{\isasymlangle}\isactrlbsub i\isactrlesub \ qmsg{\isacharparenright}\ any{\isacharparenright}{\isacharparenright}\ {\isacharparenleft}{\isasymlambda}{\isacharunderscore}\ {\isacharunderscore}\ {\isacharunderscore}{\isachardot}\ True{\isacharparenright}\ U}}
\snip{pnet_reachable_transfer}{0}{0}{\isa{{\isasymlbrakk}openproc\ np\ onp\ sr{\isacharsemicolon}\ wf{\isacharunderscore}net{\isacharunderscore}tree\ n{\isacharsemicolon}\ s{\isasymin}reachable\ {\isacharparenleft}closed\ {\isacharparenleft}pnet\ np\ n{\isacharparenright}{\isacharparenright}\ TT{\isasymrbrakk}\ {\isasymLongrightarrow}\ {\isacharparenleft}default\ {\isacharparenleft}someinit\ np\ sr{\isacharparenright}\ {\isacharparenleft}netlift\ sr\ s{\isacharparenright},\ netliftl\ sr\ s{\isacharparenright}{\isasymin}oreachable\ {\isacharparenleft}oclosed\ {\isacharparenleft}opnet\ onp\ n{\isacharparenright}{\isacharparenright}\ {\isacharparenleft}{\isasymlambda}{\isacharunderscore}\ {\isacharunderscore}\ {\isacharunderscore}{\isachardot}\ True{\isacharparenright}\ U}}
\snip{pnet_reachable_transfer_prem_1}{0}{0}{\isa{openproc\ np\ onp\ sr}}
\snip{pnet_reachable_transfer_prem_2}{0}{0}{\isa{wf{\isacharunderscore}net{\isacharunderscore}tree\ n}}
\snip{pnet_reachable_transfer_prem_3}{0}{0}{\isa{s{\isasymin}reachable\ {\isacharparenleft}closed\ {\isacharparenleft}pnet\ np\ n{\isacharparenright}{\isacharparenright}\ {\isacharparenleft}{\isasymlambda}{\isacharunderscore}{\isachardot}\ True{\isacharparenright}}}
\snip{pnet_reachable_transfer_concl}{0}{0}{\isa{{\isacharparenleft}default\ {\isacharparenleft}someinit\ np\ sr{\isacharparenright}\ {\isacharparenleft}netlift\ sr\ s{\isacharparenright},\ netliftl\ sr\ s{\isacharparenright}\\\mbox{}\hfill{}{\isasymin}oreachable\ {\isacharparenleft}oclosed\ {\isacharparenleft}opnet\ onp\ n{\isacharparenright}{\isacharparenright}\ {\isacharparenleft}{\isasymlambda}{\isacharunderscore}\ {\isacharunderscore}\ {\isacharunderscore}{\isachardot}\ True{\isacharparenright}\ U}}

\snip{someinit}{0}{0}{\isa{openproc\ np\ onp\ sr\ {\isasymLongrightarrow}\ someinit\ np\ sr\ i\ {\isacharequal}\ SOME\ x{\isachardot}\ x{\isasymin}{\isacharparenleft}fst\ {\isasymcirc}\ sr{\isacharparenright}\ {\isacharbackquote}\ init\ {\isacharparenleft}np\ i{\isacharparenright}}}
\snip{someinit_concl}{0}{0}{\isa{someinit\ np\ sr\ i\ {\isacharequal}\ SOME\ x{\isachardot}\ x{\isasymin}{\isacharparenleft}fst\ {\isasymcirc}\ sr{\isacharparenright}\ {\isacharbackquote}\ init\ {\isacharparenleft}np\ i{\isacharparenright}}}
\snip{someinit_concl_lhs}{0}{0}{\isa{someinit\ np\ sr\ i}}
\snip{someinit_concl_rhs}{0}{0}{\isa{SOME\ x{\isachardot}\ x{\isasymin}{\isacharparenleft}fst\ {\isasymcirc}\ sr{\isacharparenright}\ {\isacharbackquote}\ init\ {\isacharparenleft}np\ i{\isacharparenright}}}

\snip{openproc_init}{0}{0}{\isa{openproc\ np\ onp\ sr\ {\isasymLongrightarrow}\ {\isacharbraceleft}uu{\isacharunderscore}\ {\isacharbar}\ {\isasymexists}{\isasymsigma}\ {\isasymzeta}\ s{\isachardot}\ {\isacharunderscore}\ {\isacharequal}\ {\isacharparenleft}{\isasymsigma},\ {\isasymzeta}{\isacharparenright}\ {\isasymand}\ s{\isasymin}init\ {\isacharparenleft}np\ i{\isacharparenright}\ {\isasymand}\ {\isacharparenleft}{\isasymsigma}\ i,\ {\isasymzeta}{\isacharparenright}\ {\isacharequal}\ sr\ s\ {\isasymand}\ {\isacharparenleft}{\isasymforall}j{\isachardot}\ j\ {\isasymnoteq}\ i\ {\isasymlongrightarrow}\ {\isasymsigma}\ j{\isasymin}{\isacharparenleft}fst\ {\isasymcirc}\ sr{\isacharparenright}\ {\isacharbackquote}\ init\ {\isacharparenleft}np\ j{\isacharparenright}{\isacharparenright}{\isacharbraceright}\ {\isasymsubseteq}\ init\ {\isacharparenleft}onp\ i{\isacharparenright}}}
\snip{openproc_init_notempty}{0}{0}{\isa{openproc\ np\ onp\ sr\ {\isasymLongrightarrow}\ {\isasymforall}j{\isachardot}\ init\ {\isacharparenleft}np\ j{\isacharparenright}\ {\isasymnoteq}\ {\isasymemptyset}}}
\snip{openproc_trans}{0}{0}{\isa{{\isasymlbrakk}openproc\ np\ onp\ sr{\isacharsemicolon}\ {\isasymsigma}\ i\ {\isacharequal}\ fst\ {\isacharparenleft}sr\ s{\isacharparenright}{\isacharsemicolon}\ {\isasymsigma}{\isacharprime}\ i\ {\isacharequal}\ fst\ {\isacharparenleft}sr\ s{\isacharprime}{\isacharparenright}{\isacharsemicolon}\ {\isacharparenleft}s,\ a,\ s{\isacharprime}{\isacharparenright}{\isasymin}trans\ {\isacharparenleft}np\ i{\isacharparenright}{\isasymrbrakk}\ {\isasymLongrightarrow}\ {\isacharparenleft}{\isacharparenleft}{\isasymsigma},\ snd\ {\isacharparenleft}sr\ s{\isacharparenright}{\isacharparenright},\ a,\ ({\isasymsigma}{\isacharprime},\ snd\ {\isacharparenleft}sr\ s{\isacharprime}{\isacharparenright}){\isacharparenright}{\isasymin}trans\ {\isacharparenleft}onp\ i{\isacharparenright}}}

\snip{openproc_trans_prem_1}{0}{0}{\isa{openproc\ np\ onp\ sr}}
\snip{openproc_trans_prem_2}{0}{0}{\isa{{\isasymsigma}\ i\ {\isacharequal}\ fst\ {\isacharparenleft}sr\ s{\isacharparenright}}}
\snip{openproc_trans_prem_3}{0}{0}{\isa{{\isasymsigma}{\isacharprime}\ i\ {\isacharequal}\ fst\ {\isacharparenleft}sr\ s{\isacharprime}{\isacharparenright}}}
\snip{openproc_trans_prem_4}{0}{0}{\isa{{\isacharparenleft}s,\ a,\ s{\isacharprime}{\isacharparenright}{\isasymin}trans\ {\isacharparenleft}np\ i{\isacharparenright}}}
\snip{openproc_trans_concl}{0}{0}{\isa{{\isacharparenleft}{\isacharparenleft}{\isasymsigma},\ snd\ {\isacharparenleft}sr\ s{\isacharparenright}{\isacharparenright},\ a,\ ({\isasymsigma}{\isacharprime},\ snd\ {\isacharparenleft}sr\ s{\isacharprime}{\isacharparenright}){\isacharparenright}{\isasymin}trans\ {\isacharparenleft}onp\ i{\isacharparenright}}}

\snip{openproc_np_type}{0}{0}{\isa{ip\ {\isasymRightarrow}\ {\isacharparenleft}{\isacharprime}s,\ {\isacharprime}m\ seq{\isacharunderscore}action{\isacharparenright}\ automaton}}
\snip{openproc_onp_type}{0}{0}{\isa{ip\ {\isasymRightarrow}\ {\isacharparenleft}{\isacharparenleft}ip\ {\isasymRightarrow}\ {\isacharprime}g{\isacharparenright}\ {\isasymtimes}\ {\isacharprime}l,\ {\isacharprime}m\ seq{\isacharunderscore}action{\isacharparenright}\ automaton}}
\snip{openproc_sr_type}{0}{0}{\isa{{\isacharprime}s\ {\isasymRightarrow}\ {\isacharprime}g\ {\isasymtimes}\ {\isacharprime}l}}
\snip{par_qmsg_oreachable_prem_1}{0}{0}{\isa{{\isacharparenleft}{\isasymsigma},\ (s,\ (q,\ t)){\isacharparenright}{\isasymin}oreachable\ {\isacharparenleft}A\ {\isasymlangle}{\isasymlangle}\isactrlbsub i\isactrlesub \ qmsg{\isacharparenright}\ S\ U}}
\snip{par_qmsg_oreachable_prem_2}{0}{0}{\isa{A\ \ostepinv\ {\isacharparenleft}S,\ U\ {\isasymrightarrow}{\isacharparenright}\ {\isacharparenleft}{\isasymlambda}{\isacharparenleft}{\isacharparenleft}{\isasymsigma},\ {\isacharunderscore}{\isacharparenright},\ {\isacharunderscore},\ ({\isasymsigma}{\isacharprime},\ {\isacharunderscore}){\isacharparenright}{\isachardot}\ F\ {\isacharparenleft}{\isasymsigma}\ i{\isacharparenright}\ {\isacharparenleft}{\isasymsigma}{\isacharprime}\ i{\isacharparenright}{\isacharparenright}}}
\snip{par_qmsg_oreachable_prem_3}{0}{0}{\isa{{\isasymAnd}{\isasymxi}{\isachardot}\ F\ {\isasymxi}\ {\isasymxi}}}
\snip{par_qmsg_oreachable_prem_4}{0}{0}{\isa{{\isasymAnd}{\isasymxi}\ {\isasymxi}{\isacharprime}{\isachardot}\ E\ {\isasymxi}\ {\isasymxi}{\isacharprime}\ {\isasymLongrightarrow}\ F\ {\isasymxi}\ {\isasymxi}{\isacharprime}}}
\snip{par_qmsg_oreachable_prem_5}{0}{0}{\isa{{\isasymAnd}{\isasymsigma}\ {\isasymsigma}{\isacharprime}\ m{\isachardot}\ {\isasymlbrakk}{\isasymforall}j{\isachardot}\ F\ {\isacharparenleft}{\isasymsigma}\ j{\isacharparenright}\ {\isacharparenleft}{\isasymsigma}{\isacharprime}\ j{\isacharparenright}{\isacharsemicolon}\ R\ {\isasymsigma}\ m{\isasymrbrakk}\ {\isasymLongrightarrow}\ R\ {\isasymsigma}{\isacharprime}\ m}}

\snip{par_qmsg_oreachable_L}{0}{0}{\isa{S\ {\isacharequal}\ otherwith\ E\ {\isacharbraceleft}i{\isacharbraceright}\ {\isacharparenleft}orecvmsg\ R{\isacharparenright}}}
\snip{par_qmsg_oreachable_E}{0}{0}{\isa{U\ {\isacharequal}\ other\ F\ {\isacharbraceleft}i{\isacharbraceright}}}

\snip{par_qmsg_oreachable_conj1}{0}{0}{\isa{{\isacharparenleft}{\isasymsigma},\ s{\isacharparenright}{\isasymin}oreachable\ A\ S\ U}}
\snip{par_qmsg_oreachable_conj2}{0}{0}{\isa{{\isacharparenleft}q,\ t{\isacharparenright}{\isasymin}reachable\ qmsg\ {\isacharparenleft}recvmsg\ {\isacharparenleft}R\ {\isasymsigma}{\isacharparenright}{\isacharparenright}}}
\snip{par_qmsg_oreachable_conj3}{0}{0}{\isa{{\isasymforall}m{\isasymin}set\ q{\isachardot}\ R\ {\isasymsigma}\ m}}

\snip{par_qmsg_oreachable_Exixi'}{0}{0}{\isa{E\ {\isasymxi}\ {\isasymxi}{\isacharprime}}}
\snip{par_qmsg_oreachable_Fxixi'}{0}{0}{\isa{F\ {\isasymxi}\ {\isasymxi}{\isacharprime}}}
\snip{par_qmsg_oreachable_Fsigmaj}{0}{0}{\isa{{\isasymforall}j{\isachardot}\ F\ {\isacharparenleft}{\isasymsigma}\ j{\isacharparenright}\ {\isacharparenleft}{\isasymsigma}{\isacharprime}\ j{\isacharparenright}}}
\snip{par_qmsg_oreachable_Rsigmam}{0}{0}{\isa{R\ {\isasymsigma}\ m}}
\snip{par_qmsg_oreachable_Rsigma'm}{0}{0}{\isa{R\ {\isasymsigma}{\isacharprime}\ m}}
\snip{closed_reachable_oreachable}{0}{0}{\isa{\mbox{}\inferrule{\mbox{wf{\isacharunderscore}net{\isacharunderscore}tree\ any}\\\ \mbox{s{\isasymin}reachable\ {\isacharparenleft}closed\ {\isacharparenleft}pnet\ {\isacharparenleft}{\isasymlambda}i{\isachardot}\ ptoy\ i\ {\isasymlangle}{\isasymlangle}\ qmsg{\isacharparenright}\ any{\isacharparenright}{\isacharparenright}\ TT}}{\mbox{initmissing\ {\isacharparenleft}netgmap\ {\isacharparenleft}{\isasymlambda}{\isacharparenleft}p,\ q{\isacharparenright}{\isachardot}\ {\isacharparenleft}fst\ {\isacharparenleft}id\ p{\isacharparenright},\ snd\ {\isacharparenleft}id\ p{\isacharparenright},\ q{\isacharparenright}{\isacharparenright}\ s{\isacharparenright}{\isasymin}oreachable\ {\isacharparenleft}oclosed\ {\isacharparenleft}opnet\ {\isacharparenleft}{\isasymlambda}i{\isachardot}\ optoy\ i\ {\isasymlangle}{\isasymlangle}\isactrlbsub i\isactrlesub \ qmsg{\isacharparenright}\ any{\isacharparenright}{\isacharparenright}\ {\isacharparenleft}{\isasymlambda}{\isacharunderscore}\ {\isacharunderscore}\ {\isacharunderscore}{\isachardot}\ True{\isacharparenright}\ U}}}}

\snip{close_opnet}{0}{0}{\isa{\mbox{}\inferrule{\mbox{wf{\isacharunderscore}net{\isacharunderscore}tree\ n}\\\ \mbox{oclosed\ {\isacharparenleft}opnet\ optoy\ n{\isacharparenright}\ {\isasymTurnstile}\ {\isacharparenleft}{\isasymlambda}{\isacharunderscore}\ {\isacharunderscore}\ {\isacharunderscore}{\isachardot}\ True,\ U\ {\isasymrightarrow}{\isacharparenright}\ global\ P}}{\mbox{closed\ {\isacharparenleft}pnet\ ptoy\ n{\isacharparenright}\ {\isasymTTurnstile}\ openproc{\isachardot}netglobal\ ptoy\ id\ P}}}}
\snip{close_opnet_prem_1}{0}{0}{\isa{wf{\isacharunderscore}net{\isacharunderscore}tree\ n}}
\snip{close_opnet_prem_2}{0}{0}{\isa{oclosed\ {\isacharparenleft}opnet\ optoy\ n{\isacharparenright}\ {\isasymTurnstile}\ {\isacharparenleft}{\isasymlambda}{\isacharunderscore}\ {\isacharunderscore}\ {\isacharunderscore}{\isachardot}\ True,\ U\ {\isasymrightarrow}{\isacharparenright}\ {\isacharparenleft}{\isasymlambda}{\isacharparenleft}{\isasymsigma},\ {\isacharunderscore}{\isacharparenright}{\isachardot}\ P\ {\isasymsigma}{\isacharparenright}}}
\snip{close_opnet_concl}{0}{0}{\isa{closed\ {\isacharparenleft}pnet\ ptoy\ n{\isacharparenright}\ {\isasymTTurnstile}\ openproc{\isachardot}netglobal\ ptoy\ id\ P}}

\snip{nos_increase}{0}{0}{\isa{nos{\isacharunderscore}inc\ {\isasymxi}\ {\isasymxi}{\isacharprime}\ {\isacharequal}\ no\ {\isasymxi}\ {\isasymle}\ no\ {\isasymxi}{\isacharprime}}}
\snip{nos_increase_lhs}{0}{0}{\isa{nos{\isacharunderscore}inc\ {\isasymxi}\ {\isasymxi}{\isacharprime}}}
\snip{nos_increase_rhs}{0}{0}{\isa{no\ {\isasymxi}\ {\isasymle}\ no\ {\isasymxi}{\isacharprime}}}

\snip{msg_num_ok}{0}{0}{\isa{msg{\isacharunderscore}ok\ {\isasymsigma}\ m\ {\isacharequal}\ \textsf{case}\ m\ \textsf{of}\ Pkt\ num{\isacharprime}\ sip{\isacharprime}\ {\isasymRightarrow}\ num{\isacharprime}\ {\isasymle}\ no\ {\isacharparenleft}{\isasymsigma}\ sip{\isacharprime}{\isacharparenright}\ {\isacharbar}\ Newpkt\ nat{\isadigit{1}}\ nat{\isadigit{2}}\ {\isasymRightarrow}\ True}}
\snip{msg_num_ok_pkt}{0}{0}{\isa{msg{\isacharunderscore}ok\ {\isasymsigma}\ {\isacharparenleft}pkt\ {\isacharparenleft}data,\ src{\isacharparenright}{\isacharparenright}\ {\isacharequal}\ {\isacharparenleft}data\ {\isasymle}\ no\ {\isacharparenleft}{\isasymsigma}\ src{\isacharparenright}{\isacharparenright}}}
\snip{msg_num_ok_newpkt}{0}{0}{\isa{msg{\isacharunderscore}ok\ {\isasymsigma}\ {\isacharparenleft}newpkt\ {\isacharparenleft}data,\ dst{\isacharparenright}{\isacharparenright}}}
\snip{msg_num_ok_newpkt2}{0}{0}{\isa{msg{\isacharunderscore}ok\ {\isasymsigma}\ {\isacharparenleft}newpkt\ {\isacharparenleft}data,\ dst{\isacharparenright}{\isacharparenright}\ {\isacharequal}\ True}}

\snip{oseq_msg_num_ok}{0}{0}{\isa{optoy\ i\ \ostepinv\ {\isacharparenleft}act\ TT,\ other\ U\ {\isacharbraceleft}i{\isacharbraceright}\ {\isasymrightarrow}{\isacharparenright}\ globala\ {\isacharparenleft}{\isasymlambda}{\isacharparenleft}{\isasymsigma},\ a,\ {\isacharunderscore}{\isacharparenright}{\isachardot}\ anycast\ {\isacharparenleft}msg{\isacharunderscore}ok\ {\isasymsigma}{\isacharparenright}\ a{\isacharparenright}}}
\snip{seq_nos_increases}{0}{0}{\isa{ptoy\ i\ \stepinv\ onll\ \gammatoy\ {\isacharparenleft}{\isasymlambda}{\isacharparenleft}{\isacharparenleft}{\isasymxi},\ {\isacharunderscore}{\isacharparenright},\ {\isacharunderscore},\ {\isacharparenleft}{\isasymxi}{\isacharprime},\ {\isacharunderscore}{\isacharparenright}{\isacharparenright}{\isachardot}\ no\ {\isasymxi}\ {\isasymle}\ no\ {\isasymxi}{\isacharprime}{\isacharparenright}}}
\snip{seq_nos_increases'}{0}{0}{\isa{ptoy\ i\ \stepinv\ {\isacharparenleft}{\isasymlambda}{\isacharparenleft}{\isacharparenleft}{\isasymxi},\ {\isacharunderscore}{\isacharparenright},\ {\isacharunderscore},\ {\isacharparenleft}{\isasymxi}{\isacharprime},\ {\isacharunderscore}{\isacharparenright}{\isacharparenright}{\isachardot}\ no\ {\isasymxi}\ {\isasymle}\ no\ {\isasymxi}{\isacharprime}{\isacharparenright}}}
\snip{nhip_eq_ip}{0}{0}{\isa{ptoy\ i\ {\isasymTTurnstile}\ onl\ \gammatoy\ {\isacharparenleft}{\isasymlambda}{\isacharparenleft}{\isasymxi},\ l{\isacharparenright}{\isachardot}\ l{\isasymin}{\isacharbraceleft}PToy{\isacharminus}{\isacharcolon}{\isadigit{2}}{\isachardot}{\isachardot}PToy{\isacharminus}{\isacharcolon}{\isadigit{8}}{\isacharbraceright}\ {\isasymlongrightarrow}\ nhip\ {\isasymxi}\ {\isacharequal}\ ip\ {\isasymxi}{\isacharparenright}}}

\snip{net_global}{0}{0}{\isa{netglobal\ P\ {\isacharequal}\ {\isasymlambda}s{\isachardot}\ P\ {\isacharparenleft}default\ toy{\isacharunderscore}init\ {\isacharparenleft}netlift\ fst\ s{\isacharparenright}{\isacharparenright}}}
\snip{net_global_lhs}{0}{0}{\isa{netglobal\ P}}
\snip{net_global_rhs}{0}{0}{\isa{{\isasymlambda}s{\isachardot}\ P\ {\isacharparenleft}default\ toy{\isacharunderscore}init\ {\isacharparenleft}netlift\ fst\ s{\isacharparenright}{\isacharparenright}}}

\snip{oseq_bigger_than_next}{0}{0}{\isa{optoy\ i\ {\isasymTurnstile}\ {\isacharparenleft}otherwith\ nos{\isacharunderscore}inc\ {\isacharbraceleft}i{\isacharbraceright}\ {\isacharparenleft}orecvmsg\ msg{\isacharunderscore}ok{\isacharparenright},\ other\ nos{\isacharunderscore}inc\ {\isacharbraceleft}i{\isacharbraceright}\ {\isasymrightarrow}{\isacharparenright}\ }\\\isa{\phantom{optoy\ i\ {\isasymTurnstile}\ }{\isacharparenleft}{\isasymlambda}{\isacharparenleft}{\isasymsigma},\ {\isacharunderscore}{\isacharparenright}{\isachardot}\ no\ {\isacharparenleft}{\isasymsigma}\ i{\isacharparenright}\ {\isasymle}\ no\ {\isacharparenleft}{\isasymsigma}\ {\isacharparenleft}nhip\ {\isacharparenleft}{\isasymsigma}\ i{\isacharparenright}{\isacharparenright}{\isacharparenright}{\isacharparenright}}}
\snip{pnet_bigger_than_next}{0}{0}{\isa{opnet\ {\isacharparenleft}{\isasymlambda}i{\isachardot}\ optoy\ i\ {\isasymlangle}{\isasymlangle}\isactrlbsub i\isactrlesub \ qmsg{\isacharparenright}\ n\ {\isasymTurnstile}\ {\isacharparenleft}otherwith\ nos{\isacharunderscore}inc\ {\isacharparenleft}net{\isacharunderscore}tree{\isacharunderscore}ips\ n{\isacharparenright}\ {\isacharparenleft}oarrivemsg\ msg{\isacharunderscore}ok{\isacharparenright},\ other\ nos{\isacharunderscore}inc\ {\isacharparenleft}net{\isacharunderscore}tree{\isacharunderscore}ips\ n{\isacharparenright}\ {\isasymrightarrow}{\isacharparenright}\ global\ {\isacharparenleft}{\isasymlambda}{\isasymsigma}{\isachardot}\ {\isasymforall}i{\isasymin}net{\isacharunderscore}tree{\isacharunderscore}ips\ n{\isachardot}\ no\ {\isacharparenleft}{\isasymsigma}\ i{\isacharparenright}\ {\isasymle}\ no\ {\isacharparenleft}{\isasymsigma}\ {\isacharparenleft}nhip\ {\isacharparenleft}{\isasymsigma}\ i{\isacharparenright}{\isacharparenright}{\isacharparenright}{\isacharparenright}}}
\snip{bigger_than_next}{0}{0}{\mbox{\isa{wf{\isacharunderscore}net{\isacharunderscore}tree\ n\ {\isasymLongrightarrow}\ closed\ {\isacharparenleft}pnet\ {\isacharparenleft}{\isasymlambda}i{\isachardot}\ ptoy\ i\ {\isasymlangle}{\isasymlangle}\ qmsg{\isacharparenright}\ n{\isacharparenright}\ {\isasymTTurnstile}\ }}\\\mbox{\isa{\phantom{wf{\isacharunderscore}net{\isacharunderscore}tree\ n\ }netglobal\ {\isacharparenleft}{\isasymlambda}{\isasymsigma}{\isachardot}\ {\isasymforall}i{\isachardot}\ no\ {\isacharparenleft}{\isasymsigma}\ i{\isacharparenright}\ {\isasymle}\ no\ {\isacharparenleft}{\isasymsigma}\ {\isacharparenleft}nhip\ {\isacharparenleft}{\isasymsigma}\ i{\isacharparenright}{\isacharparenright}{\isacharparenright}{\isacharparenright}}}}
\snip{bigger_than_next_prem}{0}{0}{\isa{wf{\isacharunderscore}net{\isacharunderscore}tree\ n}}
\snip{bigger_than_next_concl}{0}{0}{\isa{closed\ {\isacharparenleft}pnet\ {\isacharparenleft}{\isasymlambda}i{\isachardot}\ ptoy\ i\ {\isasymlangle}{\isasymlangle}\ qmsg{\isacharparenright}\ n{\isacharparenright}\ {\isasymTTurnstile}\ netglobal\ {\isacharparenleft}{\isasymlambda}{\isasymsigma}{\isachardot}\ {\isasymforall}i{\isachardot}\ no\ {\isacharparenleft}{\isasymsigma}\ i{\isacharparenright}\ {\isasymle}\ no\ {\isacharparenleft}{\isasymsigma}\ {\isacharparenleft}nhip\ {\isacharparenleft}{\isasymsigma}\ i{\isacharparenright}{\isacharparenright}{\isacharparenright}{\isacharparenright}}}
\usepackage[inline]{enumitem}
\newlist{inparaenum}{enumerate*}{1}
\setlist[inparaenum,1]{label=(\arabic*),ref=\arabic*}

\urldef{\mailsa}\path|Timothy.Bourke@inria.fr|
\urldef{\mailsb}\path|{Robert.vanGlabbeek, Peter.Hoefner}@nicta.com.au|

\newenvironment{tightcenter}{%
  \setlength\topsep{4.0pt plus 0.5pt minus 1.0pt}
  \setlength\parskip{0pt}
  \begin{center}
}{%
  \end{center}
}

\newcommand{\refsec}[1]{\hyperref[sec:#1]{Section~\ref*{sec:#1}}}
\newcommand{\refsecs}[2]{\hyperref[sec:#1]{Sections~\ref*{sec:#1}}
                         \hyperref[sec:#2]{and~\ref*{sec:#2}}}
\newcommand{\reffig}[1]{\hyperref[fig:#1]{Figure~\ref*{fig:#1}}}
\newcommand{\reffigs}[2]{\hyperref[fig:#1]{Figures~\ref*{fig:#1}}
                         \hyperref[fig:#2]{and~\ref*{fig:#2}}}
\newcommand{\reffigss}[3]{\hyperref[fig:#1]{Figures~\ref*{fig:#1}},
                          \hyperref[fig:#2]{\ref*{fig:#2}},
                          \hyperref[fig:#3]{and~\ref*{fig:#3}}}
\newcommand{\reffigsss}[4]{\hyperref[fig:#1]{Figures~\ref*{fig:#1}},
                           \hyperref[fig:#2]{\ref*{fig:#2}},
                           \hyperref[fig:#3]{\ref*{fig:#3}},
                           \hyperref[fig:#4]{and~\ref*{fig:#4}}}
\newcommand{\refthm}[1]{\hyperref[thm:#1]{Theorem~\ref*{thm:#1}}}
\newcommand{\refthms}[2]{\hyperref[thm:#1]{Theorems~\ref*{thm:#1}}
                         \hyperref[thm:#2]{and~\ref*{thm:#2}}}
\newcommand{\reflem}[1]{\hyperref[thm:#1]{Lemma~\ref*{thm:#1}}}
\newcommand{\reflems}[2]{\hyperref[thm:#1]{Lemmas~\ref*{thm:#1}}
                         \hyperref[thm:#2]{and~\ref*{thm:#2}}}
\newcommand{\refdef}[1]{\hyperref[def:#1]{Definition~\ref*{def:#1}}}
\newcommand{\refdefs}[2]{\hyperref[def:#1]{Definitions~\ref*{def:#1}}
                         \hyperref[def:#1]{and~\ref*{def:#2}}}
\newcommand{\refdefr}[2]{\hyperref[def:#1]{Definitions~\ref*{def:#1}}%
                         \hyperref[def:#1]{--\ref*{def:#2}}}
\newcommand{\refeq}[1]{\hyperref[eq:#1]{(\ref*{eq:#1})}}
\newcommand{\refeqs}[2]{\hyperref[eq:#1]{(\ref*{eq:#1}}\hyperref[eq:#1]{, 
\ref*{eq:#2})}}
\newcommand{\refstep}[1]{\hyperref[step:#1]{Step~\ref*{step:#1}}}

\newcommand{\refitem}[1]{\hyperref[item:#1]{Item~\ref*{item:#1}}}
\newcommand{\refitemr}[2]{\hyperref[item:#1]{Items~\ref*{item:#1}}
                          \hyperref[item:#2]{to~\ref*{item:#2}}}
\newcommand{\refitems}[2]{\hyperref[item:#1]{Items~\ref*{item:#1}}
                          \hyperref[item:#2]{and~\ref*{item:#2}}}

\newcommand{\puncgap}{\hskip.1em}

\newcommand{\seqpsos}{\isa{seqp{\isacharunderscore}sos}}
\newcommand{\parpsos}{\isa{parp{\isacharunderscore}sos}}
\newcommand{\nodesos}{\isa{node{\isacharunderscore}sos}}
\newcommand{\pnetsos}{\isa{pnet{\isacharunderscore}sos}}
\newcommand{\cnetsos}{\isa{cnet{\isacharunderscore}sos}}
\newcommand{\oseqpsos}{\isa{oseqp{\isacharunderscore}sos}}
\newcommand{\oparpsos}{\isa{oparp{\isacharunderscore}sos}}
\newcommand{\onodesos}{\isa{onode{\isacharunderscore}sos}}
\newcommand{\opnetsos}{\isa{opnet{\isacharunderscore}sos}}
\newcommand{\ocnetsos}{\isa{ocnet{\isacharunderscore}sos}}
\newcommand{\Ri}{R\isactrlsub i}

\newcommand{\nodes}[3]{%
    $\mbox{#2}^{\mbox{\,\isastylescript{#1}}}_{\mbox{\isastylescript{#3}}}$}
\newcommand{\subnets}[2]{{#1}{\isamath{\,\shortparallel\,}}{#2}}
\newcommand{\subnetsadj}[2]{{#1}{\isamath{\hspace{.35pt}\shortparallel\,}}{#2}}
\newcommand{\subnetsit}[2]{{#1}{\isamath{\hspace{.30em}\shortparallel\,}}{#2}}

\def\NodeS\ #1\ #2\ #3{\nodes{#1}{#2}{#3}}
\renewcommand{\isasymin}{\isamath{\,\in\,}}

\newcommand{\listconcat}{{\scriptsize\ensuremath{+\!+}}}
\newcommand{\parallelcomp}{{\scriptsize\raisebox{.19ex}{\ensuremath{\,\parallel\,}}}}
\newcommand{\gammatoy}{{{\isasymGamma}\isactrlsub {\mbox{\sf\scriptsize Toy}}}}
\newcommand{\gammaqmsg}{{{\isasymGamma}\isactrlsub {\textsc{qmsg}}}}
\newcommand{\sigmatoy}{{{\isasymsigma}\isactrlsub {\textsc{toy}}}}

\newcommand{\selmsg}{{s}\isactrlsub {{msg}}}
\newcommand{\updmsg}{{u}\isactrlsub {{msg}}}
\newcommand{\selips}{{s}\isactrlsub {{ips}}}
\newcommand{\selip}{{s}\isactrlsub {{ip}}}
\newcommand{\seldata}{{s}\isactrlsub {{data}}}

\newcommand{\microstep}[1]{\isa{{\isasymleadsto}\isactrlbsub{#1}\isactrlesub}}

\newcommand{\microsteprtcl}[1]{\isa{{\isasymleadsto}\isactrlbsub{#1}\isactrlesub}\isactrlisup{\isacharasterisk}}

\newcommand{\p}[1]{\isa{p\isactrlisub{\isadigit{#1}}}}

\newcommand{\stepinv}{\isa{\isamath{\mid\!\mid\hspace{-2pt}\equiv}}}
\newcommand{\ostepinv}{\isa{\isamath{\mid\hspace{-2pt}\equiv}}}

\acrodef{AODV}{Ad hoc On-demand Distance Vector}
\acrodef{AWN}{Algebra for Wireless Networks}
\acrodef{MANET}{Mobile Ad hoc Network}
\acrodef{ITP}{Interactive Theorem Prover}
\acrodef{HOL}{Higher Order Logic}
\acrodef{WMN}{Wireless Mesh Network}
\acrodef{SOS}{Structural Operational Semantics}
\begin{document}
% Title page and abstract {-{1

\mainmatter
\title{Showing invariance compositionally for a process algebra for network protocols}

\author{Timothy Bourke\inst{1,2}%
\and Robert J.\ van Glabbeek\inst{3,4}
\and Peter H\"ofner\inst{3,4}}
\titlerunning{Showing invariance compositionally for a process algebra}
\authorrunning{Bourke, van Glabbeek, and H\"ofner}

\institute{%
Inria Paris-Rocquencourt\\
\and
Ecole normale supérieure, Paris, France\\
\and
NICTA,
Sydney, Australia\\
\and
Computer Science and Engineering, UNSW, Sydney, Australia}

\maketitle
\setcounter{footnote}{0}
\begin{abstract}
This paper presents the mechanization of a process algebra for Mobile Ad hoc 
Networks and Wireless Mesh Networks, and the development of a compositional 
framework for proving invariant properties.
Mechanizing the core process algebra in Isabelle/HOL is relatively standard, 
but its layered structure necessitates special treatment.
The control states of reactive processes, such as nodes in a network, are 
modelled by terms of the process algebra.
We propose a technique based on these terms to streamline proofs of 
inductive invariance.
This is not sufficient, however, to state and prove invariants that relate 
states across multiple processes (entire networks).
To this end, we propose a novel compositional technique for lifting global 
invariants stated at the level of individual nodes to networks of nodes.
\end{abstract}

%--   }-}1%%%%%%%%%%%%%%%%%%%%%%%%%%%%%%%%%%%%%%%%%%%%%%%%%%%%%%%%%%%%
\section{Introduction and related work}\label{sec:intro}%{-{1

The \ac{AWN} is a process algebra developed for modelling and analysing 
protocols for \acp{MANET} and \acp{WMN}~\cite[\textsection 
4]{FehnkerEtAl:AWN:2013}.
This paper reports on both its mechanization in 
Isabelle/HOL~\cite{NipkowPauWen:IsabelleTut:2002} and the development of a 
compositional framework for showing invariant properties of models.%
\footnote{The Isabelle/HOL source files can be found in the Archive of 
Formal Proofs~\cite{Bourke14}.}
The techniques we describe are a response to problems encountered during the 
mechanization of a model and proof---presented elsewhere~\cite{LICS14}---of 
an RFC-standard for routing protocols.
Despite the existence of extensive research on related 
problems~\cite{deRoeverEtAl:ConcVer:2001} and several mechanized frameworks 
for reactive systems~\cite{HeydCre:Unity:1996, ChaudhuriEtAl:TLA+:2010, 
Muller:PhD:1998}, we are not aware of other solutions that allow the 
compositional statement and proof of properties relating the states of 
different nodes in a message-passing model---at least not within the 
strictures imposed by an \ac{ITP}.

But is there really any need for yet another process algebra and associated 
framework?
\ac{AWN} provides a unique mix of communication primitives and a treatment 
of data structures that are essential for studying \ac{MANET} and \ac{WMN} 
protocols with dynamic topologies and sophisticated routing 
logic~\cite[\textsection 
1]{FehnkerEtAl:AWN:2013}. 
It supports communication primitives for one-to-one (\isa{unicast}), 
one-to-many (\isa{groupcast}), and one-to-all (\isa{broadcast}) message 
passing.
\ac{AWN} comprises distinct layers for expressing the structure of nodes and 
networks.
We exploit this structure, but we also expect the techniques proposed in 
\refsecs{proof-base}{proof-comp} to apply to similar layered modelling 
languages.
Besides this, our work differs from other mechanizations for verifying 
reactive systems, like UNITY~\cite{HeydCre:Unity:1996}, 
TLA\textsuperscript{+}~\cite{ChaudhuriEtAl:TLA+:2010}, or I/O 
Automata~\cite{Muller:PhD:1998} (from which we drew the most inspiration), 
in its explicit treatment of control states, in the form of process algebra 
terms, as distinct from data states.
In this respect, our approach is close to that of 
Isabelle/Circus~\cite{FeliachiGauWol:Circus:2012}, but it differs in
\begin{inparaenum}
\item
the treatment of operators for composing nodes, which we model directly as 
functions on automata,
\item
the treatment of recursive invocations, which we do not permit, and
\item
our inclusion of a framework for compositional proofs.
\end{inparaenum}
Other work in \acp{ITP} focuses on showing 
traditional properties of process algebras, like, for instance, the 
treatment of binders~\cite{BengtsonParrow09}, that bisimulation equivalence 
is a congruence~\cite{GothelGle:TimedCSP:2010, Hirschkoff:picalc:1997}, or 
properties of fix-point induction~\cite{TejWolff97}, while we focus on what 
has been termed `proof methodology'~\cite{FokkinkGroRen:ProcAlgProof:2004},
and develop a compositional method for showing correctness properties of 
protocols specified in a process algebra.
Alternatively, Paulson's inductive approach~\cite{Paulson:Inductive:1998} 
can be applied to show properties of protocols
specified with less generic infrastructure.
But \nolinebreak we \nolinebreak think it to be better suited to systems 
specified in a `declarative' style as opposed to the strongly operational 
models we consider.

\emph{Structure and contributions.}
\refsec{awn} describes the mechanization of \ac{AWN}.
The basic definitions are routine but the layered structure of the language 
and the treatment of operators on networks as functions on automata are 
relatively novel and essential to understanding later sections.
\refsec{proof-base} describes our mechanization of the theory of inductive 
invariants, closely following~\cite{MannaPnu:Safety}.
We exploit the structure of \ac{AWN} to generate verification conditions 
corresponding to those of pen-and-paper proofs~\cite[\textsection 
7]{FehnkerEtAl:AWN:2013}.
\refsec{proof-comp} presents a compositional technique for stating and 
proving invariants that relate states across multiple nodes.
Basically, we substitute `open' \ac{SOS} rules over the global state for the 
standard rules over local states (\refsec{omodel}), show the property over a 
single sequential process (\refsec{oinv}), `lift' it successively over 
layers that model message queueing and network communication
(\refsec{lift}), and, ultimately, `transfer' it to the original model 
(\refsec{transfer}).

%--   }-}1%%%%%%%%%%%%%%%%%%%%%%%%%%%%%%%%%%%%%%%%%%%%%%%%%%%%%%%%%%%%
\section{The process algebra \acs{AWN}} \label{sec:awn} %{-{1

\ac{AWN} comprises five layers~\cite[\textsection 4]{FehnkerEtAl:AWN:2013}.
We treat each layer as an automaton with states of a specific form and a 
given set of transition rules.
We describe the layers from the bottom up over the following sections.

\subsection{Sequential processes}\label{sec:awn:seq} %{-{2
Sequential processes are used to encode protocol logic.
Each is modelled by a \emph{(recursive) specification} \isa{\isasymGamma} of 
type \isa{'p {\isasymRightarrow} ('s, 'p, 'l) seqp}, which maps process 
names of type \isa{'p} to terms of type \isa{('s, 'p, 'l) seqp}, also
parameterized by \isa{'s}, data states, and \isa{'l}, labels.
States of sequential processes have the form \isa{(\isasymxi, p)} where 
\isa{\isasymxi} is a data state of type \isa{'s} and \isa{p} is a control 
term of type \isa{('s, 'p, 'l) seqp}.\pagebreak[1]

\begin{figure}[t]% {-{3
\subfloat[Term constructors for {\sf ('s, 'p, 'l) seqp}.\label{fig:seqp:terms}]{\tiny
\centering
  \begin{tabular}{@{}l@{\ }l@{}}
    \snippet{lseqp_assign} &
        \isa{'l {\isasymRightarrow}
             ('s {\isasymRightarrow} 's) {\isasymRightarrow}
             ('s, 'p, 'l) seqp {\isasymRightarrow}
             ('s, 'p, 'l) seqp}
    \\[.5em]
    \snippet{lseqp_guard} &
        \isa{'l {\isasymRightarrow}
             ('s {\isasymRightarrow} 's set) {\isasymRightarrow}
             ('s, 'p, 'l) seqp {\isasymRightarrow}
             ('s, 'p, 'l) seqp}
    \\[.5em]
    \snippet{lseqp_ucast}\hspace{1em} &
        \isa{'l {\isasymRightarrow}
             ('s {\isasymRightarrow} ip) {\isasymRightarrow}
             ('s {\isasymRightarrow} msg) {\isasymRightarrow}
             ('s, 'p, 'l) seqp {\isasymRightarrow}
           }\\[.2em] &\isa{
             ('s, 'p, 'l) seqp {\isasymRightarrow}
             ('s, 'p, 'l) seqp}
    \\[.5em]
    \snippet{lseqp_bcast} &
        \isa{'l {\isasymRightarrow}
             ('s {\isasymRightarrow} msg) {\isasymRightarrow}
             ('s, 'p, 'l) seqp {\isasymRightarrow}
             ('s, 'p, 'l) seqp}
    \\[.5em]
    \snippet{lseqp_gcast} &
        \isa{'l {\isasymRightarrow}
             ('s {\isasymRightarrow} ip set) {\isasymRightarrow}
             ('s {\isasymRightarrow} msg) {\isasymRightarrow}
             ('s, 'p, 'l) seqp {\isasymRightarrow}
          }\\[.2em] &\isa{
             ('s, 'p, 'l) seqp}
    \\[.5em]

    \snippet{lseqp_send} &
        \isa{'l {\isasymRightarrow}
             ('s {\isasymRightarrow} msg) {\isasymRightarrow}
             ('s, 'p, 'l) seqp {\isasymRightarrow}
             ('s, 'p, 'l) seqp}
    \\[.5em]
    \snippet{lseqp_receive} &
        \isa{'l {\isasymRightarrow}
             (msg {\isasymRightarrow} \!'s {\isasymRightarrow} \!'s) 
             {\isasymRightarrow}
             ('s, 'p, 'l) seqp {\isasymRightarrow}
             ('s, 'p, 'l) seqp}
    \\[.5em]
    \snippet{lseqp_deliver} &
        \isa{'l {\isasymRightarrow}
             ('s {\isasymRightarrow} data) {\isasymRightarrow}
             ('s, 'p, 'l) seqp {\isasymRightarrow}
             ('s, 'p, 'l) seqp}
    \\[.5em]

    \snippet{seqp_choice} &
        \isa{('s, 'p, 'l) seqp {\isasymRightarrow}
             ('s, 'p, 'l) seqp {\isasymRightarrow}
             ('s, 'p, 'l) seqp}
    \\[.5em]
    \snippet{seqp_call} &
        \isa{'p {\isasymRightarrow}
             ('s, 'p, 'l) seqp}
  \end{tabular}} %}-}3
\\
\subfloat[\ac{SOS} rules for sequential processes: examples from \seqpsos.\label{fig:seqp:sos}]{ \tiny%{-{3
\!\!\!\begin{mathpar}
        \msnippet{assignT'} \and
        \msnippet{choiceT1} \and
        \msnippet{callT} \and
        \msnippet{choiceT2} \and
        \msnippet{unicastT} \and
        \msnippet{notunicastT}
    \end{mathpar}\rule[-1em]{0mm}{20mm}\!\!\!
    } %}-}3
\caption{Sequential processes: terms and semantics\label{fig:seqp}}
\end{figure}

Process terms are built from the constructors that are shown with their 
types\footnote{Leading abstractions are omitted, for example, 
\isa{\isasymlambda l\ fa\ p.\ }\snippet{lseqp_assign} is written 
\snippet{lseqp_assign}.} in \reffig{seqp:terms}.
The inductive set \seqpsos, shown partially in \reffig{seqp:sos}, contains 
one or two \ac{SOS} rules for each constructor.
It is parameterized by a specification \isa{\isasymGamma} and relates 
triples of source states, actions, and destination states.

The `prefix' constructors are each labelled with an 
\isa{\gray{{\isacharbraceleft}l{\isacharbraceright}}}.
Labels are used to strengthen invariants when a property is only true 
in or between certain states; they have no influence on control flow (unlike 
in~\cite{MannaPnu:Safety}).
The prefix constructors are \emph{assignment}, \snippet{lseqp_assign}, which 
transforms the data state deterministically according to the function 
\isa{u} and performs a \isa{\isasymtau} action, as shown in 
\reffig{seqp:sos};
\emph{guard/bind}, \snippet{lseqp_guard}, with which we encode both guards, 
\snippet{stmt_guard}, and variable bindings, as in
\snippet{stmt_bind};\footnote{Although it strictly subsumes assignment we 
prefer to keep both.}
\emph{network synchronizations}, 
\isa{receive}/\isa{unicast}/\isa{broadcast}/\isa{groupcast}, of which the 
rules for \isa{unicast} are characteristic and shown in 
\reffig{seqp:sos}---the environment decides between a successful 
\snippet{act_unicast} and an unsuccessful \snippet{act_not_unicast}; and,
\emph{internal communications}, \isa{send}/\isa{receive}/\isa{deliver}.

The other constructors are unlabelled and serve to `glue' processes 
together: \emph{choice}, \snippet{seqp_choice}, takes the union of two 
transition sets; and, \emph{call}, \snippet{seqp_call}, affixes a term from 
the specification (\isa{{\isasymGamma}\ pn}).
The rules for both are shown in \reffig{seqp:sos}.

We introduce the specification of a simple `toy' protocol as a running 
example:
\begin{center}
\tiny
\mbox{\isa{\gammatoy\ PToy\ \isacharequal\ labelled\ PToy\  \hspace{-0.3pt}% {-{3
{\isacharparenleft}}}
\mbox{%
\renewcommand{\arraystretch}{1.8}%
\newcommand{\ptoy}[1]{\raisebox{.1ex}{%
    \isa{\scriptsize\gray{{\isacharbraceleft}PToy-:{#1}{\isacharbraceright}}}}}%
\begin{tabular}[t]{p{6.0cm}@{\hspace{.5cm}}p{1.5cm}}
\isa{
receive{\isacharparenleft}{\isasymlambda}msg{\isacharprime}\ 
{\isasymxi}{\isachardot}\ {\isasymxi}\ {\isasymlparr}\ msg\ 
{\isacharcolon}{\isacharequal}\ msg{\isacharprime}\ 
{\isasymrparr}{\isacharparenright}{\isachardot}} & \ptoy{0} \\
\isa{
{\isasymlbrakk}{\isasymlambda}{\isasymxi}{\isachardot}\ {\isasymxi}\ 
{\isasymlparr}nhip\ {\isacharcolon}{\isacharequal}\ ip\ 
{\isasymxi}{\isasymrparr}{\isasymrbrakk}} & \ptoy{1} \\
\isa{
{\isacharparenleft}\ \ \ 
{\isasymlangle}is{\isacharunderscore}newpkt{\isasymrangle}} & \ptoy{2} \\
\isa{
\ \ \ \ \ {\isasymlbrakk}{\isasymlambda}{\isasymxi}{\isachardot}\ 
{\isasymxi}\ {\isasymlparr}no\ {\isacharcolon}{\isacharequal}\ max\ 
{\isacharparenleft}no\ {\isasymxi}{\isacharparenright}\ 
{\isacharparenleft}num\ 
{\isasymxi}{\isacharparenright}{\isasymrparr}{\isasymrbrakk}} & \ptoy{3} \\
\isa{
\ \ \ \ \ 
broadcast{\isacharparenleft}{\isasymlambda}{\isasymxi}{\isachardot}\ 
pkt{\isacharparenleft}no\ {\isasymxi}{\isacharcomma}\ ip\ 
{\isasymxi}{\isacharparenright}{\isacharparenright}{\isachardot}\ 
Toy{\isacharparenleft}{\isacharparenright}
} & \ptoy{4,5} \\
\isa{
\ {\isasymoplus}\ {\isasymlangle}is{\isacharunderscore}pkt{\isasymrangle}} & 
\ptoy{2} \\
\isa{
\ \ \ \ \ {\isacharparenleft}
\ {\isasymlangle}%
{\isasymlambda}{\isasymxi}{\isachardot}\ %
if\ %
num\ {\isasymxi}\ {\isasymge}\ no\ {\isasymxi}%
\ then\ \isacharbraceleft\isasymxi\isacharbraceright\ else\ \isacharbraceleft\isacharbraceright%
{\isasymrangle}} & \ptoy{6} \\
\isa{
\ \ \ \ \ \ \ \ \ {\isasymlbrakk}{\isasymlambda}{\isasymxi}{\isachardot}\ 
{\isasymxi}\ {\isasymlparr}no\ {\isacharcolon}{\isacharequal}\ num\ 
{\isasymxi}{\isasymrparr}{\isasymrbrakk}} & \ptoy{7} \\
\isa{
\ \ \ \ \ \ \ \ \ {\isasymlbrakk}{\isasymlambda}{\isasymxi}{\isachardot}\ 
{\isasymxi}\ {\isasymlparr}nhip\ {\isacharcolon}{\isacharequal}\ sip\ 
{\isasymxi}{\isasymrparr}{\isasymrbrakk}} & \ptoy{8} \\
\isa{
\ \ \ \ \ \ \ \ \ \ \ 
broadcast{\isacharparenleft}{\isasymlambda}{\isasymxi}{\isachardot}\ 
pkt{\isacharparenleft}no\ {\isasymxi}{\isacharcomma}\ ip\ 
{\isasymxi}{\isacharparenright}{\isacharparenright}{\isachardot}\ 
Toy{\isacharparenleft}{\isacharparenright}} & \ptoy{9,10} \\
\isa{
\ \ \ \ \ {\isasymoplus}\ {\isasymlangle}%
{\isasymlambda}{\isasymxi}{\isachardot}\ %
if\ %
num\ {\isasymxi}\ {\isacharless}\ no\ {\isasymxi}%
\ then\ \isacharbraceleft\isasymxi\isacharbraceright\ else\ \isacharbraceleft\isacharbraceright%
{\isasymrangle}} & \ptoy{6} \\
\isa{
\ \ \ \ \ \ \ \ \ Toy{\isacharparenleft}{\isacharparenright}%
{\isacharparenright}%
{\isacharparenright}{\isacharparenright}\ ,} & \ptoy{11}
\end{tabular}} %}-}3
\end{center}
\noindent
where \isa{PToy} is the process name, \isa{is-newpkt} and \isa{is-pkt} are 
guards that unpack the contents of \isa{msg},
and \isa{Toy()} is an abbreviation that clears some variables before a 
\isa{call(PToy)}.
The function \isa{labelled} associates its argument \isa{PToy} paired with a 
number to every prefix constructor.
There are two types of messages: \snippet{newpkt}, from which 
\isa{is-newpkt} copies \isa{data} to the variable \isa{num}, and 
\snippet{pkt}, from which \isa{is-pkt} copies \isa{data} into \isa{num} and 
\isa{src} into \isa{sip}.

The corresponding sequential model is an automaton---a record\footnote{The 
generic record has type \snippet{automaton_type}, where the type \isa{'s} is 
the domain of states, here pairs of data records and control terms, and 
\isa{'a} is the domain of actions.} of two fields: a set of initial states 
and a set of transitions---parameterized by an address \isa{i}:
\begin{tightcenter}
\vspace{-1mm}
\snippet{ptoy_lhs} \isa{\isacharequal} \snippet{ptoy_rhs}\puncgap,
\end{tightcenter}
where \isa{toy-init\ i} yields the initial data state \snippet{toy_init}.
The last three variables are initialized to arbitrary values, as they 
are considered local---they are explicitly reinitialized before each 
\isa{call(PToy)}.
This is the biggest departure from the original definition of \ac{AWN}; it 
simplifies the treatment of \isa{call}, as we show in \refsec{cterms}, and 
facilitates working with automata where variable locality makes little sense.

%}-}2
\subsection{Local parallel composition} \label{sec:awn:par} %{-{2
\begin{figure}[b]
\vspace{-6mm}
\tiny
    \begin{mathpar}
        \msnippet{parleft} \and
        \msnippet{parright} \and
        \msnippet{parboth}
    \end{mathpar}
\vspace*{-3.7ex}
%}-}3
\caption{\acworkaround{SOS} rules for parallel processes: 
\parpsos.\label{fig:parp}}
\end{figure}

Message sending protocols must nearly always be input-enabled, that is, 
nodes should always be in a state where they can receive messages.
To achieve this, and to model asynchronous
message transmission, the protocol process is combined with a queue model, \isa{qmsg},
\pagebreak[2]
that continually appends received messages onto an internal list and offers 
to send the head message to the protocol process:\\
\snippet{ptoy_qmsg_term}.
The \emph{local parallel} operator is a function over automata:
\begin{tightcenter}
\snippet{par_comp}\puncgap.
\end{tightcenter}
The rules for \parpsos{} are shown in~\reffig{parp}.

%}-}2
\subsection{Nodes} \label{sec:awn:node} %{-{2

\begin{figure}[t]
%{-{3
\tiny
    \begin{mathpar}
        \msnippet{node_gcast} \and
        \msnippet{node_receive} \and
        \msnippet{node_arrive} \and
        \msnippet{node_connect1}
    \end{mathpar}
    \vspace*{-5ex}
%}-}3
\caption{\acworkaround{SOS} rules for nodes: examples from 
\nodesos.\label{fig:node}}
 \vspace*{-3ex}
\end{figure}

At the node level, a  local process~\isa{np} is wrapped in a layer that 
records its address~\isa{i} and tracks the set of neighbouring node 
addresses, initially~\isa{\Ri}:
\begin{tightcenter}
\snippet{node_comp}\puncgap.
\end{tightcenter}
Node states are denoted \snippet{net_state_nodes}.
\reffig{node} presents rules typical of \nodesos.
Output network synchronizations, like \isa{groupcast}, are filtered by the 
list of neighbours to become \isa{*cast} actions.
The \snippet{act_arrive} action---in \reffig{node} instantiated as 
\isa{{\isasymemptyset}{\isasymnot}{\isacharbraceleft}i{\isacharbraceright}{\isacharcolon}arrive{\isacharparenleft}m{\isacharparenright}}, 
and 
\isa{{\isacharbraceleft}i{\isacharbraceright}{\isasymnot}{\isasymemptyset}{\isacharcolon}arrive{\isacharparenleft}m{\isacharparenright}}---is 
used to model a message \isa{m} received by nodes in \isa{H} and not by 
those in \isa{K}.
The \snippet{act_connect} adds node \isa{i'} to the set of neighbours of 
node \isa{i}; \snippet{act_disconnect} works similarly.

%}-}2
\subsection{Partial networks} \label{sec:awn:pnet} %{-{2

Partial networks are specified as values of type \snippet{net_tree}, that 
is, as a node \snippet{pnet_node_term} with address \isa{i} and a set of 
initial neighbours \isa{\Ri}, or a composition of two \snippet{net_tree}s 
\snippet{pnet_par_term}.
The function \isa{pnet} maps such a value, together with the 
process~\isa{np\ i} to execute at each node~\isa{i}, here parameterized by 
an address, to an automaton:
\begin{tightcenter}
\begin{tabular}{lcl}
\snippet{pnet1_lhs} & \isa{\isacharequal} & \snippet{pnet1_rhs}\\
\snippet{pnet2_lhs} & \isa{\isacharequal} & \snippet{pnet2_rhs}\puncgap,
\end{tabular}
\end{tightcenter}
The states of such automata mirror the tree structure of the network term; 
we denote composed states \snippet{net_state_subnets}.
This structure, and the node addresses, remain constant during an execution.
These definitions suffice to model an example three node network of toy 
processes:
\begin{tightcenter}
\snippet{eg1_pnet}\puncgap.
\end{tightcenter}

\reffig{pnet} presents rules typical of \pnetsos.
There are rules where only one node acts, like the one shown for 
\isa{\isasymtau}, and rules where all nodes act, like those for \isa{*cast} 
and \isa{arrive}.
The latter ensure---since \isa{qmsg} is always ready to 
\snippet{act_receive}---that a partial network can always perform an 
\snippet{act_arrive} for any combination of \isa{H} and \isa{K} consistent 
with its node addresses, but that pairing with an \snippet{act_cast} restricts 
the possibilities to the one consistent with the destinations in \isa{R}.

\begin{figure}[t]
%{-{3
    \begin{mathpar}
        \msnippet{pnet_cast1} \and
        \msnippet{pnet_arrive} \hfill
        \msnippet{pnet_tau1}
    \end{mathpar}
    \vspace*{-2mm}
%}-}3
\caption{\acworkaround{SOS} rules for partial networks: examples from 
\pnetsos.\label{fig:pnet}}
\end{figure}

%}-}2
\subsection{Complete networks} \label{sec:awn:cnet} %{-{2

The last layer closes a network to further interactions with an environment; 
the \isa{*cast} action becomes a \isa{\isasymtau} and \snippet{act_arrive} 
is forbidden:
\begin{tightcenter}
\snippet{closed'} \isa{\isacharequal} \snippet{closed_term}\puncgap.
\end{tightcenter}
The rules for \cnetsos{} are straight-forward and not presented here.

%}-}2
%--   }-}1%%%%%%%%%%%%%%%%%%%%%%%%%%%%%%%%%%%%%%%%%%%%%%%%%%%%%%%%%%%%
\section{Basic invariance}\label{sec:proof-base} %{-{1
% {-{2

This paper only considers proofs of invariance, that is, properties of 
reachable states.
The basic definitions are classic~\cite[Part III]{Muller:PhD:1998}.

\begin{definition}[reachability]\label{def:reachable}
Given an automaton~\isa{A} and an assumption~\isa{I} over actions, 
\isa{reachable\ A\ I} is the smallest set defined by the rules:
\vspace{-1ex}
\begin{mathpar}
\msnippet{reachable_init}
\and
\msnippet{reachable_step}
\end{mathpar}
\end{definition}

\begin{definition}[invariance]\label{def:invariant}
Given an automaton~\isa{A} and an assumption~\isa{I}, a predicate~\isa{P} is 
\emph{invariant}, denoted \snippet{invariant_lhs}, iff 
\snippet{invariant_rhs}.
\end{definition}

\noindent
We state reachability relative to an assumption on (input) actions \isa{I}.
When \isa{I} is \snippet{TT_rhs}, we write simply 
\snippet{invariant_TT_lhs}.

\begin{definition}[step invariance]\label{def:step-invariant}
Given an automaton~\isa{A} and an assumption~\isa{I}, a predicate~\isa{P} is 
\emph{step invariant}, denoted \snippet{step_invariant_lhs}, iff\\
\centerline{\snippet{step_invariant_rhs}\puncgap.}
\end{definition}

Our invariance proofs follow the compositional strategy recommended 
in~\cite[\textsection 
1.6.2]{deRoeverEtAl:ConcVer:2001}.
That is, we show properties of sequential process automata using the 
induction principle of~\refdef{reachable}, and then apply 
generic proof rules to successively lift such properties over each of the 
other layers.
The inductive assertion method, as stated in rule \textsc{inv-b} 
of~\cite{MannaPnu:Safety}, requires a finite set of transition schemas, 
which, together with the obligation on initial states yields a set of sufficient 
verification conditions.
We develop this set in \refsec{cterms} and use it to derive the main proof 
rule presented in \refsec{showinv} together with some examples.

%---  }-}2------------------------------------------------------------
\subsection{Control terms}\label{sec:cterms} %{-{2

Given a specification \isa{\isasymGamma} over finitely many process 
names, we can 
generate a finite set of verification conditions because transitions from 
\isa{('s, 'p, 'l) seqp} terms always yield subterms of terms in 
\isa{\isasymGamma}.
But, rather than simply consider the set of all subterms, we prefer to 
define a subset of `control terms' that reduces the number of verification 
conditions, avoids tedious duplication in proofs, and corresponds with the 
obligations considered in pen-and-paper proofs.
The main idea is that the \isa{\isasymoplus} and \isa{call} operators serve 
only to combine process terms: they are, in a sense, executed recursively by 
\seqpsos{} to determine the actions that a term offers to its environment.
This is made precise by defining a relation between sequential process 
terms.

\begin{definition}[{\microstep{\isasymGamma}}]\label{def:microsteps}
For a (recursive) specification \isa{\isasymGamma}, let 
\microstep{\isasymGamma} be the smallest relation such that
\snippet{microstep_choiceI1},
\snippet{microstep_choiceI2}, and
\snippet{microstep_callI}.
\end{definition}

\noindent
We write \microsteprtcl{\isasymGamma} for its reflexive transitive closure.
We consider a specification to be \emph{well formed}, when the inverse of 
this relation is well founded:
\begin{tightcenter}
\snippet{wellformed}\puncgap.
\end{tightcenter}
Most of our lemmas only apply to well formed specifications, since otherwise 
functions over the terms they contain cannot be guaranteed to terminate.
Neither of these two specifications is well formed:
\isa{\isasymGamma\isactrlisub{a}(1)\ \isacharequal\ p\ {\isasymoplus}\ 
call(1)};
\isa{\isasymGamma\isactrlisub{b}(n)\ \isacharequal\ call(n + 1)}.

We will also need a set of `start terms'---the subterms that can act 
directly.

\begin{definition}[sterms]\label{def:sterms}
Given a \snippet{wellformed_gamma} and a sequential process term \isa{p}, 
\snippet{sterms_other_lhs} is the set of maximal elements related to \isa{p} 
by the reflexive transitive closure of the \microstep{\isasymGamma} 
relation\footnote{This characterization is equivalent to 
\snippet{sterms_maximal_microstep_rhs}.
Termination follows from \snippet{wellformed_gamma}, that is,
\snippet{sterms_termination} for all \isa{p}.}:

\begin{tabular}{lcl}
  \snippet{sterms_choice_concl_lhs} &\isa{\isacharequal}& \snippet{sterms_choice_concl_rhs}, \\
  \snippet{sterms_call_concl_lhs}   &\isa{\isacharequal}& \snippet{sterms_call_concl_rhs}, and,\\
  \snippet{sterms_other_lhs} &\isa{\isacharequal}& 
  \snippet{sterms_other_rhs} otherwise\puncgap.
\end{tabular}
\end{definition}

\noindent
We also define `local start terms' by
\snippet{stermsl_choice} and otherwise 
\snippet{stermsl_other}
to permit the sufficient syntactic condition that
a specification \isa{\isasymGamma} is well formed if 
\snippet{wf_no_direct_calls_prem_1}.

Similarly to the way that start terms act as direct sources of transitions, 
we define `derivative terms' giving possible active destinations of 
transitions.

\begin{definition}[dterms]\label{def:dterms}
Given a \snippet{wellformed_gamma} and a sequential process term \isa{p}, 
\isa{dterms p} is defined by:

\begin{tabular}{lcl}
\snippet{dterms_choice_concl_lhs}  &\isa{\isacharequal}& \snippet{dterms_choice_concl_rhs}, \\
\snippet{dterms_call_concl_lhs}    &\isa{\isacharequal}& \snippet{dterms_call_concl_rhs}, \\
\snippet{dterms_other2_concl_lhs}  &\isa{\isacharequal}& \snippet{dterms_other2_concl_rhs}, \\
\snippet{dterms_unicast_concl_lhs} &\isa{\isacharequal}& \snippet{dterms_unicast_concl_rhs},
and so on.
\end{tabular}
\end{definition}

\noindent
These derivative terms overapproximate the set of reachable \isa{sterms}, 
since they do not consider the truth of guards nor the willingness of 
communication partners. 

These auxiliary definitions lead to a succinct definition of the set of 
control terms of a specification.

\begin{definition}[cterms]\label{def:cterms}
For a specification $\Gamma$, \isa{cterms} is the smallest set where:
\vspace{-1ex}
\begin{mathpar}
\msnippet{cterms_SI}
\and
\msnippet{cterms_DI}
\and
\end{mathpar}
\end{definition}

\noindent
It is also useful to define a local version independent of any 
specification.

\begin{definition}[ctermsl]\label{def:ctermsl}
Let \isa{ctermsl} be the smallest set defined by:

\begin{tabular}{lcl}
\snippet{ctermsl_choice_lhs} &\isa{\isacharequal}& \snippet{ctermsl_choice_rhs}, \\
\snippet{ctermsl_call_lhs}   &\isa{\isacharequal}& \snippet{ctermsl_call_rhs}, \\
\snippet{ctermsl_other2_lhs} &\isa{\isacharequal}& \snippet{ctermsl_other2_rhs}, and so on. \\
\end{tabular}
\end{definition}

\noindent
Including \isa{call} terms ensures that \snippet{stermsl_ctermsl_prem} 
implies \snippet{stermsl_ctermsl_concl}, which facilitates proofs.
For \snippet{wellformed_gamma}, \isa{ctermsl} allows an alternative 
definition of \isa{cterms},
\begin{equation}\label{eq:cterms_def'}
\mbox{\snippet{cterms_def'_concl}\puncgap.}
\end{equation}
While the original definition is convenient for developing the meta-theory, 
due to the accompanying induction principle, this one is more useful for 
systematically generating the set of control terms of a specification, and 
thus, we will see, sets of verification conditions.
And, for \snippet{wellformed_gamma}, we have as a corollary
\begin{equation}\label{eq:cterms_subterms}
\mbox{\snippet{cterms_subterms_concl}\puncgap,}
\end{equation}
where \isa{subterms}, \isa{not-call}, and \isa{not-choice} are defined in 
the obvious way.

We show that \isa{cterms} over-approximates the set of reachable control 
states.

\begin{lemma}\label{thm:seq_reachable_in_cterms}
For
\snippet{seq_reachable_in_cterms_prem_1} and
automaton~\isa{A}
where
\snippet{seq_reachable_in_cterms_prem_2}
and
\snippet{seq_reachable_in_cterms_prem_3},
if \snippet{seq_reachable_in_cterms_prem_4} and 
\snippet{seq_reachable_in_cterms_prem_5} then
\snippet{seq_reachable_in_cterms_concl}.
\end{lemma}

\noindent
The predicate \snippet{control_within} serves to state that the initial 
control state is within the specification.
%---  }-}2------------------------------------------------------------
\subsection{Basic proof rule and invariants}\label{sec:showinv} %{-{2

Using the definition of invariance (\refdef{invariant}), we can state a 
basic property of an instance of the toy process:
\begin{equation}\label{eq:basicinv}
\mbox{\snippet{nhip_eq_ip}\puncgap,}
\end{equation}
This invariant states that between the lines labelled \isa{PToy-:2} and 
\isa{PToy-:8},
that is,
after the assignment of \isa{PToy-:1} until before the assignment of 
\isa{PToy-:8},
the values of \isa{nhip} and \isa{ip} are equal;
\snippet{onl_lhs}, defined as \snippet{onl_rhs}, extracts labels from 
control states.%
\footnote{Using labels in this way is standard, see, for 
instance,~\cite[Chap. 1]{MannaPnu:Safety}, or the `assertion networks' of
~\cite[\textsection 2.5.1]{deRoeverEtAl:ConcVer:2001}.
Isabelle rapidly dispatches all the uninteresting cases.}
Invariants like these are solved using a procedure whose soundness is 
justified as a theorem.
The proof exploits \refeq{cterms_def'} and \reflem{seq_reachable_in_cterms}.

\begin{theorem}\label{thm:seq_invariant_ctermsI} %{-{3
To prove \snippet{seq_invariant_ctermsI_concl}, where
\snippet{seq_invariant_ctermsI_prem_1},
\snippet{seq_invariant_ctermsI_prem_3},
\snippet{seq_invariant_ctermsI_prem_2}, and
\snippet{seq_invariant_ctermsI_prem_4}, it suffices
\begin{description}[leftmargin=4.1em,style=nextline]

\item[(init)]
for arbitrary
\snippet{seq_invariant_ctermsI_init_1} and
\snippet{seq_invariant_ctermsI_init_2}, to show
\snippet{seq_invariant_ctermsI_init_3}, and,

\item[(step)]
for arbitrary
\snippet{seq_invariant_ctermsI_trans_1}, but 
\snippet{seq_invariant_ctermsI_trans_2}, and
\snippet{seq_invariant_ctermsI_trans_3},
given that
\snippet{seq_invariant_ctermsI_trans_8} for some
\snippet{seq_invariant_ctermsI_trans_7},
to assume
\snippet{seq_invariant_ctermsI_trans_4} and
\snippet{seq_invariant_ctermsI_trans_9},
and then for any \isa{(\isasymxi\isacharprime,\ q)} such that
\snippet{seq_invariant_ctermsI_trans_5}
and
\snippet{seq_invariant_ctermsI_trans_6},
to show
\snippet{seq_invariant_ctermsI_trans_10}.
\end{description}
\end{theorem} %}-}3

\noindent
Here, \snippet{simple_labels}: each control term must have exactly one 
label, that is, \isa{\isasymoplus} terms must be labelled consistently.

We incorporate this theorem into a 
tactic that \begin{inparaenum}

\item
applies the introduction rule,

\item
replaces \snippet{seq_invariant_ctermsI_trans_1} by a disjunction over the 
values of \isa{pn},

\item
applies \refdef{ctermsl} and repeated simplifications of \isa{\isasymGamma}s 
and eliminations on disjunctions to generate one subgoal (verification 
condition) for each control term,

\item \label{step:elimder}
replaces control term derivatives, the subterms in \refdef{dterms}, by fresh 
variables, and, finally,

\item
tries to solve each subgoal by simplification.

\end{inparaenum}
\refstep{elimder} replaces potentially large control terms by their 
(labelled) heads, which is important for readability and prover performance.
The tactic takes as arguments a list of existing invariants to include after 
having applied the introduction rule and a list of lemmas for trying to 
solve any subgoals that survive the final simplification.
There are no schematic variables in the subgoals and we benefit greatly from
Isabelle's \textsc{parallel\_goals} tactical~\cite{Wenzel:ParITP:2013}.

In practice, one states an invariant, applies the tactic, and examines the 
resulting goals.
One may need new lemmas for functions over the data state or explicit proofs 
for difficult goals.
That said, the tactic generally dispatches the uninteresting goals, and 
the remaining ones typically correspond with the cases treated explicitly in 
manual proofs~\cite{LICS14}.

For step invariants, we show a counterpart to 
\refthm{seq_invariant_ctermsI}, and declare it to the tactic.
Then we can show, for our example, that the value of \isa{no} never 
decreases:
\begin{tightcenter}
\snippet{seq_nos_increases'}\puncgap.
\end{tightcenter}

%---  }-}2------------------------------------------------------------
%--   }-}1%%%%%%%%%%%%%%%%%%%%%%%%%%%%%%%%%%%%%%%%%%%%%%%%%%%%%%%%%%%%
\section{Open invariance}\label{sec:proof-comp} %{-{1
% {-{2

The analysis of network protocols often requires `inter-node' invariants, 
like
\begin{multline}\label{eq:bigger_than_next}
\vspace{-2pt}\snippet{bigger_than_next}\puncgap,
\end{multline}
\vspace{-2pt}

\vskip-1.16em
\noindent%
which states that, for any \isa{net-tree} with disjoint node addresses 
(\isa{wf-net-tree\ n}), the value of \isa{no} at a node is never greater 
than its value at the `next hop'---the address in \isa{nhip}.
This is a property of a global state \isa{\isasymsigma} mapping addresses to 
corresponding data states.
Such a global state is readily constructed with:
\begin{tightcenter}
\snippet{netglobal},

\snippet{default}, and
\medskip

\begin{tabular}{lcl}
\snippet{netlift1_lhs} & \isa{\isacharequal} & \snippet{netlift1_rhs} \\
\snippet{netlift2_lhs} & \isa{\isacharequal} & 
\snippet{netlift2_rhs}\puncgap.
\end{tabular}
\end{tightcenter}
\pagebreak

\noindent
The applications of \isa{fst} elide the state of \isa{qmsg} and the 
protocol's control state.\footnote{The formulation here is a technical 
detail: \isa{sr} corresponds to \isa{netlift} as \isa{np} does to 
\isa{pnet}.}

While we can readily state inter-node invariants of a complete model, 
showing them compositionally is another issue.
\refsecs{omodel}{oinv} present a way to state and prove such invariants at 
the level of sequential processes---that is, with only~\isa{ptoy\ i} left 
of the turnstile.
\refsecs{lift}{transfer} present, respectively, rules for lifting such 
results to network models and for recovering invariants 
like~\refeq{bigger_than_next}.

%---  }-}2------------------------------------------------------------
\subsection{The open model}\label{sec:omodel} %{-{2

Rather than instantiate the \isa{'s} of \isa{('s, 'p, 'l) seqp} with 
elements \isa{\isasymxi} of type \isa{state}, our solution introduces a 
global state \isa{\isasymsigma} of type \snippet{sigma_type}.
This necessitates a stack of new \ac{SOS} rules that we call the \emph{open 
model}; \reffig{open} shows some representatives. 

The rules of \oseqpsos{} are parameterized by an address \isa{i} and 
constrain only that entry of the global state, either to say how it changes 
(\snippet{oassignT_prem_1}) or that it does not 
(\snippet{ounicastT_prem_1}).
The rules for \oparpsos{} only allow the first sub-process to constrain 
\isa{\isasymsigma}.
This choice is disputable: it precludes comparing the states of \isa{qmsg}s 
(and any other local filters) across a network, but is also simplifies the 
mechanics and use of this layer of the framework.\footnote{The treatment of 
the other layers is completely independent of this choice.}
The sets \onodesos{} and \opnetsos{} need not be parameterized since they 
are generated inductively from lower layers.
Together they constrain subsets of elements of \isa{\isasymsigma}.
This occurs naturally for rules like those for \isa{arrive} and \isa{*cast}, 
where the synchronous communication serves as a conjunction of constraints 
on sub-ranges of \isa{\isasymsigma}. But for others that normally only 
constrain a single element, like those for \isa{\isasymtau}, assumptions 
(\snippet{onode_tau_prem_2}) are introduced here and later dispatched 
(\refsec{transfer}).
The rules for \ocnetsos{}, not shown, are similar---elements not addressed 
within a model may not change.

The stack of operators and model layers described in \refsec{awn} is 
refashioned to use the new transition rules and to distinguish the global 
state, which is preserved as the \isa{fst} element across layers, from the 
local state elements which are combined in the \isa{snd} element as before.

For instance, a sequential instance of the toy protocol is defined as
\begin{tightcenter}
\snippet{optoy_lhs} \isa{\isacharequal} \snippet{optoy_rhs}\puncgap,
\end{tightcenter}
combined with the standard \isa{qmsg} process using the operator
\vspace{-1mm}
\begin{align*}
\snippet{opar_comp'}\mbox{\ ,}
\end{align*}
\vspace*{-5mm}

\noindent and lifted to the node level via the open node constructor
\begin{tightcenter}
\snippet{onode_comp}\puncgap.
\end{tightcenter}
Similarly, to map a \snippet{net_tree} term to an open model we define:

\smallskip
\begin{tabular}{@{\hspace{-5.1mm}}l@{\,}c@{\,}l}
\centering
\snippet{opnet1_lhs} & \isa{\isacharequal} & \snippet{opnet1_rhs} \\
\snippet{opnet2_lhs} & \isa{\isacharequal} & \snippet{opnet2_rhs}\puncgap.
\end{tabular}

This definition is non-empty only for well-formed \snippet{net_tree}s 
(\isa{net-ips} gives the set of node addresses in the state of a 
partial network).
Including such a constraint within the open model, rather than as a separate 
assumption like
the \isa{wf-net-tree\ n} in \refeq{bigger_than_next},
eliminates an annoying technicality from the inductions described in 
\refsec{lift}.
As with the extra premises in the open \ac{SOS} rules, we can freely adjust 
the open model to facilitate proofs but each `encoded assumption' 
becomes an obligation to be discharged in the transfer lemma of 
\refsec{transfer}.

\begin{figure}[t]
\subfloat[Sequential processes: examples from \oseqpsos.\label{fig:oseqp:sos}]{ %{-{3
\tiny
    \hspace{-2.5pt}\begin{mathpar}
        \msnippet{oassignT}\hfill
        \msnippet{ochoiceT1} \\
        \msnippet{ounicastT} 
    \end{mathpar}\hspace{-2.5pt}
} %}-}3

\subfloat[Parallel processes: example from \oparpsos.\label{fig:oparp:sos}]{ %{-{3
\tiny
     \hspace{-2.5pt}\begin{mathpar}
        \msnippet{oparboth}
    \end{mathpar} \hspace{-2.5pt}
} %}-}3

\subfloat[Nodes: examples from \onodesos.\label{fig:onode:sos}]{ %{-{3
\tiny
     \hspace{-2.5pt}\begin{mathpar}
        \msnippet{onode_receive}\hfill
         \msnippet{onode_tau}
    \end{mathpar} \hspace{-2.5pt}
} %}-}3

\subfloat[Partial networks: example from \opnetsos.\label{fig:opnet:sos}]{ %{-{3
\tiny
    \hspace{-2.5pt}\begin{mathpar}
        \msnippet{opnet_arrive}
    \end{mathpar}\hspace{-2.5pt}
} %}-}3
\vspace{-1mm}
\caption{\ac{SOS} rules for the open model 
(cf.~\reffigsss{seqp}{parp}{node}{pnet})\label{fig:open}}
\vspace{-3mm}
\end{figure}

An operator for adding the last layer is also readily defined by
\begin{tightcenter}
\snippet{oclosed'} \isa{\isacharequal} \snippet{oclosed_term}\puncgap,
\end{tightcenter}
giving all the definitions necessary to turn a standard model into an open 
one.

%---  }-}2------------------------------------------------------------
\subsection{Open invariants}\label{sec:oinv} %{-{2

The basic definitions of reachability and invariance, 
\refdefr{reachable}{step-invariant}, apply to open models, but constructing 
a compositional proof requires considering the effects of both synchronized 
and interleaved actions of possible environments.

\begin{definition}[open reachability]\label{def:oreachable}
Given an automaton~\isa{A} and assumptions~\isa{S} and~\isa{U} over, 
respectively, synchronized and interleaved actions, \isa{oreachable\ A\ S\ 
U} is the smallest set defined by the rules:
\begin{mathpar}
\msnippet{oreachable_init}
\and
\msnippet{oreachable_other'}
\\\vspace{-3mm}
\msnippet{oreachable_local'}
\end{mathpar}
\end{definition}

\pagebreak

\noindent
In practice, we use restricted forms of the assumptions \isa{S} and \isa{U}, 
respectively,
\begin{align}
\mbox{\snippet{otherwith_def}}\puncgap,\label{eq:otherwith}\\
\mbox{\snippet{other_def}}\puncgap.\label{eq:other}
\end{align}
The former permits the restriction of possible environments (\isa{E}) and 
also the extraction of information from shared actions (\isa{I}).
The latter restricts (\isa{F}) the effects of interleaved actions, which may 
only change non-local state elements.%

\begin{definition}[open invariance]\label{def:oinvariant}
Given an automaton~\isa{A} and assumptions~\isa{S} and~\isa{U} over, 
respectively, synchronized and interleaved actions, a predicate~\isa{P} is 
an \emph{open invariant}, denoted \snippet{oinvariant_lhs}, iff 
\snippet{oinvariant_rhs}.
\end{definition}

\noindent
It follows easily that existing invariants can be made open: most invariants 
can be shown in the basic context but still exploited in the more 
complicated one.

\begin{lemma}\label{thm:open_seq_invariant}
Given an invariant
\snippet{open_seq_invariant_prem_1}
where \snippet{open_seq_invariant_prem_4},
and any \isa{F},
there is an open invariant
\snippet{open_seq_invariant_concl'}
where \snippet{open_seq_invariant_prem_3},
provided that
\snippet{open_seq_invariant_prem_2'}.
\end{lemma}

\noindent
Open step invariance and a similar transfer lemma are defined similarly.
The meta theory for basic invariants is also readily adapted, in particular,

\begin{theorem}\label{thm:oseq_invariant_ctermsI} %{-{3
To show \snippet{oseq_invariant_ctermsI_concl}, in addition to the 
conditions and the obligations \textbf{(init)} and \textbf{(step)} of 
\refthm{seq_invariant_ctermsI}, suitably adjusted, it suffices,
\begin{description}[leftmargin=4.1em,style=nextline]

\item[(env)]
for arbitrary \snippet{oseq_invariant_ctermsI_trans_1}
and \snippet{oseq_invariant_ctermsI_trans_2},
to assume both
\snippet{oseq_invariant_ctermsI_trans_3} and
\snippet{oseq_invariant_ctermsI_trans_4},
and then to show
\snippet{oseq_invariant_ctermsI_trans_5}.
\end{description}
\end{theorem} %}-}3

\noindent
This theorem is declared to the tactic described in \refsec{showinv} and 
proofs proceed as before, but with the new obligation to show invariance 
over interleaved steps.

We finally have sufficient machinery to state (and prove) Invariant 
\refeq{bigger_than_next} at the level of a sequential process:
\begin{equation}\label{eq:oseq_bigger_than_next}
\begin{tabular}{l}
\snippet{oseq_bigger_than_next}\mbox{ ,}
\end{tabular}
\end{equation}
where \snippet{nos_increase},
\isa{orecvmsg} applies its given predicate to \isa{receive} actions and 
is otherwise true,
\snippet{msg_num_ok_pkt}, and \snippet{msg_num_ok_newpkt2}.
So, given that the variables \isa{no} in the environment never decrease and that 
incoming \isa{pkt}s reflect the state of the sender, there is a relation 
between the local node and the next hop.
Similar invariants occur in proofs of realistic protocols~\cite{LICS14}.

%---  }-}2------------------------------------------------------------
\subsection{Lifting open invariants}\label{sec:lift} %{-{2

The next step is to lift Invariant \refeq{oseq_bigger_than_next} over each 
composition operator of the open model.
We mostly present the lemmas over \isa{oreachable}, rather than those for 
open invariants and step invariants, which follow more or less directly.

The first lifting rule treats composition with the \isa{qmsg} process.
It mixes \isa{oreachable} and \isa{reachable} predicates: the former for the 
automaton being lifted, the latter for properties of \isa{qmsg}.
The properties of \isa{qmsg}---only received messages are added to the queue 
and sent messages come from the queue---are shown using the techniques of 
\refsec{proof-base}.

\begin{lemma}[qmsg lifting]\label{thm:par_qmsg_oreachable} %{-{3
Given
\snippet{par_qmsg_oreachable_prem_1},
where predicates
\snippet{par_qmsg_oreachable_L} and
\snippet{par_qmsg_oreachable_E}, and provided
\begin{inparaenum}
\item
\snippet{par_qmsg_oreachable_prem_2},\label{eq:qmsglift:F}
\item
for all
\isa{\isasymxi,\ \isasymxi\isacharprime},
\snippet{par_qmsg_oreachable_Exixi'} implies
\snippet{par_qmsg_oreachable_Fxixi'},\label{eq:qmsglift:EF}
\item
for all
\isa{\isasymsigma,\ \isasymsigma',\ m},
\snippet{par_qmsg_oreachable_Fsigmaj}
and
\snippet{par_qmsg_oreachable_Rsigmam}
imply
\snippet{par_qmsg_oreachable_Rsigma'm}, and,\label{eq:qmsglift:RR}
\item
\isa{F} is reflexive,\label{eq:qmsglift:Frefl}
\end{inparaenum}
then
\snippet{par_qmsg_oreachable_conj1} and
\snippet{par_qmsg_oreachable_conj2}, and furthermore
\snippet{par_qmsg_oreachable_conj3}.
\end{lemma} %}-}3

\noindent
The key intuition is that every message \isa{m} received, queued, and sent 
by \isa{qmsg} satisfies \isa{R\ \isasymsigma\ m}.
The proof is by induction over \isa{oreachable}.
The \isa{R}'s are preserved when the external environment acts independently 
\refeqs{qmsglift:RR}{qmsglift:Frefl}, when it acts synchronously 
\refeq{qmsglift:EF}, and when the local process acts 
\refeqs{qmsglift:F}{qmsglift:RR}.

The rule for lifting to the node level adapts assumptions on \isa{receive} 
actions (\isa{orecvmsg}) to \isa{arrive} actions (\isa{oarrivemsg}).

\begin{lemma}[onode lifting]\label{thm:node_proc_reachable} %{-{3
If, for all \isa{\isasymxi} and \isa{\isasymxi\isacharprime},
\snippet{node_proc_reachable_prem_2_prem}
implies
\snippet{node_proc_reachable_prem_2_concl},
then given
\snippet{node_proc_reachable_prem_1}
it follows that
\snippet{node_proc_reachable_concl}.
\end{lemma} %}-}3

\noindent
The sole condition is needed because certain node-level actions---namely 
\isa{connect}, \isa{disconnect}, and \snippet{act_not_arrive}---synchronize 
with the environment (giving \isa{E\ \isasymxi\ \isasymxi'}) but appear to 
`stutter' (requiring \isa{F\ \isasymxi\ \isasymxi'}) relative to the 
underlying process.

The lifting rule for partial networks is the most demanding.
The function \snippet{net_tree_ips}, giving the set of addresses in a 
\snippet{net_tree}, plays a key role.

\begin{lemma}[opnet lifting]\label{thm:subnet_oreachable} %{-{3
Given
\snippet{subnet_oreachable_prem_1},
where
\snippet{subnet_oreachable_prem_1_L},
\snippet{subnet_oreachable_prem_1_E},
and
\isa{E} and \isa{F} are reflexive,
for arbitrary \isa{p\ i} of the form
\snippet{subnet_oreachable_prem_4_p'},
\snippet{subnet_oreachable_prem_4_p},
and similar step invariants for
\snippet{subnet_oreachable_prem_5_pred} and
\snippet{subnet_oreachable_prem_6_pred},
then it follows that both
\snippet{subnet_oreachable_concl_1}
and \snippet{subnet_oreachable_concl_2},
where
\snippet{S1} and \snippet{U1} are over \p{1}, and
\snippet{S2} and \snippet{U2} are over \p{2}.
\end{lemma} %}-}3

\noindent
The proof is by induction over \isa{oreachable}.
The initial and interleaved cases are trivial.
For the local case, given open reachability of \isa{(\isasymsigma, s)} and 
\isa{(\isasymsigma, t)} for \p{1} and \p{2}, respectively, and 
\snippet{subnet_oreachable_proof_step}, we must show open reachability of
\isa{(\isasymsigma\isacharprime, s\isacharprime)} and 
\isa{(\isasymsigma\isacharprime, t\isacharprime)}.
The proof proceeds by cases of \isa{a}.
The key step is to have stated the lemma without introducing cyclic 
dependencies between (synchronizing) assumptions and (step invariant) 
guarantees.
For a synchronizing action like \isa{arrive}, \refdef{oreachable} requires 
satisfaction of \snippet{S1} to advance in \p{1} and of \snippet{S2} to 
advance in \p{2}, but the assumption \isa{S} only holds for addresses 
\snippet{subnet_oreachable_proof_jL}.
This is why the step invariants required of nodes only assume
\snippet{subnet_oreachable_proof_oarrivemsg} of the environment, rather than 
an \isa{S} over node address \isa{\{i\}}.
This is not unduly restrictive since the step invariants provide guarantees 
for individual local state elements and not between network nodes.
The assumption \snippet{subnet_oreachable_proof_oarrivemsg} is never cyclic: 
it is either assumed of the environment for paired \isa{arrive}s, or 
trivially satisfied for the side that \isa{*cast}s.
The step invariants are lifted from nodes to partial networks by induction 
over \snippet{net_tree}s.
For non-synchronizing actions, we exploit the extra guarantees built into 
the open \ac{SOS} rules.

The rule for closed networks is similar to the others.
Its important function is to eliminate the synchronizing assumption (\isa{S} 
in the lemmas above), since messages no longer arrive from the environment.
The conclusion of this rule has the form required by the transfer 
lemma of the next section.

%}-}2
\subsection{Transferring open invariants}\label{sec:transfer} %{-{2

The rules in the last section extend invariants over sequential processes, 
like that of~\refeq{oseq_bigger_than_next}, to arbitrary, open network 
models.
All that remains is to transfer the extended invariants to the standard 
model.
We do so using a locale~\cite{KammullerWenPau:Locales:1999} 
\snippet{pnet_reachable_transfer_prem_1} where \isa{np} has type
\snippet{openproc_np_type},
\isa{onp} has type \snippet{openproc_onp_type},
and \isa{sr} has type \snippet{openproc_sr_type}.
The automata use the actions of \refsec{awn:seq} with arbitrary messages 
(\isa{'m\ seq-action}).

The \isa{openproc} locale relates an automaton \isa{np} to a corresponding 
`open' automaton \isa{onp}, where \isa{sr} splits the states of the former 
into global and local components.
Besides two technical conditions on initial states, this relation 
requires assuming \snippet{openproc_trans_prem_2},
\snippet{openproc_trans_prem_3} and
\snippet{openproc_trans_prem_4},
and then showing
\snippet{openproc_trans_concl}---that is, that \isa{onp} simulates 
\isa{np}.
For our running example, we show \isa{openproc\ ptoy\ optoy\ id}, and then 
lift it to the composition with \isa{qmsg}, using
a generic relation on \isa{openproc} locales.

\begin{lemma}[transfer]\label{thm:close_opnet}
Given \isa{np}, \isa{onp}, and \isa{sr} such that 
\snippet{pnet_reachable_transfer_prem_1}, then for any 
\snippet{pnet_reachable_transfer_prem_2} and 
\snippet{pnet_reachable_transfer_prem_3}, it follows that\\
\snippet{pnet_reachable_transfer_concl}.
\end{lemma}

\noindent
This lemma uses two \isa{openproc} constants:
\snippet{someinit_concl_lhs} chooses an arbitrary initial state from 
\isa{np} (\snippet{someinit_concl_rhs}), and
\begin{tightcenter}
\begin{tabular}{lcl}
\snippet{netliftl1_lhs} & \isa{\isacharequal} & \snippet{netliftl1_rhs} \\
\snippet{netliftl2_lhs} & \isa{\isacharequal} & 
\snippet{netliftl2_rhs}\puncgap. \\[.7em]
\end{tabular}
\end{tightcenter}

The proof of the lemma `discharges' the assumptions incorporated into the 
open \ac{SOS} rules.
An implication from an open invariant on an open model to an invariant on 
the corresponding standard model follows as a corollary.

\paragraph{Summary.}

The technicalities of the lemmas in this and the preceding section are 
essential for the underlying proofs to succeed.
The key idea is that through an open version of \ac{AWN} where automaton 
states are segregated into global and local components, one can reason 
locally about global properties, but still, using the so called transfer and 
lifting results, obtain a result over the original model.

%}-}2
%--   }-}1%%%%%%%%%%%%%%%%%%%%%%%%%%%%%%%%%%%%%%%%%%%%%%%%%%%%%%%%%%%%
\section{Concluding remarks} %{-{1

We present a mechanization of a modelling language for 
\ac{MANET} and \ac{WMN} protocols, including a streamlined adaptation of 
standard theory for showing invariants of individual reactive processes, 
and a novel and compositional framework for lifting such results to network 
models.
The framework allows the statement and proof of inter-node properties.
We think that many elements of our approach would apply to similarly 
structured models in other formalisms.

It is reasonable to ask whether the basic model presented in 
\refsec{awn} could not simply be abandoned in favour of the open model of 
\refsec{omodel}.
But we believe that the basic model is the most natural way of describing 
what \ac{AWN} means, proving semantic properties of the language, showing 
`node-only' invariants, and, potentially, for showing refinement relations.
Having such a reference model allows us to freely incorporate assumptions 
into the open \ac{SOS} rules, knowing that their soundness must later be 
justified.

\medskip
\noindent{\textbf{\em The \ac{AODV} case study.}}
The framework we present in this paper was successfully applied 
in the mechanization of a proof of loop freedom~\cite[\textsection 
7]{FehnkerEtAl:AWN:2013} of the \ac{AODV} protocol~\cite{rfc3561},
a widely-used routing protocol designed for \acp{MANET}, and one of the four 
protocols currently standardized by the IETF \ac{MANET} working group.
The model has about~$100$ control locations across $6$ different processes, and
uses about~$40$ functions to 
manipulate the data state.
The main property (loop freedom) roughly states that `a data packet is never 
sent round in circles without being delivered'.
To establish this property, we proved around 
$400$ lemmas.
Due to the complexity of the protocol logic and the length of the 
proof, we present the 
details elsewhere~\cite{LICS14}.
The case study shows that the presented framework can be applied to 
verification tasks of industrial relevance.

%--   }-}1%%%%%%%%%%%%%%%%%%%%%%%%%%%%%%%%%%%%%%%%%%%%%%%%%%%%%%%%%%%%
\medskip\noindent\emph{Acknowledgments.} %{-{1
We thank G.\ Klein and M.\ Pouzet for support and complaisance, 
and M.\ Daum for participation in discussions.
Isabelle/jEdit~\cite{Wenzel:jEdit:2012}, 
Sledge\-hammer~\cite{BlanchetteBohPau:Sledgehammer:2011}, parallel processing~\cite{Wenzel:ParITP:2013}, and the TPTP 
project~\cite{Sutcliffe:TPTP} were invaluable.

NICTA is funded by the Australian Government through the Department of 
Communications and the Australian Research Council through the ICT Centre of 
Excellence Program. 

\vspace{-1mm}
%--   }-}1%%%%%%%%%%%%%%%%%%%%%%%%%%%%%%%%%%%%%%%%%%%%%%%%%%%%%%%%%%%%
\bibliographystyle{abbrv}
\bibliography{paper}
\end{document}